\documentclass[11pt]{article}
\pdfoutput=1
\usepackage{jheppub}
\usepackage{tikz}
\usetikzlibrary{calc}
\usepackage{graphicx,subcaption}

\newcommand\fft[2]{{\frac{#1}{#2}}}
\newcommand\ft[2]{{\textstyle\frac{#1}{#2}}}
\newcommand\nn{\nonumber}
\renewcommand{\Re}{\operatorname{Re}}
\renewcommand{\Im}{\operatorname{Im}}
\newcommand{\Tr}{\operatorname{Tr}}

\newtheorem{lemma}{Lemma}
\newtheorem{conjecture}{Conjecture}
\begin{document}
%%%%%

\preprint{LCTP-19-32, UUITP-50/19}

%\preprint{}

\title{\boldmath Asymptotic growth of the 4d $\mathcal N=4$ index and partially deconfined phases}

\author[a]{Arash Arabi Ardehali,}
\affiliation[a]{Department of Physics and Astronomy, Uppsala University,\\
Box 516, SE-751 20 Uppsala, Sweden}
\author[b]{Junho Hong,}
\author[b]{and James T. Liu}
\affiliation[b]{Leinweber Center for Theoretical Physics, Randall Laboratory of Physics,\\
The University of Michigan, Ann Arbor, MI 48109-1040, USA}

\emailAdd{ardehali@physics.uu.se}\emailAdd{junhoh@umich.edu}
\emailAdd{jimliu@umich.edu}

\abstract{We study the Cardy-like asymptotics of the 4d $\mathcal N=4$ index and demonstrate the existence of partially deconfined phases where the asymptotic growth of the index is not as rapid as in the fully deconfined case.  We then take the large-$N$ limit after the Cardy-like limit and make a conjecture for the leading asymptotics of the index.  While the Cardy-like behavior is derived using the integral representation of the index, we demonstrate how the same results can be obtained using the Bethe ansatz type approach as well.  In doing so, we discover new non-standard solutions to the elliptic Bethe ansatz equations including continuous families of solutions for $SU(N)$ theory with $N\ge3$.  We argue that the existence of both standard and continuous non-standard solutions has a natural interpretation in terms of vacua of $\mathcal N=1^*$ theory on $\mathbb R^3\times S^1$.}

\maketitle \flushbottom

%%%%%%%%%%%%%%%%%%%%
\section{Introduction}\label{sec:intro}
%%%%%%%%%%%%%%%%%%%%

For the first time it has become possible to study a
black hole-counting index \cite{Romelsberger:2005eg,Kinney:2005ej} both in a Cardy-like
limit \cite{Choi:2018hmj,Honda:2019cio,ArabiArdehali:2019tdm} and in
a large-$N$ limit \cite{Benini:2018ywd}. In the present work we
further study the asymptotics of this so-called 4d $\mathcal{N}=4$ index
\cite{Kinney:2005ej}, finding that the two limits shed light not
only on each other, but also on new black objects in the dual
AdS$_5$ theory.

Our work is motivated by the important recent discovery
\cite{Benini:2018ywd,Choi:2018vbz} that by varying the fugacity
parameters of the index, its \emph{large-$N$ asymptotics} exhibits a
Hawking-Page-type deconfinement transition
\cite{Hawking:1982dh,Witten:1998zw} from a multi-particle
phase---already observed in \cite{Kinney:2005ej}---to a
long-anticipated black hole phase
\cite{Gutowski:2004ez,Gutowski:2004yv,Kunduri:2006ek,Chong:2005da,Cvetic:2005zi,Hosseini:2017mds}.

Here we argue that by varying the chemical potentials in the index,
its \emph{Cardy-like asymptotics} displays ``infinite-temperature''
Roberge-Weiss-type first-order phase transitions \cite{Roberge:1986mm} between the
fully-deconfined phase associated to black holes
\cite{Gutowski:2004ez,Gutowski:2004yv,Kunduri:2006ek,Chong:2005da,Cvetic:2005zi,Hosseini:2017mds},
and confined or partially-deconfined phases, with the latter
possibly associated to new multi-center black objects. Guided by this
Cardy-limit analysis, we revisit the large-$N$ asymptotics of the
index and argue that by including some previously neglected
contributions in the asymptotic analysis of \cite{Benini:2018ywd} it
is possible to see the partially-deconfined phases in the
large-$N$ limit as well.

In the rest of this introduction, we present a more precise
description of our framework, as well as a brief outline of our new technical results. Section~\ref{sec:phases} spells out our terminology regarding various ``phases'' of the index. There we outline a correspondence with $\mathcal{N}=1^\ast$ theory, which turns out to yield surprisingly powerful insight into the Bethe Ansatz approach discussed later in the paper. In Section~\ref{sec:Cardy} we
study the Cardy-like limit of the index using its expression as an
integral over holonomy variables
\cite{Kinney:2005ej,Spiridonov:2010qv}, extending previous partial
results by one of us in \cite{ArabiArdehali:2019tdm}. For the
SU$(2)$ case we explain that varying the chemical potentials
triggers an ``infinite-temperature'' Roberge-Weiss-type transition
\cite{Roberge:1986mm} between a confined phase where the center-symmetric
(i.e.\ $\mathbb{Z}_2$-symmetric) holonomy configuration dominates the
index, and a deconfined phase where two center-breaking holonomy
configurations take over. For $N=3$ we establish a similar behavior with a $\mathbb{Z}_3$ center-breaking pattern, while for $N=4$ we encounter a partially deconfined phase with a $\mathbb{Z}_4\to\mathbb{Z}_2$ center-breaking pattern. We also consider the SU$(N>4)$ cases, and in particular
argue that taking the large-$N$ limit \emph{after} the Cardy-like
limit should yield various partially-deconfined infinite-temperature
phases. Our investigation of this double-scaling limit leads up to a conjecture for the leading asymptotics of the index as displayed in (\ref{eq:doubleScalingConjecture}). This is the main result of our paper: the asymptotics of the index is expressed in (\ref{eq:doubleScalingConjecture}) as a sum over infinitely many exponential contributions, each presumably corresponding to a gravitational saddle in the bulk.

In Section~\ref{sec:BA} we study the index using its
expression as a sum over solutions to a system of elliptic Bethe Ansatz Equations (eBAEs)
\cite{Closset:2017bse,Benini:2018mlo}. First we review the Bethe
Ansatz formula for the 4d $\mathcal N=4$ index. Then we study the
Cardy-like asymptotics of the index in this approach. It turns
out that compatibility with the same partially-deconfined infinite-temperature
phases observed in the integral approach requires the existence of new solutions to
the eBAEs, which are not covered
in \cite{Hong:2018viz}. We find a vast number of such new solutions in this work; in most cases
numerically, in some low-rank cases asymptotically, and in one case exactly! The exact solution comes from the remarkable correspondence with $\mathcal{N}=1^\ast$ theory, discussed in section~\ref{sec:phases}. The correspondence also gives powerful insight into \emph{continua} of eBAE solutions which exist for $N\ge3$. The existence of such continua of solutions implies in fact that the Bethe Ansatz formula in its
current form \cite{Closset:2017bse,Benini:2018mlo} as a finite sum
is incomplete for $N>2$, and calls for an
integration with a so-far unknown measure, which we leave
unresolved. Finally we move on to the large-$N$ limit of the index,
extending previous results by Benini and Milan in
\cite{Benini:2018ywd}. Section~\ref{sec:discussion}
summarizes our main findings by placing them in the context of recent literature, and also outlines a few important related
directions for future research. The appendices elaborate on some
technical details used in the main text.

%%%%%
\subsection{Setup}
%%%%%
The 4d $\mathcal{N}=4$ index \cite{Kinney:2005ej}
\begin{equation}
    \begin{split}
        \mathcal{I}(p,q,y_{1,2,3})=\mathrm{Tr}\left[(-1)^F
        p^{J_1}q^{J_2}y_1^{Q_1}y_2^{Q_2}y_3^{Q_3}\right],
    \end{split}\label{eq:indexDefN=4}
\end{equation}
is expected to be a meromorphic function of five complex parameters
$p,q,y_{1,2,3}$ subject to $y_1y_2y_3=pq$, on the domain
$|p|,|q|\in(0,1)$ and $y_{1,2,3}\in\mathbb{C}^\ast$---\emph{c.f.}
\cite{Spiridonov:2010qv,Rains:2005}. For simplicity, throughout this
paper we restrict ourselves to the special case where $y_1,y_2$ are
on the unit circle; alternatively, we define $\sigma$, $\tau$, $\Delta_a$
through $p=e^{2\pi i \sigma}$, $q=e^{2\pi i\tau}$, $y_a=e^{2\pi i
\Delta_a}$ and take
$\Delta_{1,2}\in\mathbb{R}$. Then, for the SU$(N)$ case, the index
can be evaluated as the following elliptic hypergeometric integral \cite{Spiridonov:2010qv}:
\begin{equation}
	\mathcal{I}(p,q,y_{1,2,3})=\frac{\big((p;p)(q;q)\big)^{N-1}}{N!}\prod_{a=1}^{3}\Gamma_e^{N-1}\big(y_a\big)\oint\prod_{j=1}^{N-1}\frac{\mathrm{d}z_j}{2\pi i z_j}\prod_{1\le i,j\le N}^{i\neq j}\frac{\prod_{a=1}^{3}\Gamma_e\big(y_a \fft{z_i}{z_j}\big)}{\Gamma_e\big(\fft{z_i}{z_j}\big)},\label{eq:EHI}
\end{equation}
with the unit-circle contour\footnote{More precisely, the
unit-circle contour works if one uses an $i\varepsilon$-type
prescription of the form $\Delta_{1,2}\in\mathbb{R}+i0^+$, or one
lets $y_{1,2}$ approach the unit circle from inside in the
Cardy-like limit (as in \cite{ArabiArdehali:2019tdm}). If
$\Delta_{1,2}$ are kept strictly real, then the contour should be
slightly deformed. This seems to be a technicality of no
significance for our purposes in the present work though, so we
neglect it for the rest of this paper.} for the $z_j=e^{2\pi i
x_j}$, while $\prod_{j=1}^N z_j=1$; the $x_j$ variables (satisfying
$\sum_{j=1}^N x_j\in\mathbb{Z}$) will be referred to as \emph{the
holonomies}. The two special functions $(\cdot;\cdot)$ and
$\Gamma_e(\cdot)\equiv\Gamma(\cdot;p,q)$ are respectively the \emph{Pochhammer symbol} and
the \emph{elliptic gamma function} \cite{Ruijsenaars:1997}:
\begin{align}
    (p;q)&:=\prod_{k=0}^{\infty}(1-p q^k),\label{eq:PochDef}\\
    \Gamma(z;p,q)&:=\prod_{j,k\ge 0}\frac{1-z^{-1}p^{j+1}q^{k+1}}{1-z
    p^{j}q^{k}}.\label{eq:GammaDef}
\end{align}
Finally, we assume $p,q\notin\mathbb{R}$ and
$\Delta_{1,2,3}\notin\mathbb{Z}$, so that the index exhibits fast
asymptotic growth---\emph{c.f.}
\cite{Ardehali:2015bla,ArabiArdehali:2019tdm}.

Further defining $b,\beta$ through $\tau=\frac{i\beta
b^{-1}}{2\pi}$, $\sigma=\frac{i\beta b}{2\pi}$, the Cardy-like
\cite{Cardy:1986ie} limit of our interest corresponds to
\cite{Choi:2018hmj}
\begin{equation}
\text{\emph{the CKKN limit:}}\quad |\beta|\to 0,\ \text{with\
}b\in\mathbb{R}_{>0},\ \Delta_{a}\in\mathbb{R}\setminus\mathbb{Z},\
0<|\arg\beta|<\frac{\pi}{2} \
\text{fixed}.\label{eq:CKKNlimit}
\end{equation}
Throughout this paper, unless otherwise stated, by the ``high-temperature'' or ``Cardy-like'' limit, we always mean the CKKN limit (\ref{eq:CKKNlimit}). 

It turns out \cite{Honda:2019cio,ArabiArdehali:2019tdm} that $b$
does not control the leading asymptotics of the index in the CKKN
limit, and $\arg\beta$ controls its qualitative behavior
only through its sign. On the other hand $\Delta_3$ is redundant
thanks to the ``balancing condition'' $y_1y_2y_3=pq$. Therefore we
end up with only two \emph{control-parameters} $\Delta_{1,2}$ for
each sign of $\arg\beta$.

\begin{figure}[t]
\centering
    \includegraphics[scale=.6]{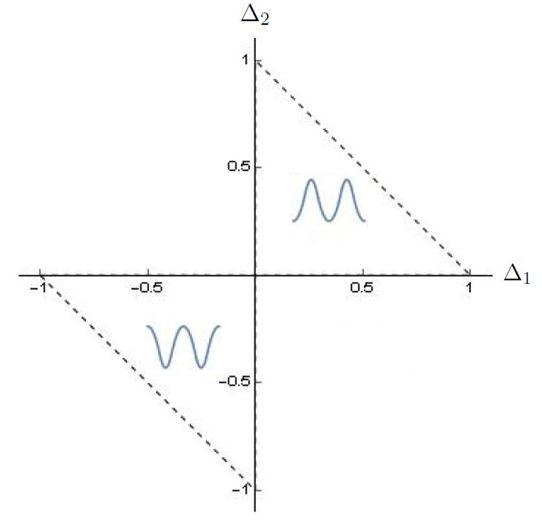}
\caption{The qualitative behavior of the pairwise potential for the
holonomies, as a function of the pair's separation, for fixed
$\Delta_{1,2}$ and fixed $\arg\beta\in(0,\pi/2)$, in the two
complementary regions
$-1<\Delta_{1},\Delta_{2},-1-\Delta_1-\Delta_2<0$ (lower-left) and
$0<\Delta_{1},\Delta_{2},1-\Delta_1-\Delta_2<1$ (upper-right) of the
space of the control-parameters $\Delta_{1,2}$ (taken to be inside
$\mathbb{R}$). The M and W wings switch places if
$\arg\beta$ is taken to be inside $(-\pi/2,0)$
instead---\emph{c.f.} \cite{ArabiArdehali:2019tdm}.
\label{fig:CatastSimp}}
\end{figure}

Since $\Delta_{1,2}$ are defined mod $\mathbb{Z}$, we can focus on a
fundamental domain. It turns out to be useful
\cite{ArabiArdehali:2019tdm} to take the fundamental domain to
consist of the two wings
$0<\Delta_{1},\Delta_{2},1-\Delta_1-\Delta_2<1$ (upper-right) and
$-1<\Delta_{1},\Delta_{2},-1-\Delta_1-\Delta_2<0$ (lower-left) of a
butterfly in the $\Delta_1$-$\Delta_2$ plane. When
$\arg\beta>0$, the effective pairwise potential for the
holonomies in the Cardy-like limit, given explicitly in
(\ref{eq:QhPot}) below, is M-shaped on the upper-right wing of the
butterfly, while on the other wing it is W-shaped---and conversely
for $\arg\beta<0$. See Figure~\ref{fig:CatastSimp} for a representation of the fundamental domain along with the M and W wings. Throughout
this paper, the M wing (\emph{resp.}\ W wing) denotes the part of the
fundamental domain where the effective pairwise potential for the
holonomies is M-shaped (\emph{resp.}\ W-shaped).

%%%%%
%%%%%

The asymptotics of the index on the M wings was obtained in \cite{ArabiArdehali:2019tdm}. On the M wings because of the shape of the pairwise potential the holonomies condense in the Cardy-like limit, and the ``saddle-points'' with $x_{ij}(:=x_i-x_j)=0$ dominate the matrix-integral expression (\ref{eq:EHI}) for the index. As reviewed in subsection~\ref{sec:Cardy:M-wing} the resulting asymptotics allows making contact with the entropy $S_{\mathrm{BH}}(J_{1,2},Q_a)$ of the bulk BPS black holes. On the other hand, the asymptotics on the W wings has been an open problem. In this work we discover a host of interesting phenomena, most importantly partial deconfinement, on the W wings.\\

As a complementary approach to that based on the integral representation (\ref{eq:EHI}) of the index, we also study it via the Bethe Ansatz type formula of \cite{Closset:2017bse,Benini:2018mlo}. In this approach we limit ourselves for simplicity to $p=q$, in which case the formula takes the form
\begin{equation}
	\mathcal{I}(q,q,y_{1,2,3})\overset{?}{=}\sum_{\hat{u}\in\mathrm{eBAEs}}I(\hat{u};\Delta_{1,2,3},\tau),\label{eq:superSimpleBAE}    
\end{equation}
for some rather elaborate special function $I$ spelled out in subsection~\ref{sec:BA:review}, involving the elliptic gamma function. Here $\hat{u}\in\mathrm{eBAEs}$ means that we have to sum over the solutions to the elliptic Bethe Ansatz equations of the SU$(N)$ $\mathcal{N}=4$ theory spelled out in section~\ref{sec:BA:review}, involving Jacobi theta functions. Prior to the present work, only a set of isolated solutions to the eBAEs were known. These were derived in \cite{Hong:2018viz}, and we will refer to them as ``the standard solutions''. They are labeled by three non-negative integers $\{m,n,r\}$ subject to $mn=N$, $0\le r<n$, and correspond to perfect tilings of the torus with modular parameter $\tau$. In this work we discuss various new solutions, which we refer to as ``non-standard''. In particular, we will argue that for $N>2$ there are continua of such solutions. The Bethe Ansatz formula for the index then breaks down, and needs to be reformulated to incorporate such continua; hence the question mark above the equal-sign in (\ref{eq:superSimpleBAE}).

Despite the said shortcoming of the Bethe Ansatz approach for $N>2$, we will still utilize it in section~\ref{sec:BA} by temporarily neglecting the continua of eBAE solutions. This way we study the CKKN limit (with $b=1$) of the index, and will compare the result with that obtained in section~\ref{sec:Cardy} from the integral expression.

Up to the same caveat, we will also utilize the Bethe Ansatz formula in section~\ref{sec:BA} to study the large-$N$ limit of the index. Taking the Cardy-like limit \emph{after} the large-$N$ limit, we obtain an answer that we will compare with the asymptotics of the index in the Cardy-like \emph{before} the large-$N$ limit, analyzed in section~\ref{sec:Cardy}.

%%%%%
\subsection{Outline of the new technical results}
%%%%%
For readers interested in specific technical results, here we provide a list of the new findings of the present paper, with reference to the appropriate section where they are discussed.

\begin{itemize}
    \item \textbf{Relation between the $\mathcal N=1^\ast$ theory and the $\mathcal N=4$ theory.} 
    
    The correspondence between vacua of the $\mathcal{N}=1^\ast$ theory and solutions to the $\mathcal{N}=4$ eBAEs is spelled out in subsection~\ref{subsec:correspondenceWithN=1*}. It leads to Conjecture~2 in section~\ref{sec:BA} stating that for $N\ge(l+1)(l+2)/2$, there are $l$-complex-dimensional continua of solutions to the SU$(N)$ $\mathcal{N}=4$ eBAEs.
    
    \item \textbf{The Cardy-like asymptotics of the index.} 
    
    We studied the Cardy-like asymptotics of the index for generic $\Delta_{1,2}\in\mathbb{R}$ and $\arg\beta\neq0$ using the elliptic hypergeometric integral form in section \ref{sec:Cardy}. For $N=2,3,4$, it is presented for the first time in subsections~\ref{subsec:SU(2)Cardy} and \ref{subsec:SU(N>2)Cardy}. Some earlier studies had considered only the $x_{ij}=0$ saddle-point in the matrix-integral, which is \emph{incorrect} on the W wings. For $N\to\infty$, a lower bound for the index is obtained in (\ref{eq:dSummedLargeN}), wherein $C_\mathrm{max}$ can be taken to infinity. This lower bound along with Lemma~1 establishes that the index is partially deconfined all over the W wings in the double scaling limit. This finding encourages Conjecture~1 that the said lower bound is actually optimal and therefore gives the large-$N$ \emph{after} the Cardy-like asymptotics of the index. Using this asymptotic expression for the index and assuming  $Q_a\in C\mathbb{Z}$, we find critical points of the Legendre transform of (logarithm of) the index yielding micro-canonical entropies $S_C(J_{1,2},Q_a)=S_\mathrm{BH}(J_{1,2},Q_a)/C$ for $C=2,3,4,5$; these presumably correspond to entropies of new (possibly multi-center) black objects in the bulk. (What happens to the micro-canonical entropy for $C>5$ is not clear to us; see the comment at the end of section~\ref{sec:Cardy}.)
    
    We studied the same index using the Bethe Ansatz form in subsection \ref{sec:BA:Cardy}. Compatibility with the results from the elliptic hypergeometric integral form implies the existence of new eBAE solutions that were not covered in \cite{Hong:2018viz}. We indeed found such solutions numerically (analytically in the Cardy-like limit) for some simple cases and reproduced the lower bound (\ref{eq:dSummedLargeN}). Conjecture~1 would imply that in the Bethe Ansatz approach the other eBAE solutions, which we have not fully figured out, will not contribute to the leading Cardy-like asymptotics of the index at large $N$. 
    
    \item \textbf{``Non-standard'' eBAE solutions.} 
    
    We discuss various new SU$(N)$ eBAE solutions that were not covered in \cite{Hong:2018viz}, referred to as ``non-standard'' solutions. In subsection~\ref{subsec:SU(2)non-standard}, we employ elementary elliptic function theory to establish the existence of one such solution (two if we count the different signs) for $N=2$, and present its asymptotics. In subsection~\ref{subsec:SU(3)nonstandard} we discuss numerical evidence that for $N=3$ there is a one-complex-dimensional continuum of eBAE solutions. We further discuss this continuum in the low- and high-temperature limits. It turns out that a member of this continuum can be captured exactly (i.e. at finite temperature). This is thanks to the correspondence of subsection~\ref{subsec:correspondenceWithN=1*} with the $\mathcal{N}=1^\ast$ theory, which allows us to borrow a result of Dorey \cite{Dorey:1999sj}. We present analytic evidence that this exact non-standard solution is indeed a member of a one-complex-dimensional continuum. This is achieved via a beautiful three-term theta function identity presented as Lemma~2, which establishes that the associated Jacobian factor of the eBAE solution vanishes. Finally we discuss numerical evidence for Conjecture~2 for $N\leq10$: in particular, we found numerical evidence for two and three complex dimensional continua of solutions to the SU$(N)$ eBAEs for $N=6$ and $N=10$ respectively. These findings imply that the Bethe Ansatz formula for the 4d $\mathcal{N}=4$ index is valid in its currently available form only for $N=2$; for higher $N$ it needs to be reformulated to take the continua of Bethe roots into account. 

    \item \textbf{The large-$N$ asymptotics of the index.}
    
    In subsection \ref{sec:BA:large-N} we estimated the large-$N$ limit of the index, extending previous results in \cite{Benini:2018ywd}. In particular, the leading Cardy-like asymptotics of the improved large-$N$ limit of the index turns out to match the large-$N$ \emph{after} the Cardy-like asymptotics of the index (\ref{eq:doubleScalingConjecture}), with a couple of subtle issues discussed in details in the main text. This suggests that the asymptotic behavior of the index in the double-scaling limit is captured by Conjecture~1 regardless of the order of the Cardy-like limit and the large-$N$ limit.

\end{itemize}

%%%%%%%%%%%%%%%%%%%%%
\section{High-temperature phases of the 4d \texorpdfstring{$\mathcal{N}=4$}{N=4} index}\label{sec:phases}

The 4d $\mathcal{N}=4$ index (just as any Romelsberger index \cite{Romelsberger:2005eg} for that matter) can be computed as a partition function on a primary Hopf surface \cite{Assel:2014paa} with complex-structure moduli $\tau,\sigma$. To gain intuition on these moduli, we work with $b,\beta$ instead, defined through $\tau={i\beta
b^{-1}}/{2\pi}$, $\sigma={i\beta b}/{2\pi}$. When $b,\beta$ are positive real numbers, the Hopf surface is $S_b^3\times S^1$, with a direct product metric, and with $\beta=2\pi r_{S^1}/r_{S^3}$, while $b$ becomes the squashing parameter of the three-sphere. Then, in analogy with thermal quantum physics, one can interpret the $S^1$ as the Euclidean time circle, and hence think of $\beta$ as inverse-temperature in units of $r_{S^3}$. 

More generally, the complex-structure moduli of the Hopf surface could be such that $\beta$ becomes complex. Then the Hopf surface is still topologically $S^3\times S^1$, but metrically it is not a direct product anymore. This situation would correspond to having a ``\emph{complex temperature}''.

From a field theory perspective, allowing $\beta$ to become complex simply amounts to extending the territory of exploration, with potentially new behaviors of the index to be discovered in the extended domain. For example, as we will recollect in subsection~\ref{subsec:relationToPrevious}, general Romelsberger indices seem to exhibit a much faster and much more universal Cardy-like growth in subsets of the complex-$\beta$ domain \cite{Kim:2019yrz,Cabo-Bizet:2019osg}.

From a holographic perspective, on the other hand, complexifying $\beta$ finds a distinctly significant meaning through its relation with rotation in the bulk---c.f. \cite{Assel:2014paa,Genolini:2016ecx}. This relation arises because the non-direct product geometry of the boundary can be filled in only with rotating spacetimes. This observation in turn explains why studies of the Cardy-like limit of the 4d $\mathcal{N}=4$ index prior to CKKN \cite{Choi:2018hmj} found a much slower growth than that required by the bulk black holes: earlier studies had focused on real $\beta$, while the bulk BPS black holes have rotation and require complex $\beta$. (As we will recollect in subsection~\ref{subsec:relationToPrevious}, a second important novelty of the limit studied by CKKN was considering complex $y_k$.)

In this work we study various ``phases of the 4d $\mathcal{N}=4$ index at high (complex) temperatures''. What we mean by this is as follows. First of all, since the Hopf surface the index corresponds to is compact, in order to have a notion of ``phase'' (associated to various ``saddle-points'' dominating the partition function) we need to take some limit of the index. When a large-$N$ limit is taken, one can speak of high- (complex-) temperature phases of the index when $|\beta|$ is small enough---smaller than some finite critical value for instance. For finite $N$ on the other hand, to have a notion of phase we go to ``infinite- (complex-) temperature'' $|\beta|\to0$ (with $|\arg\beta|\in(0,\pi/2)$ fixed). We can then classify various behaviors of the index in those limits as various ``phases''. The control-parameters $\Delta_{1,2}$ in turn would often allow Roberge-Weiss type \cite{Roberge:1986mm} transitions between such phases.\\

There appear to be two particularly natural classification schemes in the present context, and we now explain both in some detail. In particular, two different notions of \emph{partial deconfinement} arise from the following classifications. When discussing partial deconfinement in the following sections, it should be clear from the context which of the two notions we are referring to. (A ``sub-matrix deconfinement'', different from partial deconfinement in the senses elaborated on below, has been
recently discussed in
\cite{Berenstein:2018lrm,Hanada:2018zxn,Hanada:2019czd,Hanada:2019kue}.)

\subsection{Classification via center symmetry}
A first classification scheme arises if following in the footsteps of Polyakov
\cite{Polyakov:1978vu} one considers patterns of center-symmetry breaking by the dominant holonomy configurations in the index.

In section~\ref{sec:Cardy} we will discuss various
center-breaking patterns in the Cardy-like limit of the index. The Cardy-like limit is analogous to the infinite-temperature limit of thermal
partition functions. We will speak of partial
deconfinement in a sense similar to that of Polyakov, when a dominant holonomy
configuration breaks the $\mathbb{Z}_N$ center to a subgroup
(possibly an approximate one for large $N$, in a sense elaborated on in subsection~\ref{sec:Cardy:large-N}). For $C>1$ a divisor of $N$, a
useful order-parameter for \emph{a single} critical holonomy
configuration $\boldsymbol{x}^\ast$ is the $C$-th power of the Polyakov loop $
\Tr P^C|_{\boldsymbol{x}^\ast}=\sum_{j=1}^N e^{2\pi i C
x^\ast_j}$, which condenses (i.e. becomes nonzero) in a phase where
$\mathbb{Z}_N\to \mathbb{Z}_C$. We say the index is in a ``$C$-center phase'' if a dominant holonomy configuration is $C$-centered, with $C$ packs of $N/C$ condensed (i.e. collided) holonomies distributed uniformly on the circle, so that $
\left|\Tr P^C|_{\boldsymbol{x}^\ast}\right|=N$; summing over \emph{all}
critical holonomies recovers the center symmetry of course: $
\sum_{\boldsymbol{x}^\ast}\Tr P^C|_{\boldsymbol{x}^\ast}=0$.
Partial deconfinement in this sense has been discussed earlier (see
e.g. \cite{Ogilvie:2007tj})  in the more conventional thermal,
non-supersymmetric context with $\mathbb{R}^3$ as the spatial
manifold; there a sharper notion of an order-parameter exists,
since on non-compact spatial manifolds cluster decomposition forbids
the analog of summing over all $\boldsymbol{x}^\ast$.

The one-center phase, as one would expect, is the fully deconfined one where all the holonomies condense at a given value (either $0$, or $\frac{1}{N}$, ..., or $\frac{N-1}{N}$, due to the SU$(N)$ constraint). However, in principle this is not the only pattern for a full breaking of the $\mathbb{Z}_N$ center. For example, a random distribution of the holonomies on the circle would also completely break the center. When the dominant holonomy configurations $\boldsymbol{x}^\ast$ break the $\mathbb{Z}_N$ center completely (or more generally to $\mathbb{Z}_C$), but are not one-centered (or more generally $C$-centered), we say the high-temperature phase of the index is ``non-standard''; the Polyakov loop then may or may not condense (and more generally $
\left|\Tr P^C|_{\boldsymbol{x}^\ast}\right|<N$). We will encounter such non-standard phases in the Cardy-like limit of the $\mathcal{N}=4$ index for $N=5,6$ in section~\ref{sec:Cardy}; they correspond to the holes in Figure~\ref{fig:N=5and6}, and in the Bethe Ansatz approach they would arise when non-standard eBAE solutions take over the index in the Cardy-like limit.

In section~\ref{sec:BA} we will demonstrate how partial deconfinement
in a similar sense can occur in the large-$N$
limit of the index as well. The large-$N$ limit will be analyzed for $\tau=\sigma$ via the Bethe Ansatz approach, where the behavior of the index depends on which solution of the eBAEs dominates the large-$N$ limit. Such solutions can be thought of as complexified holonomy configurations, whose $\tau$-independent parts correspond to the Polyakov loops. At finite $\beta$, they also have $\tau$-dependent parts though, that are analogous to 't~Hooft loops. There is hence also an analog of ``magnetic'' $\mathbb{Z}_N$ center at finite $\beta$, which has in fact appeared in the Bethe Ansatz context already in \cite{Hong:2018viz}. Therefore the high-temperature phases at finite $\beta$ and large $N$, can be classified via subgroups of $\mathbb{Z}_N\times \mathbb{Z}_N$, in a picture that is in a sense dual to 't~Hooft's classification of phases of SU$(N)$ gauge theories \cite{tHooft:1977nqb,tHooft:1979rtg,tHooft:1981bkw}. We now proceed to expand on this duality---or correspondence---below.

\subsubsection{Correspondence with vacua of compactified \texorpdfstring{$\mathcal{N}=1^\ast$}{N=1*} theory}\label{subsec:correspondenceWithN=1*}

There is a regime of parameters where we expect close connection between high-temperature phases of the $\mathcal{N}=4$ index and low-energy phases of the $\mathcal{N}=1^\ast$ theory on $\mathbb{R}^3\times S^1$. This is the regime where $i$) $\beta\to0$ and $\beta\in\mathbb{R}^{>0}$, such that in the $r_{S^1}\to0$ ``direct channel'' one is probing high-temperature phases on $S^3\times S^1$, while in the $r_{S^3}\to\infty$ ``crossed channel'' one is probing low-energy phases on $\mathbb{R}^3\times S^1$; $ii$) the chemical potentials $\Delta_{1,2,3}$ are small enough that their periodicity and balancing condition are not significant. Even then, the $\Delta_k$ are \emph{real} masses for the adjoint chiral multiplets of compactified $\mathcal{N}=4$ theory, while $\mathcal{N}=1^\ast$ theory has \emph{complex} masses for its adjoint chirals. Nevertheless, based on the channel-crossing argument one might expect at least some resemblance between potential high-temperature phases of the $\mathcal{N}=4$ index and possible low-energy phases of compactified $\mathcal{N}=1^\ast$ theory, and interestingly enough closer inspection reveals not just a resemblance, but a precise quantitative correspondence, aspects (though not all) of which extend even to finite complex $\beta$ and arbitrary $\Delta_{1,2}\in\mathbb{R}$.

First, the proper identification between the complex-structure modulus $\tau$ of the $\mathcal{N}=4$ index and the complexified gauge coupling $\tilde{\tau}$ of the $\mathcal{N}=1^\ast$ theory seems to be as follows\footnote{For an analogous duality swapping the gauge coupling and the inverse temperature see \cite{Peskin:1977kp}. See also \cite{ArabiArdehali:2019zac} where a similar correspondence is discussed in the $\tau\to i0^+$ limit for generic $\mathcal{N}=1$ gauge theories with a U(1)$_R$ symmetry.}:
\begin{equation}
\tilde{\tau}\ \longleftrightarrow\ -\frac{1}{\tau}.\label{eq:tausRelate}
\end{equation}
Alternatively, the electric and magnetic loops are swapped in the two pictures---i.e. the Polyakov loops in the direct channel correspond to the 't~Hooft loops in the crossed channel and the 't~Hooft loops in the direct channel correspond to the Wilson loops in the crossed channel. The identification (\ref{eq:tausRelate}) can be motivated through the crossed-channel relation between the two pictures, but a more satisfactory derivation of it would be desirable. Once the identification is accepted though, one can compare the vacua of compactified $\mathcal{N}=1^\ast$ theory as determined via Dorey's elliptic superpotential \cite{Dorey:1999sj}, with the possible high-temperature phases of the $\mathcal{N}=4$ index as determined via solutions to the elliptic BAEs.

A particularly interesting aspect of the correspondence which survives at finite complex $\beta$ and finite $\Delta_{1,2,3}$, is the connection between the massive phases of the compactified $\mathcal{N}=1^\ast$, and the standard solutions to the $\mathcal{N}=4$ eBAEs:
\begin{equation}
\text{massive vacua}\ \longleftrightarrow\ \text{standard eBAE solutions},
\end{equation}
valid for arbitrary $N$. Specifically, the vacuum associated to the subgroup $F'_r$ of $\mathbb{Z}_N\times\mathbb{Z}_N$ generated by $(0,n),(\frac{N}{n},r)$ in the Donagi-Witten terminology \cite{Donagi:1995cf}, corresponds to the $\{\frac{N}{n},n,r\}$ standard eBAE solution \cite{Hong:2018viz} spelled out in subsection~\ref{sec:BA:review} below. Moreover, just as the massive phases are permuted via $S$ duality in $\mathcal{N}=1^\ast$ theory, the standard solutions to the $\mathcal{N}=4$ eBAEs are permuted via an SL($2,\mathbb{Z}$) acting on $\tau$ \cite{Hong:2018viz}.

Most strikingly for our purposes, the Coulomb phase of the SU$(3)$ $\mathcal{N}=1^\ast$ theory corresponds to a continuous set of non-standard solutions to the SU$(3)$ $\mathcal{N}=4$ eBAEs. This yields in particular an \emph{exact} non-standard SU$(3)$ eBAE solution via the corresponding $\mathcal{N}=1^\ast$ vacuum given by Dorey \cite{Dorey:1999sj}. We will discuss this exact non-standard solution in section~\ref{sec:BA}. The continua that this exact non-standard eBAE solution is part of is found analytically both in the high-temperature ($|\tau|\ll1$) and in the low-temperature ($|\tau|\gg1$) limits, as well as numerically in the intermediate regime with generic $\tau$; see subsection~\ref{subsec:SU(3)nonstandard}. We suspect that more generally for any $N>2$, for general $\tau$ in the upper-half plane, and at least for an appropriate range of $\Delta_{1,2}$, there is a correspondence
\begin{equation}
\text{Coulomb vacua}\ \longleftrightarrow\ \text{continua of non-standard eBAE solutions},
\end{equation}
though the precise map might quite non-trivially depend on the chosen $\Delta_{1,2}$.
A further bridge, due to Dorey \cite{Dorey:1999sj}, is expected to connect the vacua on $\mathbb{R}^3\times S^1$ to those on $\mathbb{R}^4$.\footnote{The correspondence put forward in the present subsection essentially boils down to one between stationary points of elliptic Calogero-Moser Hamiltonians \cite{Dorey:1999sj} associated to compactified $\mathcal{N}=1^\ast$, and a subset of solutions to the $\mathcal{N}=4$ elliptic Bethe Ansatz equations. So although mathematically intriguing (with potential connections to \cite{Gorsky:1994dj,Gorsky:2013xba}), it might not bear conceptual lessons for QFT. Dorey's correspondence with $\mathbb{R}^4$ \cite{Dorey:1999sj} on the other hand, seems to require navigating rather deep waters of quantum gauge theory to make complete sense of. We leave a more thorough investigation of these connections to future work.} Based on this correspondence and available knowledge (see e.g. \cite{Bourget:thesis}) on semi-classical Coulomb vacua on $\mathbb{R}^4$, we expect that for $N\ge(l+1)(l+2)/2$ there are $l$-complex-dimensional continua of eBAE solutions for the SU$(N)$ $\mathcal{N}=4$ theory. We have numerically checked that this expectation pans out (at finite $\tau$) for $N=4$ through $10$ as well: the $N=4,5$ cases, just like for $N=3$, contain one-complex-dimensional continua of non-standard eBAE solutions, while in the $N=6$ case for the first time a two-complex-dimensional continuum of solutions arises.  This persists for $N=7,8,9$, and then a new three-complex-dimensional family of solutions appears at $N=10$.  These continua of solutions present a serious difficulty for the Bethe Ansatz formula for the 4d $\mathcal{N}=4$ index \cite{Closset:2017bse,Benini:2018mlo}, which is derived assuming only isolated eBAE solutions. We will comment more on this point in section~\ref{sec:BA}.\\

An example of phenomena arising for finite complex $\tau$ or large $\Delta_{1,2}$ that are outside the regime of validity of the correspondence is presented by the isolated non-standard SU$(2)$ eBAE solution $u_\Delta$ discussed in subsection~\ref{sec:BA:non-standard}. For $\Delta_{1,2}$ so large that the lower branch of (\ref{highT:N=2:flowend}) becomes relevant, even in the $\beta\to0$ limit the non-standard solution $u_\Delta$ does not correspond to an SU$(2)$ $\mathcal{N}=1^\ast$ vacuum.

\subsection{Classification via asymptotic growth}
Deconfinement in a second sense can be associated with the
asymptotic growth of $\mathrm{Re}\log\mathcal{I}$ (or more generally
$\mathrm{Re}\log$ of a partition function, with the Casimir-energy
piece removed), either as $|\beta|\to0$ for
finite $N$, or as $N\to\infty$. This is a more general sense as it does not rely on a center symmetry. In the case of the 4d $\mathcal{N}=4$ index, an $\mathcal{O}(1/|\beta|^2)$
growth as $|\beta|\to0$, or an $\mathcal{O}(N^2)$ growth as
$N\to\infty$ \cite{Witten:1998zw} could count as ``deconfinement'' in this second sense.

So far
these criteria do not distinguish between the fully-deconfined and
partially-deconfined phases, as classified in the first sense. A more refined classification is possible in the present context though, due to the presence of the first-order Roberge-Weiss type transitions \cite{Roberge:1986mm}. Let us consider the
specific case of the SU$(4)$ $\mathcal{N}=4$ theory in the Cardy-like limit as an illustrative example. In
Figure~\ref{fig:N=4} we see three high-temperature ``phases'' of the index separated by first-order transitions. Each phase can be labeled according to its fastest growth, say by its
\begin{equation}
s:=\sup(\lim_{|\beta|\to0}|\beta|^2\mathrm{Re}\log\mathcal{I}),
\end{equation}
which is proportional to the maximum height of its corresponding curve in Figure~\ref{fig:N=4}. Then we get an ordering of the phases with $s>0$: although no longer a fully-deconfined or a confined phase in an absolute sense, we still get a ``maximally deconfined'' and (possibly) a ``minimally deconfined'' phase in a relative sense, as well as other ``partially deconfined'' phases in the middle. Phases with $s\le0$ might more appropriately be called ``non-deconfined''.

Figure~\ref{fig:N=4} displays a clear correlation between this and the previous sense of deconfinement: the curves corresponding
to larger center-breaking have higher maxima and therefore faster maximal growth. Also, note that the blue curve which would correspond to a confined (i.e. center-preserving) phase in the Polyakov sense, is associated to a ``minimally deconfining'' phase in the sense of asymptotic growth. On the other hand, the non-standard phases arising for $N\ge5$, exhibit intermediate asymptotic growth, so would be partially deconfined in the second sense, even though their dominant holonomy configurations might break the center completely.

In the large-$N$ limit, besides a similar ordering of the deconfined phases via
\begin{equation}
\tilde{s}:=\sup(\lim_{N\to\infty}\mathrm{Re}\log\mathcal{I}/N^2),
\end{equation}
there is also a useful notion of a ``confined'' phase \cite{Witten:1998zw} where $\mathrm{Re}\log\mathcal{I}=\mathcal{O}(N^0)$. In similar problems, there could of course be various other phases with intermediate scaling as well.

%%%%%%%%%%%%%%%%%%%%%
\section{Cardy-like asymptotics of the index}\label{sec:Cardy}

Following \cite{ArabiArdehali:2019tdm}, the elliptic gamma functions
can be expanded in the CKKN Cardy-like limit (\ref{eq:CKKNlimit}), so that the index (\ref{eq:EHI}) simplifies as
\begin{equation}
    \begin{split}
        \mathcal{I}(p,q,y_{1,2,3})\xrightarrow{\text{in the CKKN limit}}
        \int_{-1/2}^{1/2} e^{-2\pi i\frac{Q_h(\boldsymbol{x};\Delta_a)}{\tau\sigma}}\ \mathrm{d}^{N-1}\,x,
    \end{split}\label{eq:EHICardy}
\end{equation}
with the integral over the $N-1$ independent holonomies
corresponding to the maximal torus of SU$(N)$.  Here
\begin{equation}
    \begin{split}
    Q_h(\boldsymbol{x};\Delta_{1,2}):=\frac{1}{12}\sum_{a=1}^3\left((N-1)\kappa(\Delta_a)+\sum_{1\le
    i<j\le N}\kappa(\Delta_a+(x_i-x_j))+\kappa(\Delta_a- (x_i-x_j))\right),\label{eq:QhDef}
    \end{split}
\end{equation}
where
\begin{equation}
    \kappa(x):=\{x\}(1-\{x\})(1-2\{x\})\qquad\mbox{with}\qquad \{x\}:=x-\lfloor
    x\rfloor.
    \label{eq:kappa}
\end{equation}
Note in particular that $\kappa(x)$ is compatible with the unit
periodicity of the holonomies. Note also that on the RHS of
(\ref{eq:QhDef}) every $\Delta_3$ can be replaced with
$-\Delta_1-\Delta_2$; this is because the balancing condition
$y_1y_2y_3=pq$ fixes $\Delta_3=\tau+\sigma-\Delta_1-\Delta_2$
$\mathrm{mod\ }\mathbb{Z}$, and since we are interested in the
leading Cardy-like asymptotics, we can neglect $\tau$ and $\sigma$
in $\Delta_3$. (To capture the subleading effects a generalization
of $\kappa(x)$ to complex domain is needed
\cite{ArabiArdehali:2019tdm}---c.f. (\ref{eq:kappa:tau}) below.)

Since we are interested in the $|\tau\sigma|\to0$ limit, the
integral in (\ref{eq:EHICardy}) is dominated by the global maximum
of the real part of the exponent, or alternatively the global
minimum of $\mathrm{Re}(iQ_h/\tau\sigma)$.  Moreover, this limit is
well defined since $\kappa(x)$ is continuous and bounded and the
integration domain is compact.  As a result, the leading asymptotic
behavior of the index is given by
\begin{equation}
    \begin{split}
        \mathcal{I}(p,q,y_{1,2,3})\xrightarrow{\text{in the CKKN limit}}
        e^{-2\pi i\frac{Q_h(\boldsymbol{x}^\ast;\Delta_a)}{\tau\sigma}},
    \end{split}\label{eq:EHIsimplified}
\end{equation}
where $\boldsymbol{x}^\ast$ is the holonomy configuration corresponding to the global minimum%
\footnote{If there are degenerate minima, $\boldsymbol{x}^\ast$ can
be taken to correspond to any one of them, as the added degeneracy
factor is subleading in the CKKN limit.}.
Taking the parametrization $\tau=\fft{i\beta b^{-1}}{2\pi}$ and
$\sigma=\fft{i\beta b}{2\pi}$, and noting that $Q_h$ is a real
function, we see that $\boldsymbol{x}^\ast$ corresponds to the
global minimum of
\begin{equation}
    \mathcal V_{\mathrm{eff}}:=-\sin(2\arg\beta)Q_h(\boldsymbol{x};\Delta_a).
\end{equation}
From the $\boldsymbol{x}$-dependent part of $Q_h$ we see that minimizing $\mathcal{V}_\mathrm{eff}$ is equivalent to minimizing a potential of the form
$\sum_{1\le i< j\le N}V^Q(x_{ij};\arg\beta,\Delta_{1,2})$,
with the pairwise part explicitly reading
\begin{equation}
    \begin{split}
    V^Q(x_{ij};\arg\beta,\Delta_{1,2})=-\mathrm{sign}(\arg\beta)\cdot\sum_{a=1}^3\left(\kappa(\Delta_a+ x_{ij})+\kappa(\Delta_a- x_{ij})\right).\label{eq:QhPot}
    \end{split}
\end{equation}
Figure~\ref{fig:CatastSimp} above shows the qualitative behavior of
this pairwise potential.

%%%%%
\subsection{Behavior on the M wings}\label{sec:Cardy:M-wing}
%%%%%
As Figure~\ref{fig:CatastSimp} shows, on the M wings the pairwise potential (\ref{eq:QhPot}) is minimized
at $x_{ij}=0$.\footnote{This was found in
\cite{ArabiArdehali:2019tdm} by numerically scanning the space of
the control-parameters. In
\cite{Honda:2019cio} an analytic proof was suggested for an M-type behavior all over the
parameter-space when $\arg\beta>0$; however, as pointed out
in the Added Note of \cite{ArabiArdehali:2019tdm}, the proof
actually applies only to the upper-right wing of
Figure~\ref{fig:CatastSimp}, and the oddity of the potential under
$\Delta_{1,2}\to-\Delta_{1,2}$ establishes in fact the W-type behavior on
the lower-left wing when $\arg\beta>0$. The two wings of
course switch places for $\arg\beta<0$.} Consequently
the overall potential $\mathcal V_{\mathrm{eff}}$ also takes on its
global minimum when all holonomies are identical.  Taking SU$(N)$
into account, there are $N$ possible configurations, namely all
$x_i=k/N$ with $k=0,1,\ldots,N-1$.  These configurations are
dominant in the Cardy-like limit and completely break the
$\mathbb{Z}_N$ center. Here we have complete deconfinement
\cite{ArabiArdehali:2019tdm} (see also
\cite{Choi:2018hmj,Honda:2019cio}), as the index exhibits
``maximal'' asymptotic growth
\begin{equation}
    \mathcal I(p,q,y_{1,2,3})\sim\exp\left(-\fft{i\pi}{6\tau\sigma}(N^2-1)\sum_{a=1}^3\kappa(\Delta_a)\right)=\exp\left(-i\pi (N^2-1)
    \frac{\Delta_1 \Delta_2 \Delta^{(\pm1)}_3}{\tau\sigma}\right),
\label{eq:Mgrowth}
\end{equation}
with $\Delta^{(\pm1)}_3=\pm1-\Delta_1-\Delta_2$, where the sign
should be taken to be the same as that of $\arg\beta$. (More precisely, it is after appropriate
tuning of $\Delta_{1,2}$ that the maximal asymptotic growth is
achieved in this fully deconfined phase; see Figure~\ref{fig:compareDs}.)

This ``grand-canonical'' asymptotics in (\ref{eq:Mgrowth}), when
translated to the micro-canonical ensemble, yields the expected
entropy $S_{\mathrm{BH}}$ of the bulk AdS$_5$ black holes
\cite{Hosseini:2017mds,Cabo-Bizet:2018ehj}. (The original work
\cite{Hosseini:2017mds} showed this last statement for the minus
sign, and \cite{Cabo-Bizet:2018ehj} later established it for the
plus sign as well.)

Our focus in this work is on the W wings though, to which we now
turn.

%%%%%
\subsection{Behavior on the W wings}\label{sec:Cardy:W-wing}
%%%%%
The W wings are characterized by the feature that the minimum of the
pairwise potential (\ref{eq:QhPot}) is displaced away from
$x_{ij}=0$.  In particular, it is located either at
$x_{ij}=1/2\mod1$ or in a flat region around this point.  The issue
now is that, except for special case of SU$(2)$, it is impossible to
center the differences $x_{ij}$ around $1/2\mod1$ for all $i$ and
$j$.  As a result, the global minimum of $V_{\mathrm{eff}}$ cannot
correspond to the individual minima of all the individual pairwise
potentials, and the extremization problem then becomes quite
challenging. For this reason, it has been an open problem to find
the Cardy-like asymptotics of the index on the W wings (\emph{c.f.}
Problem 1 in Section 5 of \cite{ArabiArdehali:2019tdm}). In this
section we completely address the problem for $N\le 4,$ and take
steps towards addressing it for $N>4$.

%%%%%
\subsubsection{\texorpdfstring{SU$(2)$}{SU(2)} and infinite-temperature confinement/deconfinement transition}\label{subsec:SU(2)Cardy}
%%%%%
The W-wing behavior is easy to determine for the SU$(2)$ case, as
the minimum at $x_{12}=1/2$ along with the SU$(2)$ condition
$x_1+x_2=0$ is trivially solved by the ``confining'' holonomy
configuration $x_1=-x_2=1/4$. This leads to the W-wing asymptotics
\begin{equation}
    \mathcal{I}\xrightarrow{\text{W wings}}
    \exp\left(-\frac{i\pi}{6\tau\sigma}\sum_{a=1}^3\big(\kappa(\Delta_a)+2\kappa(\Delta_a+1/2)\big)\right).\label{eq:N=2Wwing}
\end{equation}
Note that $i)$ the chosen dominant holonomy configuration on the W wings
respects the $\mathbb{Z}_2$ center symmetry generated by $x_i\to
x_i+1/2$, and $ii)$ as already discussed in
\cite{ArabiArdehali:2019tdm}, the fastest asymptotic growth of the
index on the W wings is slower than the fastest asymptotic growth on
the M wings, as expected.

%%%%%
\subsubsection{\texorpdfstring{SU$(N)$}{SU(N)} for finite \texorpdfstring{$N>2$}{N>2}}\label{subsec:SU(N>2)Cardy}
%%%%%
For $N>2$ it is no longer possible to have all $x_{ij}$ equal to
$1/2$, and finding the global minimum of the effective potential
becomes a difficult problem.  Nevertheless, we can obtain lower
bounds on the asymptotic growth by examining special sets of
holonomy configurations. In particular, we note that
\begin{equation}
    \begin{split}
        \mathcal{I}(p,q,y_{1,2,3})\gtrsim
        e^{-2\pi i\frac{Q_h(\boldsymbol{x}_0;\Delta_a)}{\tau\sigma}},
    \end{split}
\end{equation}
for any set of holonomies $\boldsymbol{x}_0$ on the maximal torus.
This bound is saturated when $\boldsymbol{x}_0=\boldsymbol{x}^\ast$,
but is suboptimal otherwise.  Our goal is then to pick a family of
configurations $\{\boldsymbol{x}_{0,i}\}$ and optimize over this
family. Because the potential is exponentiated, we can in fact write
\begin{equation}
    \begin{split}
        \mathcal{I}(p,q,y_{1,2,3})\gtrsim\sum_i
        e^{-2\pi i\frac{Q_h(\boldsymbol{x}_{0,i};\Delta_a)}{\tau\sigma}},
    \end{split}
\end{equation}
which is a convenient way to package the lower bound on the
asymptotic growth.

The choice of holonomy configurations to optimize over will of
course determine how optimal the bound will be.  As a compromise
between simplicity and robustness of the estimate, we consider the
family based on grouping the $N$ holonomies into packs of $d$
collided holonomies ($x_1=x_2=\cdots=x_d$,
$x_{d+1}=x_{d+2}=\cdots=x_{2d}$, etc.) where $d$ is a divisor of
$N$. There are a total of $N/d$ distinct packs, and they are then
distributed uniformly on the periodic interval $[-1/2,1/2]$ in such
a way that they satisfy the SU$(N)$ condition $\sum_j
x_j\in\mathbb{Z}$.  This latter condition gives rise to $d$ discrete
configurations (which we collectively denote by $\boldsymbol{x}_d$),
signalling a partial breaking $\mathbb{Z}_N\to\mathbb{Z}_{N/d}$ of
the center. These special configurations were shown in \cite{Cabo-Bizet:2019osg} to be saddle point solutions for real holonomies in the large-$N$ limit. Alternatively, they arise as the
hyperbolic (or ``high-temperature'') reduction of the set of
eigenvalue configurations found in \cite{Hong:2018viz}; we will
comment more on this point below.

For a given divisor $d$, since there are $C:=N/d$ packs distributed evenly on the periodic
unit interval, the spacing between packs is $1/C$.  As a result, the
configuration $\boldsymbol{x}_d$  yields
\begin{equation}
    \begin{split}
    Q_h(\boldsymbol{x}_d;\Delta_{1,2})=\frac{1}{12}\sum_{a=1}^3\left((N-1)\kappa(\Delta_a)+N(d-1)\kappa(\Delta_a)+d^2\sum_{J=1}^{C-1}J(\kappa(\Delta_a+\frac{J}{C})+\kappa(\Delta_a-\frac{J}{C}))\right).\label{eq:divisorQh}
    \end{split}
\end{equation}
Here the second term is the contribution of the $d(d-1)/2$ collided
holonomy pairs inside a single pack, with $x_{ij}=0$, and the third
term is the contribution of the holonomy pairs between the different
packs, with $x_{ij}=J/C$.

To simplify (\ref{eq:divisorQh}) further, we use the remarkable
identity
\begin{equation}
    \sum_{J=1}^{n-1}J\left(\kappa(\Delta_a+\frac{J}{n})+\kappa(\Delta_a-\frac{J}{n})\right)=\frac{\kappa(n\Delta_a)}{n}-n\kappa(\Delta_a),\label{eq:simplifyWidentity}
\end{equation}
which can be derived with Mathematica's aid. Applying this to
(\ref{eq:divisorQh}) and substituting in $d=N/C$ then gives
\begin{equation}
    \mathcal{I}\gtrsim\sum_{C|N}\exp\left(-\frac{i\pi}{6\tau\sigma}\sum_{a=1}^3\Big(\frac{N^2}{C^3}\kappa(C\Delta_a)-\kappa(\Delta_a)\Big)\right).\label{eq:dSummedLowerBound}
\end{equation}
The symbol $\gtrsim$ emphasizes that the RHS is only a lower
bound on the asymptotic growth of $\mathcal{I}$ in the CKKN limit.
While we are mainly interested in the W wings, derivation of this
bound is independent of the wings.  In particular, this bound is
optimal on the M wings, where the optimal term is given by $C=1$,
corresponding to the condensation of all holonomies into a single
pack.  On the W wings, however, except for $N=2$ where it reduces to
(\ref{eq:N=2Wwing}), the bound (\ref{eq:dSummedLowerBound}) is not
necessarily optimal, as we will argue below.

\begin{figure}[t]
\centering
    \includegraphics[scale=.6]{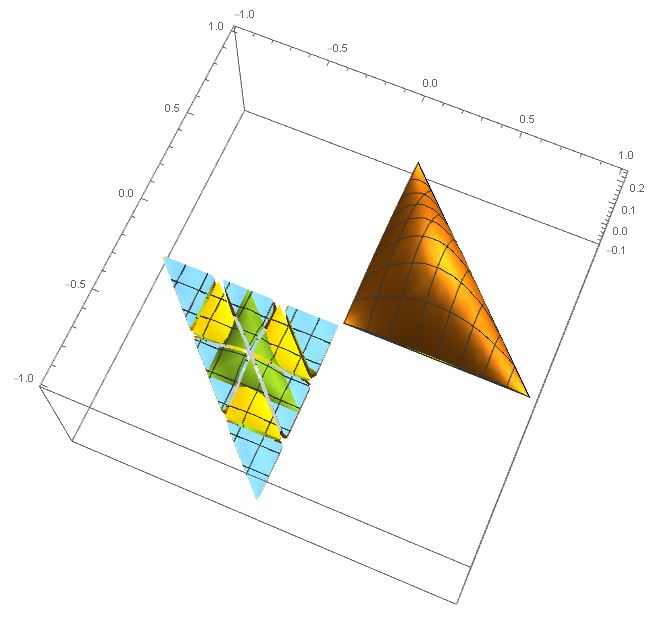}
\caption{The functions $C^{-3}\sum_{a=1}^3\kappa(C\Delta_a)$ for
$C=1$ (brown), $C=2$ (green), $C=3$ (yellow), and $C=6$ (blue).
\label{fig:compareDs}}
\end{figure}

Although the bound (\ref{eq:dSummedLowerBound}) is written as a sum
over `trial' configurations, generically, depending on where exactly
we are on the W wings, only one term would dominate the sum. The
question of which divisor $C$ provides the strongest bound then
boils down to the comparison of
\begin{equation}
    \fft1{C^3}\sum_{a=1}^3\kappa(C\Delta_a),\label{eq:dComparison}
\end{equation}
for various divisors $C$ of $N$.  For $N$ a prime number, there are
only two divisors, namely $C=1$ and $C=N$.  In this case, the answer
is simple: the ``confined'' $C=N$ term dominates the sum in
(\ref{eq:dSummedLowerBound}) on the W wings, while the ``fully
deconfined'' $C=1$ term dominates on the M wings. For composite $N$,
however, other divisors (besides $1$ and $N$) may give the dominant
contribution to the sum in (\ref{eq:dSummedLowerBound}) on the W
wings. For example, for $N=6$, Figure~\ref{fig:compareDs} shows that
while on the M wing the fully condensed $C=1$ term is dominant, on
the W wing there are regions where the other divisors take over the
sum in (\ref{eq:dSummedLowerBound}).

%%%%%
%%%%%

As already emphasized, the sum in (\ref{eq:dSummedLowerBound}) gives
us a lower bound, but not necessarily the true asymptotic growth of
the index. Nevertheless, we conjecture that at least on subsets of
the regions where various divisors become dominant in
(\ref{eq:dSummedLowerBound}), the corresponding term in the sum
actually gives the true asymptotics. In other words, that for any
finite $N$ there are confining or partially deconfining phases on
the W wings.

For small values of $N$ we can attack the extremization problem
numerically of course, and make more precise statements. This is
what we will do next. For example, for $N=6$ we establish that on a
subset of the region in Figure~\ref{fig:compareDs} where the green
curve takes over, the dominant holonomy configuration is indeed the
partially-deconfining ($\mathbb{Z}_6\to \mathbb{Z}_2$) configuration
$\boldsymbol{x}_{d=3}$, that on a subset of the ``yellow region''
the dominant holonomy configuration is the partially-deconfining
($\mathbb{Z}_6\to \mathbb{Z}_3$) configuration
$\boldsymbol{x}_{d=2}$, and that on a subset of the ``blue region''
the dominant holonomy configuration is the confining
($\mathbb{Z}_6$-symmetric) configuration $\boldsymbol{x}_{d=1}$.

A remarkable surprise of the numerical investigation discussed below
is that for $N=3,4$ the bound (\ref{eq:dSummedLowerBound}) is in
fact optimal! Therefore (\ref{eq:dSummedLowerBound}) gives the exact
leading-order asymptotics of the index in these cases. For larger
values of $N$ on the other hand, the numerical analysis shows that
there are indeed regions on the W wings where the bound
(\ref{eq:dSummedLowerBound}) is not optimal.

%%%%%
\subsubsection*{SU$(3)$: infinite-temperature confinement/deconfinement transition}
%%%%%
Our numerical investigation shows that for $N=3$ the bound
(\ref{eq:dSummedLowerBound}) is optimal to within two parts in
$10^{15}$---which is essentially the machine precision. We therefore
conclude that the exact leading asymptotics of the index in this
case reads
\begin{equation}
    \mathcal{I}_{\mathrm{SU}(3)}\sim e^{-\frac{8i\pi}{6\tau\sigma}\sum_{a=1}^3 \kappa(\Delta_a)}+e^{-\frac{i\pi}{6\tau\sigma}\sum_{a=1}^3(\frac{1}{3}\kappa(3\Delta_a)-\kappa(\Delta_a))}.\label{eq:N=3exactAsy}
\end{equation}
Hence, just as in the SU$(2)$ case, we have infinite-temperature
confinement/deconfinement transitions moving from the W wings to the
M wings. On the M wings the first term on the RHS of
(\ref{eq:N=3exactAsy}) dominates, and on the W wings the second
term.

Before moving on to richer cases, once again we emphasize that in
the present paper we are studying the asymptotics on generic points
of the parameter-space. On non-generic points where
$\Delta_a\in\mathbb{Z}$ or $\tau,\sigma\in i\mathbb{R}^+$, the
asymptotic growth would be slower, and a more involved analysis is
required; c.f. section~3 of \cite{ArabiArdehali:2019tdm}.

%%%%%
\subsubsection*{SU$(4)$: infinite-temperature partial deconfinement}
\begin{figure}[t]
\centering
    \includegraphics[scale=.6]{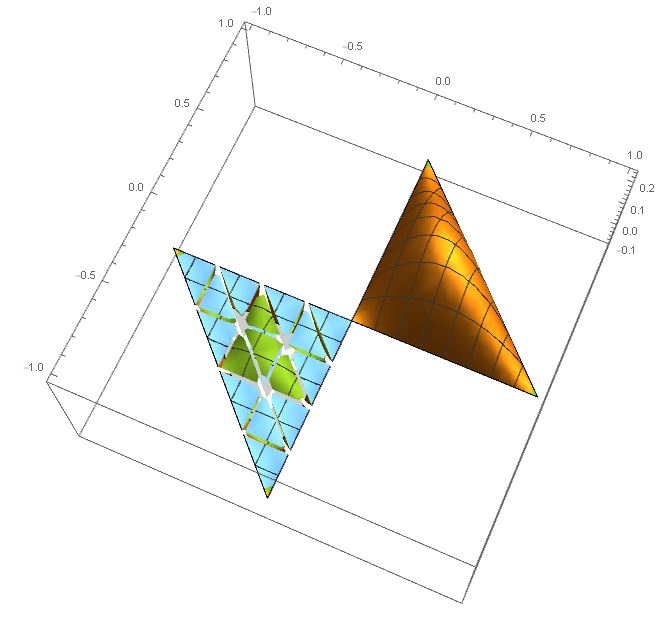}
\caption{The functions $C^{-3}\sum_{a=1}^3\kappa(C\Delta_a)$ for
$C=1$ (brown), $C=2$ (green), and $C=4$ (blue). The take-over of the
green curve signifies the partially deconfined phase in that region
when $N=4$. \label{fig:N=4}}
\end{figure}

%%%%%
In this case as well, the numerical investigation shows that the
bound (\ref{eq:dSummedLowerBound}) is optimal to within two parts in
$10^{15}$. We therefore conclude that the exact leading asymptotics
of the index for $N=4$ reads
\begin{equation}
    \mathcal{I}_{\mathrm{SU}(4)}\sim e^{-\frac{15i\pi}{6\tau\sigma}\sum_{a=1}^3\kappa(\Delta_a)}+e^{-\frac{i\pi}{6\tau\sigma}\sum_{a=1}^3(2\kappa(2\Delta_a)-\kappa(\Delta_a))}+e^{-\frac{i\pi}{6\tau\sigma}\sum_{a=1}^3(\frac{1}{4}\kappa(4\Delta_a)-\kappa(\Delta_a))}.\label{eq:N=4exactAsy}
\end{equation}
The first and the third terms on the RHS come respectively from the
fully-deconfined ($C=1$) and the confined ($C=N=4$) holonomy
configurations. But here we have also the first instance of
infinite-temperature partial deconfinement in the superconformal
index: the middle term on the RHS of (\ref{eq:N=4exactAsy}) takes
over on the middle triangle of the W wings as shown in
Figure~\ref{fig:N=4}. This term corresponds to $C=2$, and signals a
$\mathbb{Z}_4\to\mathbb{Z}_2$ breaking of the center symmetry in the
Cardy-like limit. This qualifies as a \emph{partially deconfined
phase}, not only because of its partial-breaking pattern of the
center symmetry, but also because of its ``partial liberation of the
constituents'' as signified by the fact that the height of the green
curve lies between those of the blue ($C=4$, confined) curve and the
brown ($C=1$, fully-deconfined) curve.

%%%%%
%%%%%

\subsubsection*{SU$(5)$ and SU$(6)$: insufficiency of the divisor configurations}

\begin{figure}[t]
\centering
    \includegraphics[scale=.6]{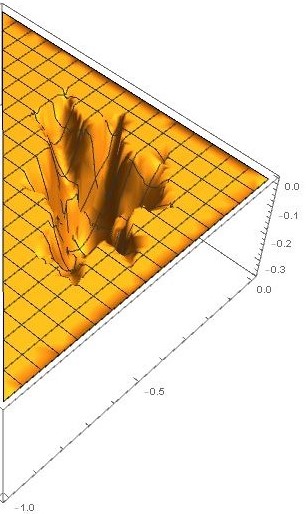}
    \hspace{2cm}
    \includegraphics[scale=.6]{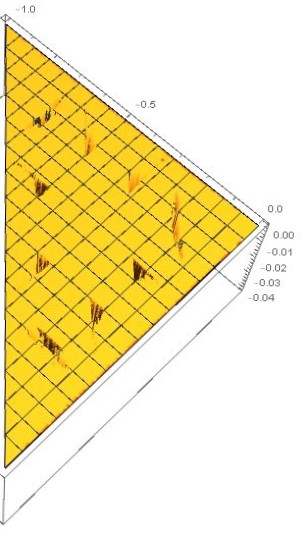}
\caption{The difference (scaled by a factor of 12) between the
numerically maximized $Q_h$, and the $Q_h$ maximized over the
divisor configurations $\boldsymbol{x}_d$, on the $\arg\beta>0$ W wing, for
$N=5$ on the left, and for $N=6$ on the right. When the result is
zero, it means the divisor configurations are maximizing $Q_h$
(hence minimizing $\mathcal{V}_{\mathrm{eff}}$). Note the big hole
in the middle for $N=5$, and the small holes for $N=6$, signalling
the failure of the divisor configurations to maximize $Q_h$ (hence
to minimize $\mathcal{V}_{\mathrm{eff}}$). \label{fig:N=5and6}}
\end{figure}

In these cases the numerical analysis shows that there are regions
on the W wings where none of the divisor configurations
$\boldsymbol{x}_d$ minimizes $\mathcal{V}_{\mathrm{eff}}$. Fixing
$\arg\beta>0$ for concreteness, we see from
Figure~\ref{fig:N=5and6} that in the SU$(5)$ case there is a
relatively large such region, but for SU$(6)$ there are rather small
subsets of the W wing where this happens. Hence the bound
(\ref{eq:dSummedLowerBound}) seems much more efficient in the SU$(6)$
case. This is to be expected of course, as there are three
contributing trial configurations ($C=2,3,6$) on the W wings when
$N=6$, while there is only one such configuration ($C=5$) when
$N=5$.

%%%%%
%%%%%

%%%%%
\subsubsection{Taking the \texorpdfstring{large-$N$}{large-N} limit}\label{sec:Cardy:large-N}
%%%%%
The bound on the asymptotic growth of the index,
(\ref{eq:dSummedLowerBound}), was derived using a family of holonomy
configurations based on divisors $C$ of $N$.  This bound can of
course be improved by enlarging the family of trial configurations.
One way to do this is to divide the $N$ holonomies into $C$ collided
packs with the packs evenly distributed on the periodic interval for
all integer $C=1,2,\ldots,N$.  In general, each pack cannot have the
same number of holonomies unless $C$ is a divisor of $N$.
Nevertheless, we can make the packs nearly uniform by first
distributing $\lfloor N/C\rfloor$ holonomies into each of the $C$
packs.  This leaves $N\bmod C$ holonomies left over, which can then
be distributed in some prescribed manner in the packs.  This set of
trial configurations would in principle improve the bound given in
(\ref{eq:dSummedLowerBound}).  However, the resulting bound would be
sensitive to the particular distribution of the left over $N\bmod C$
holonomies, and can no longer be expressed in such a compact manner.

Although the refined bound that is obtained by splitting the
eigenvalues into $C$ packs for all integers $C$ does not admit a
simple expression for finite $N$, it nevertheless simplifies in the
large-$N$ limit, at least for the leading order growth of the index.
The idea here is that, instead of taking $C=1,2,\ldots,N$, we cut
off the set of trial configurations at some large but finite
$C_{\mathrm{max}}$ that is independent of $N$.  For a given $C$, we
then start with $C$ packs of $\lfloor N/C\rfloor$ holonomies and
compute $\mathcal V_{\mathrm{eff}}$ for this subset of $C\lfloor
N/C\rfloor$ holonomies.  This is of course incomplete, but we can
add in the remaining pairwise potentials, (\ref{eq:QhPot}), between
these uniform holonomies and the $N\bmod C$ remaining ones (as well
as those among the remaining holonomies themselves).  These
interactions between $\mathcal{O}(C)$ objects and $\mathcal{O}(N)$
objects (as well as those among the remaining $\mathcal{O}(C)$
objects) add a correction of at most $\mathcal{O}(N)$ since we keep
the cutoff on $C$ fixed.  Alternatively, starting with $C\lfloor
N/C\rfloor$ instead of $N$ holonomies also leads to a correction of
the same order.  As a result, the leading $\mathcal O(N^2)$ behavior
of the index is captured by (\ref{eq:dSummedLowerBound}) with the
modification that the sum is taken over all integers up to the
cutoff:
\begin{equation}
    \mathcal{I}_{N\to\infty}\gtrsim\sum_{C=1}^{C_{\mathrm{max}}}\exp\left(-\frac{i\pi N^2}{\tau\sigma}\sum_{a=1}^3\frac{\kappa(C\Delta_a)}{6C^3}\right).\label{eq:dSummedLargeN}
\end{equation}
Since we have dropped terms of $\mathcal O(N)$ or smaller, this
asymptotic bound is only valid when considering the $\mathcal
O(N^2/\tau\sigma)$ growth of the index.  In this case, in fact
because the bound applies for any finite
$C_{\mathrm{max}}\in\mathbb{N}$, we can remove the
$C_{\mathrm{max}}$ cutoff and instead take the sum to infinity.

Note that the large-$N$ bound, (\ref{eq:dSummedLargeN}), confirms
that the finite $N$ bound, (\ref{eq:dSummedLowerBound}), is not
optimal in general!  For $N$ a large prime, for example, the finite $N$ bound
would consist of a sum over only the $C=1$ and $C=N$ terms, with the
$C=N$ (or ``confining'') term winning on the W wings.  But we know
(\emph{c.f.} Figure~\ref{fig:compareDs}) that for large enough $N$,
at least on subsets of the W wings, the $C=2,3,\dots$ terms in the
large-$N$ bound (\ref{eq:dSummedLargeN}) dominate over the confining
term.  This is of course a simple result of enlarging the set of
trial configurations to include more general $C$ collided packs of
holonomies for all integer $C$, whether $C$ divides $N$ or not.

Returning to the large-$N$ analysis, we see that as long as at least one term in (\ref{eq:dSummedLargeN}) has a
positive real part in the exponent, the index will exhibit $\mathcal
O(N^2)$ growth in the Cardy-like limit.  This corresponds to either
full deconfinement when the $C=1$ term dominates, or partial
deconfinement when some $C>1$ term dominates. As discussed above,
the $C=1$ term always dominates in the M wings, even at finite $N$,
with the resulting behavior given by (\ref{eq:Mgrowth}). On the
other hand, the situation is more elaborate in the W wings. Let us
fix $\arg\beta<0$ for concreteness; then the W wing consists
of all $\Delta_{1,2}$ subject to
$0<\Delta_1,\Delta_2,1-\Delta_1-\Delta_2<1$. The question then becomes
whether for any such $\Delta_{1,2}$ we can find a $C\in\mathbb{N}$ such
that $\sum_a \kappa(C\Delta_a)<0$. We now argue that this is the case! In fact since $\kappa(C\Delta_a)$
is periodic under $\Delta_{1,2}\to\Delta_{1,2}+1/C$, we can simply
focus on the square $0<\Delta_{1,2}<1/C$. Now, it follows from the scaling $\Delta_a\to\Delta_a/C$ that on this
square the sign of $\sum_a \kappa(C\Delta_a)$ is positive (resp.
negative) if the representatives $\{C\Delta_{1,2}\}/C$ of
$\Delta_{1,2}$ on the square $0<\Delta_{1,2}<1/C$ lie on the lower
triangle with vertices $(0,0),(0,1/C),(1/C,0)$ (resp. the upper
triangle with vertices $(0,1/C),(1/C,0),(1/C,1/C)$). Hence the question
boils down to \emph{whether we can find a $C$ such that the
representatives are on the upper triangle} where
$\{C\Delta_{1}\}/C+\{C\Delta_{2}\}/C>1/C$. (The interested reader might find that a simple drawing of the said triangles would render the previous sentences obvious.) The following lemma
answers this question in the positive.

\begin{lemma}
For every pair of real numbers $x,y$ subject to $0<x,y,1-x-y<1$,
there exists a natural number $C>1$ such that $\{Cx\}+\{Cy\}>1$.
\end{lemma}
An elementary proof of this lemma can be found in the
appendix.\footnote{We learned the proof, as well as the following
remark regarding the Kronecker-Weyl theorem, from David~E~Speyer, a
mathematician at University of Michigan.} Here we instead point out
that it follows from a much stronger result, often
associated\footnote{See
https://mathoverflow.net/questions/162875/reference-for-kronecker-weyl-theorem-in-full-generality.}
to the names Kronecker and Weyl, that if there is no integer
relationship between $x,y$ (i.e. no solution to $ax+by+c=0$ in
integers other than $(a,b,c)=(0,0,0)$) then the points
$(\{Cx\},\{Cy\})$ are \emph{dense} in the unit torus. (In our case,
even if there is such an integer relationship between
$\Delta_1,\Delta_2$, we can always establish our desired result by
applying the Kronecker-Weyl theorem to $\Delta_1,\Delta_2+\epsilon$,
with a small-enough $\epsilon$ chosen such that there is no integer
relationship between $\Delta_1,\Delta_2+\epsilon$.)

Similar arguments apply when $\arg\beta>0$. We thus
conclude that for all points strictly inside the W wings, there
exists a natural number $C>1$ such that the exponent of the ``$C$-th
bound'' in (\ref{eq:dSummedLargeN})
\begin{equation}
    \mathcal{I}_{N\to\infty}\gtrsim\exp\left(-\frac{i\pi N^2}{\tau\sigma}\sum_{a=1}^3\frac{\kappa(C\Delta_a)}{6C^3}\right),\label{eq:oneTermBound}
\end{equation}
has positive real part, and hence the index is partially deconfined.

A ``non-deconfined'' behavior (i.e. $o(N^2)/\tau\sigma$ growth for
$\log \mathcal{I}$ as $N\to\infty$ after the Cardy-like limit) might
appear in the non-generic situations where $\arg\beta=0$
(c.f. section~3 of \cite{ArabiArdehali:2019tdm}), or
$\Delta_a\in\mathbb{Z}$. In such cases, subdominant terms of $O(N)$
or smaller may be important in order to fully pin down the behavior
of the index.

With some optimism, this genericity of partial deconfinement on the W wings can be taken as a sign that it would be consistent to conjecture that the large-$N$ bound (\ref{eq:dSummedLargeN}),
with the cut-off removed, gives not just a lower bound but the
\emph{actual leading asymptotics} of the index.
\begin{conjecture}
The leading asymptotics of the superconformal index (\ref{eq:EHI})
of the 4d $\mathcal{N}=4$ theory with SU$(N)$ gauge group, in the
CKKN limit (\ref{eq:CKKNlimit}), simplifies as $N\to\infty$ to
\begin{equation}
    \mathcal{I}_{N\to\infty}\sim\sum_{C=1}^{\infty}\exp\left(-\frac{i\pi N^2}{\tau\sigma}\sum_{a=1}^3\frac{\kappa(C\Delta_a)}{6C^3}\right),
    \label{eq:doubleScalingConjecture}
\end{equation}
with the error such that logarithms of the two sides differ by
$o(N^2/\tau\sigma)$.
\end{conjecture}
This conjecture is motivated in part by the following two observations: $i)$
that in the $N\to\infty$ limit there are infinitely many trial
configurations, and hence increasing chance of their sufficiency;
$ii)$ that already for $N$ as small as $6$, as witnessed by
Figure~\ref{fig:N=5and6}, the divisor configurations go a long way
towards minimizing $\mathcal{V}_{\mathrm{eff}}$ on the W wings.

Note that Lemma~1 implies that for generic $\Delta_{1},\Delta_{2}$ \emph{at least one} of the exponentials on the right-hand side of (\ref{eq:doubleScalingConjecture}) has absolute value greater than one. The stronger Kronecker-Weyl theorem implies that there are actually \emph{infinitely many} such exponentials on the right-hand side of (\ref{eq:doubleScalingConjecture}). Therefore the infinite sum in (\ref{eq:doubleScalingConjecture}) is not actually convergent, but should be considered as an asymptotic expression, with only one term in the infinite sum dominating for generic $\Delta_{1},\Delta_{2}$.

%%%%%
\subsubsection*{Entropy of the partially deconfined phases}
%%%%%
Let us now focus on the $C$-th term in
(\ref{eq:doubleScalingConjecture}), and see what entropy it bears.\footnote{Even if Conjecture~1 turns out to be incorrect for generic $\Delta_{1,2}$, it might very well be correct in the vicinity of the critical points we find below, and this would be enough for the following entropy calculation to be valid.}
First, we rewrite the $C$-th term explicitly as
\begin{equation}
    \exp\bigg(-\fft{\pi iN^2}{\tau\sigma}\sum_{a=1}^3\fft{\kappa(C\Delta_a)}{6C^3}\bigg)=\exp\bigg(-\fft{\pi i N^2}{\tau\sigma}\fft{\langle C\Delta_1\rangle\langle C\Delta_2\rangle\langle C\Delta_3\rangle}{C^3}\bigg),\label{C-th}
\end{equation}
where we have defined $\langle C\Delta_a\rangle$ as
\begin{equation}
    \langle C\Delta_a\rangle=\begin{cases}
    \{C\Delta_a\}-1 & \text{if }\{C\Delta_1+C\Delta_2\}=\{C\Delta_1\}+\{C\Delta_2\}-1;\\
    \{C\Delta_a\} & \text{if }\{C\Delta_1+C\Delta_2\}=\{C\Delta_1\}+\{C\Delta_2\}.
    \end{cases}
\end{equation}
The corresponding entropy $S_C(J_{1,2},Q_a)$ is obtained by
performing a Legendre transform of (\ref{C-th}), which requires
adding $-2\pi i(\sigma J_1+\tau J_2+\sum_a\Delta_aQ_a)$ in the
exponent, and then extremizing. The first step can be written explicitly as
\begin{align}
    &\exp[\hat S_C(J_{1,2},Q_a;\Delta_a,\sigma,\tau)]\nn\\
    &=\exp\left[-\fft{2\pi i}{C}\left(\fft{N^2}{2(C\tau)(C\sigma)}\langle C\Delta_1\rangle\langle C\Delta_2\rangle\langle C\Delta_3\rangle+(C\sigma)J_1+(C\tau)J_2+\sum_{a=1}^3\langle C\Delta_a\rangle Q_a\right)\right],\label{Entropy:fct}
\end{align}
where we have replaced $C\Delta_a$ with $\langle C\Delta_a\rangle$
in the last term; this replacement is allowed assuming $Q_a\in C\mathbb Z$, which we do for simplicity, because then since the difference between $C\Delta_a$ and $\langle C\Delta_a \rangle$ is an integer it would not change the exponential when multiplied by $-2\pi i Q_a/C$.
Extremizing the function $\hat S_C(J_{1,2},Q_a;\Delta_a,\sigma,\tau)$ in (\ref{Entropy:fct}) with respect to
the chemical potentials $\Delta_a,\sigma,\tau$ under the constraint
$\sum_a\Delta_a-\sigma-\tau\in\mathbb Z$ determines the entropy as
\begin{equation}
    S_C(J_{1,2},Q_a)=\hat S_C(J_{1,2},Q_a;\Delta_a^C,\sigma^C,\tau^C),
\end{equation}
where $\Delta_a^C$, $\sigma^C$, and $\tau^C$ denote the critical
points under the aforementioned constraint. With an appropriate relation between $J_{1,2},Q_a$ the resulting entropy will be a real number \cite{Choi:2018hmj}, and hence acceptable.

Note that the expression inside the round bracket in
(\ref{Entropy:fct}) with general $C$ reduces to the expression with
$C=1$ under $\langle C\Delta_a\rangle\to\langle\Delta_a\rangle$,
$C\sigma\to\sigma$, and $C\tau\to\tau$. In other words, this simple
replacement maps the present problem to that of the $C=1$
(single-center) black hole entropy problem. We thus find
\begin{equation}
    S_C(J_{1,2},Q_a)=S_{\mathrm{BH}}(J_{1,2},Q_a)/C,\label{C:entropy}
\end{equation}
as alluded to in section \ref{sec:intro}.

Using the same property, we can also figure out the critical points
with general $C$ directly from the known ones with $C=1$. Since our
analysis is restricted to real $\Delta_{1,2}$, however, we should
focus on the equal-charge case where all $Q_a$'s are equal to each
other (\emph{c.f.} the end of section 2 in
\cite{ArabiArdehali:2019tdm}). In that case, the critical points
with $C=1$ have been already known as
\begin{equation}
    \langle\Delta_a^{C=1}\rangle=\pm\fft13\quad\Leftrightarrow\quad (\Delta_1^{C=1},\Delta_2^{C=1})=\Big(\pm\fft{1}{3},\,\pm\fft{1}{3}\Big),
\end{equation}
with the signs the same as that of $\arg\beta$. The critical points with general $C$ are then determined as
\begin{align}
    \langle C\Delta_a^C\rangle=\pm\fft13\quad\Leftrightarrow\quad (\Delta_1^C,\Delta_2^C)=\Big(\pm\fft{1}{3C}+\fft{j}{C},\,\pm\fft{1}{3C}+\fft{k}{C}\Big),\label{C:critical}
\end{align}
where $j,k$ are arbitrary integers. For $C=2$ as an example, fixing $\arg(\beta)>0$ for concreteness,
we have the critical point at $(\Delta_1,\Delta_2)=(-\fft13,-\fft13)$ on the W wing. The interested reader is encouraged to locate the $C=3$ critical points in Figure~\ref{fig:compareDs}.

An interesting
question is whether at the critical point the $C$-th term in
(\ref{eq:dSummedLargeN}) is indeed dominant; otherwise the entropy derivation would not be self-consistent. Curiously, a numerical
investigation shows that the answer is positive for $C\le5$, and negative for general $C>5$.
The interpretation of this result is not yet clear to us.

%%%%%
\section{Comparison with the Bethe Ansatz type approach}\label{sec:BA}
%%%%%

It was argued in \cite{Closset:2017bse,Benini:2018mlo} that the
index (\ref{eq:indexDefN=4}) can be rewritten as a Bethe Ansatz type
formula.  One advantage of this reformulation is that the integral
over the Coulomb branch in (\ref{eq:EHI}) is replaced by a sum over
solutions to a set of Bethe Ansatz like equations.  This was the
approach used in \cite{Benini:2018ywd} to obtain the black hole
microstate counting in the large-$N$ limit.  Here we briefly review
the Bethe Ansatz approach (BA approach) to the index and then
demonstrate how the partially deconfined phases identified in the
previous section emerge in this approach.

%%%%%
\subsection{The Bethe Ansatz type expression for the index}\label{sec:BA:review}
%%%%%

For simplicity, we restrict to $p=q$ (i.e.
$\tau=\sigma=\fft{i\beta}{2\pi}$) in the index.  In this case, the
Bethe Ansatz type formula reads \cite{Benini:2018mlo,Benini:2018ywd}
\begin{align}
    \mathcal I(q,q,y_{1,2,3})=\alpha_N(\tau)\sum_{\hat u\in\mathrm{eBAEs}}\mathcal Z(\hat u;\Delta_a,\tau)H(\hat u;\Delta_a,\tau)^{-1},\label{eq:index:BA}
\end{align}
where $\alpha_N(\tau)=\fft{1}{N!}\prod_{k=1}^\infty(1-e^{2\pi
ik\tau})^{2(N-1)}$ and $\hat u=\{u_1,\cdots,u_N\}$ denotes all
possible solutions to the following system of elliptic Bethe Ansatz
equations (eBAEs),
\begin{equation}
    1=Q_i(\hat u;\Delta_a,\tau):=e^{2\pi i(\lambda+3\sum_{j}u_{ij})}\prod_{j=1}^N\fft{\theta_0(u_{ji}+\Delta_1;\tau)\theta_0(u_{ji}+\Delta_2;\tau)\theta_0(u_{ji}-\Delta_1-\Delta_2;\tau)}{\theta_0(u_{ij}+\Delta_1;\tau)\theta_0(u_{ij}+\Delta_2;\tau)\theta_0(u_{ij}-\Delta_1-\Delta_2;\tau)},\label{eq:BAE}
\end{equation}
under the SU$(N)$ constraint $\sum_{i=1}^Nu_i\in\mathbb Z+\tau\mathbb Z$ and the
abbreviation $u_{ij}=u_i-u_j$. The third chemical potential
$\Delta_3$ is constrained via
$\Delta_3=2\tau-\Delta_1-\Delta_2~(\mathrm{mod}~\mathbb Z)$ as in
the previous section. Note that $\lambda$ is a free parameter
independent of $i$. Here we have introduced $\mathcal Z(\hat
u;\Delta_a,\tau)$ and $H(\hat u;\Delta_a,\tau)$ as
\begin{subequations}
    \begin{align}
    \mathcal Z(\hat u;\Delta_a,\tau)&=\left(\prod_{a=1}^3\tilde{\Gamma}^{N-1}(\Delta_a;\tau,\tau)\right)\prod_{i,j=1\,(i\neq j)}^{N}\frac{\prod_{a=1}^{3}\tilde{\Gamma}(u_{ij}+\Delta_a;\tau,\tau)}{\tilde{\Gamma}(u_{ij};\tau,\tau)},\label{eq:Z}\\
    H(\hat u;\Delta_a,\tau)&=\det\left[\fft{1}{2\pi i}\fft{\partial(Q_1,\cdots,Q_N)}{\partial(u_1,\cdots,u_{N-1},\lambda)}\right],\label{eq:H}
    \end{align}\label{ZH}%
\end{subequations}
and the elliptic functions $\theta_0(u;\tau)$ and $\tilde
\Gamma(u;\tau,\sigma)$ are defined as $(z=e^{2\pi i u},q=e^{2\pi i
\tau},p=e^{2\pi i \sigma})$
\begin{align}
    \theta_0(u;\tau)&=(1-z)\prod_{k=1}^\infty(1-zq^k)(1-z^{-1}q^k),\label{def:theta0}\\
    \tilde\Gamma(u;\sigma,\tau)&=\Gamma(z;p,q).\label{def:Gamma}
\end{align}
Note that $H(\hat u;\Delta_a,\tau)$ has to be evaluated at the
solutions to the eBAEs after taking the partial derivatives of
$Q_i$'s with respect to $u_i$'s.

Since the right-hand side of (\ref{eq:index:BA}) is summed over all possible
solutions to the eBAEs (\ref{eq:BAE}), the first step towards the
computation of the index (\ref{eq:index:BA}) is to find the most
general solutions to the eBAEs (\ref{eq:BAE}).  Here we find it
convenient to use 
the relation 
\begin{equation}
\begin{split}
	\theta_1(u;\tau)&=-ie^{\fft{\pi i\tau}{4}}(e^{\pi iu}-e^{-\pi iu})\prod_{k=1}^\infty(1-e^{2\pi ik\tau})(1-e^{2\pi i(k\tau+u)})(1-e^{2\pi i(k\tau-u)})\\
	&=ie^{\fft{\pi i\tau}{4}}e^{-\pi iu}\prod_{k=1}^\infty(1-e^{2\pi ik\tau})\theta_0(u;\tau),\label{theta1:product}
\end{split}
\end{equation}
to rewrite the eBAEs (\ref{eq:BAE}) in terms of the Jacobi
theta function $\theta_1(u;\tau)$ as
\begin{equation}
    1=Q_i=e^{2\pi i\lambda}\prod_{j=1}^N\fft{\theta_1(u_{ji}+\Delta_1;\tau)}{\theta_1(u_{ij}+\Delta_1;\tau)}\fft{\theta_1(u_{ji}+\Delta_2;\tau)}{\theta_1(u_{ij}+\Delta_2;\tau)}\fft{\theta_1(u_{ji}-\Delta_1-\Delta_2;\tau)}{\theta_1(u_{ij}-\Delta_1-\Delta_2;\tau)},\label{BAE}
\end{equation}
These eBAEs are a set of highly non-linear equations, and it seems
rather challenging to find the most general solutions.  However,
in the form (\ref{BAE}), these equations coincide with those for the topologically twisted index of $\mathcal
N=4$ SYM on $T^2\times S^2$ \cite{Hosseini:2016cyf}. In \cite{Hong:2018viz} oddity and quasi-periodicity of $\theta_1(u;\tau)$ with respect to the first argument
\begin{equation}
\begin{split}
	\theta_1(u;\tau)&=-\theta_1(-u;\tau),\\
	\theta_1(u+l+k\tau;\tau)&=(-1)^le^{-2\pi iku}e^{-\pi ik^2\tau}\theta_1(u;\tau),\qquad (k,l\in\mathbb{Z})
\label{theta1:property}
\end{split}
\end{equation}
were used to find a large set
of solutions to (\ref{BAE}). 
These
solutions are denoted in terms of three non-negative integers
$\{m,n,r\}$ with $N=mn$ and $r=0,1,\ldots,n-1$, and have the $u_i$'s
distributed as
\begin{align}
    \hat u_{\{m,n,r\}}=\left\{u_{\hat j\hat k}=\bar u+\fft{n\hat j+r\hat k}{N}+\fft{\hat k}{n}\tau\bigg|N=mn,~0\leq r<n,~(\hat j,\hat k)\in\mathbb Z_m\times\mathbb Z_n\right\},\label{BAE:sols:standard}
\end{align}
where $\bar u$ is determined by the SU$(N)$ constraint $\sum_{\hat
j}\sum_{\hat k}u_{\hat j\hat k}\in\mathbb{Z}+\tau\mathbb{Z}$.  In essence, these $\{m,n,r\}$
solutions correspond to regular distributions of the $N$ holonomies
over the fundamental domain of the torus specified by $(1,\tau)$%
\footnote{When $\gcd(m,n,r)=1$, these distributions are equivalent to those labeled by $(m',n')$ with $\hat u_{(m',n')}=\{u_i=\bar u+(m'\tau+n')i/N|\gcd(m',n',N)=1,0\le i<N\}$ considered in the saddle point analysis of \cite{Cabo-Bizet:2019eaf}, with $m={N}/{n}=\gcd(N,m')$ and $r=n'b~(\mathrm{mod}~n)$ where the integers $a,b$ are determined by $1=na+(m'/m)b$.}.
We will refer to these as the \emph{standard} solutions to the eBAEs
(\ref{eq:BAE}) and refer to the $u_i$'s as holonomies in analogy to
the holonomies $x_i$ in the integral representation of the index.

It turns out, however, that the standard solutions,
(\ref{BAE:sols:standard}), are in fact not the most general
solutions to the eBAEs.  In a way, this is not particularly
surprising because of the non-linear nature of the equations.  We
will refer to the additional solutions that do not fall into the
class of (\ref{BAE:sols:standard}) as \emph{non-standard} solutions.
Such solutions will not correspond to a periodic tiling of the
fundamental domain and moreover may depend on the chemical
potentials $\Delta_k$.  The Bethe Ansatz form for the index is then
a sum over standard and non-standard solutions, which we denote
schematically as
\begin{align}
    \mathcal I(q,q,y_{1,2,3})=\sum_{n|N}\sum_{r=0}^{n-1}I_{\{N/n,n,r\}}(\Delta_a,\tau)+\sum_{\mathrm{non\mathchar`-standard}~\hat u}I_{\hat u}(\Delta_a,\tau).\label{eq:index:BA:2}
\end{align}
Note that the contribution from the standard solutions to the index,
dubbed as $I_{\{m,n,r\}}(\Delta_a,\tau)$ in
(\ref{eq:index:BA:2}), can be written explicitly as
\begin{align}
    &I_{\{m,n,r\}}(\Delta_a,\tau)\nn\\
    &=\fft{\alpha_N(\tau)\prod_{a=1}^3\tilde\Gamma^{N-1}(\Delta_a;\tau,\tau)}{H(\hat u_{\{m,n,r\}};\Delta_a,\tau)}\prod_{(\hat j,\hat k),(\hat j',\hat k')\in\mathbb Z_m,\mathbb Z_n}^{(\hat j,\hat k)\neq(\hat j',\hat k')}\fft{\prod_{a=1}^3\tilde\Gamma(\fft{\hat j-\hat j'}{m}+\fft{\hat k-\hat k'}{n}(\tau+\fft{r}{m})+\Delta_a;\tau,\tau)}{\tilde\Gamma(\fft{\hat j-\hat j'}{m}+\fft{\hat k-\hat k'}{n}(\tau+\fft{r}{m});\tau,\tau)},\label{eq:Zmnr}
\end{align}
by substituting (\ref{BAE:sols:standard}) into
(\ref{eq:index:BA},\,\ref{eq:Z}).

%%%%%
\subsection{The Cardy-like limit of the index}\label{sec:BA:Cardy}
%%%%%

In Section~\ref{sec:Cardy}, we were able to bound the asymptotic
growth of the index in the Cardy-like limit based on a set of trial
holonomy configurations with the $N$ holonomies distributed among
$C$ packs of collided holonomies.  For finite $N$, we took $C$ to be
a divisor of $N$ so that all packs contain an equal number $N/C$ of
holonomies and arrived at the bound (\ref{eq:dSummedLowerBound}).
This bound can be improved in the large-$N$ limit by removing the
requirement that $N/C$ is an integer; the $\mathcal O(N^2)$ behavior
of the index is then governed by (\ref{eq:dSummedLargeN}). In this
subsection we reproduce the same results in the BA approach.

%%%%%
\subsubsection{Standard solutions and the asymptotic bound (\ref{eq:dSummedLowerBound})}\label{sec:bound}
%%%%%
At first it may not be obvious how (\ref{eq:dSummedLowerBound}) can
be obtained in the BA approach.  However, the connection can be made
by taking the $\tau\to0$ limit of the standard solutions for the
holonomies, given in (\ref{BAE:sols:standard}). They reduce in this
\emph{hyperbolic} (or ``high-temperature'') limit to
\begin{align}
    \hat u_{\{m,n,r\}}=\left\{u_{\hat j\hat k}=\bar u+\fft{n\hat j+r\hat k}{N}\bigg|(\hat j,\hat k)\in\mathbb Z_m\times\mathbb Z_n\right\}.\label{holonomy:highT}
\end{align}
This distributes the $N$ holonomies in the unit interval spaced in
multiples of $1/N$.  In fact, it is not difficult to see that
periodicity of the holonomies ensures that (\ref{holonomy:highT}) is
equivalent to the $N$ holonomies distributed evenly into $C$
distinct packs of $N/C$ collided holonomies with
\begin{equation}
    C=\fft{N}{\gcd(n,r)}.
    \label{eq:CardyC}
\end{equation}
(In the special case $r=0$ we define $\gcd(n,0):=n$.) This indeed corresponds directly to the holonomy configurations
considered in Section~\ref{sec:Cardy} which led to the asymptotic
bound, (\ref{eq:dSummedLowerBound}), for the finite-$N$ index in the
Cardy-like limit.

To derive (\ref{eq:dSummedLowerBound}) in the BA approach more
explicitly, we compute the contributions from the standard solutions
to the index, namely $I_{\{m,n,r\}}(\Delta_a,\tau)$ given
in (\ref{eq:Zmnr}), in the Cardy-like limit. The result, whose derivation we now outline, confirms that they
saturate the asymptotic bound in (\ref{eq:dSummedLowerBound}). First, since we are interested in the $1/|\tau|^2$ leading behavior of $\log\mathcal{I}$, the contribution
$\log\alpha_N(\tau)$ from the prefactor is negligible. Next, the Jacobian determinant $H(\hat u_{\{m,n,r\}};\Delta_a,\tau)$ involves derivatives of theta functions which are of order $\mathcal O(e^{-1/|\tau|})$ \cite{Hong:2018viz} in general\footnote{We did not rule out the possibility that for some non-generic values of $\Delta_{1,2}$ the Jacobian determinant evaluated for a standard solution can be exactly zero. This would be rather \emph{unnatural}, as the standard solutions are expected to be isolated. At any rate, we exclude such pathological $\Delta_{1,2}$ from our consideration.} and therefore $-\log H(\hat u_{\{m,n,r\}};\Delta_a,\tau)$ does not contribute to the $1/|\tau|^2$ leading order of $\log\mathcal{I}$. Hence the leading contribution to $\log\mathcal{I}$ comes entirely from $\log\mathcal
Z_{\{m,n,r\}}(\Delta_a,\tau)$, which can be estimated using the
following asymptotic formula for the elliptic gamma function:
\begin{align}
    \log\tilde\Gamma(u;\tau,\tau)=-\fft{\pi i}{6\tau^2}\kappa_\tau(u)+\mathcal O\left(\fft{1}{|\tau|}\right),\label{Gamma:asymptotic}
\end{align}
where $\kappa_\tau(u)$ is defined as
\begin{align}
    \kappa_\tau(x)=\{x\}_\tau(1-\{x\}_\tau)(1-2\{x\}_\tau)\quad\mathrm{with}\quad\{x\}_\tau=x-\lfloor\Re x+\tan(\arg\beta)\Im x\rfloor.\label{eq:kappa:tau}
\end{align}
This $\kappa_\tau(x)$ is a generalization of $\kappa(x)$ introduced
in (\ref{eq:kappa}) to complex domain \cite{ArabiArdehali:2019tdm}.
Note that $\{x\}_\tau$ and $\kappa_\tau(x)$ reduce repectively to $\{x\}$ and
$\kappa(x)$ for $x\in\mathbb R$.

The $1/|\tau|^2$-leading order behavior of the index is then
obtained from the asymptotic expansion of (\ref{eq:Z})
\begin{equation}
\log I(\hat u;\Delta_a,\tau)=-\fft{\pi i}{6\tau^2}
\sum_a\left((N-1)\kappa_\tau(\Delta_a)+\sum_{i\ne
j}\kappa_\tau(\hat{u}_{ij}+\Delta_a)\right)+\mathcal O(1/\tau),
\label{eq:logZcardy}
\end{equation}
where we made use of the relation
$\kappa_\tau(u)+\kappa_\tau(-u)=0$.  The connection to the integral
approach to the index is now manifest, as this reproduces
$Q_h(\mathbf{x};\Delta)$ in (\ref{eq:QhDef}) obtained in the CKKN
limit of (\ref{eq:EHI}).  Of course, here, the holonomies are not
integrated over, but rather are taken to satisfy the eBAEs,
(\ref{eq:BAE}) or equivalently (\ref{BAE}).

In the hyperbolic limit, the standard solutions all reduce to
(\ref{holonomy:highT}), which is equivalent to dividing them into
$C$ distinct packs of collided holonomies.  Substituting any one of
these solutions into (\ref{eq:logZcardy}) then necessarily gives the
same result that was previously obtained in the CKKN limit.  In
particular, taking $\Delta_1,\Delta_2\in\mathbb R$ and noting that
$\kappa_\tau(x)$ reduces to $\kappa(x)$ for real arguments, we
obtain
\begin{align}
    \log I_{\{m,n,r\}}(\Delta_a,\tau)&=-\fft{\pi i}{6\tau^2}\sum_{a=1}^3\left(\fft{N^2}{C^3}\kappa(C\Delta_a)-\kappa(\Delta_a)\right)+\mathcal O\left(1/\tau\right),\label{eq:Zmnr:highT}
\end{align}
where $C=N/\gcd(n,r)$.

In the Cardy-like limit, the Bethe Ansatz form of the index,
(\ref{eq:index:BA:2}), then takes the form
\begin{align}
    \mathcal I(q,q,y_{1,2,3})&=\sum_{C|N}d(C)\exp\left(-\fft{\pi i}{6\tau^2}\sum_{a=1}^3\left(\fft{N^2}{C^3}\kappa(C\Delta_a)-\kappa(\Delta_a)\right)+\mathcal O\left(1/\tau\right)\right)\nn\\
    &\quad+\sum_{\mathrm{non\mathchar`-standard}~\hat u}I_{\hat u}(\Delta_a,\tau),
\end{align}
where we have made explicit the degeneracy factor $d(C)$ counting the number of distinct
standard solutions that gives rise to a given $C$, though it of course does
not contribute to the $1/|\tau|^2$ leading order.  If we discard the
contributions of any possible non-standard solutions, then this
saturates the asymptotic bound obtained earlier as
(\ref{eq:dSummedLowerBound}).  Recall that, in the integral
approach, the bound was obtained by taking a set of trial
configurations corresponding to $C$ equal packs of collided
holonomies.  Here, the Bethe Ansatz approach uses the same set of
holonomy configurations, so it is no surprise that the final
asymptotic expression for the index is the same.  However, in this
approach, the holonomy configurations are not just trial
configurations, but are exact solutions to the eBAEs, and remain
exact solutions even away from the Cardy-like limit.

%%%%%
\subsubsection{Non-standard solutions and the improved asymptotic bound (\ref{eq:dSummedLargeN})}\label{sec:improved:bound}
%%%%%

As demonstrated in the CKKN limit of the elliptic hypergeometric
integral, the basic asymptotic bound, (\ref{eq:dSummedLowerBound}),
can be improved by enlarging the set of trial configurations to
encompass $C$ nearly uniform packs of collided holonomies for all
integers $C$, not necessarily dividing $N$.  In the BA approach,
however, we cannot arbitrarily choose \textit{trial} configurations;
rather we are limited to solutions of the eBAEs.  Since the standard
solutions only allow for values of $C$ that divide $N$, they cannot
generate the improved bound, (\ref{eq:dSummedLargeN}).  This
strongly suggests that the set of standard solutions is in fact
incomplete and that we need non-standard solutions resembling
arbitrary $C$ packs where $C$ does not divide $N$.

As an example, consider the case $C=2$.  When $N$ is even, this is
realized by a standard solution, which corresponds in the Cardy-like
limit to $N/2$ holonomies taking the value $u_i=0$ and another $N/2$
holonomies taking the value $u_i=1/2$ (up to a constant $\bar u$
shift enforcing the SU$(N)$ condition).  Away from the Cardy-like
limit, this splits into two eBAE solutions, the first corresponding
to $\{m,n,r\}=\{2,N/2,0\}$, and the second to $\{1,N,N/2\}$.  In
both cases, the $N/2$ holonomies in each pack are evenly distributed
along the periodic $\tau$ direction, although in the second solution
the two packs are offset by $\tau/N$ while in the first solution
they are not.

The more interesting case is how $C=2$ is realized when $N$ is odd.
The expectation here is that there must be a non-standard solution
where the holonomies are split into two packs.  Although we cannot
equally divide an odd number of holonomies, we can imagine grouping
$(N+1)/2$ of them into one pack and $(N-1)/2$ of them in the other.
While the first pack has one additional holonomy, this should become
unimportant in the large-$N$ limit.  Nevertheless, we demonstrate
that such a non-standard solution exists, even for finite $N$.

Before describing this non-standard $C=2$ solution, recall that we
assumed $\Delta_1,\Delta_2\in\mathbb R$ in section \ref{sec:Cardy}
so we make the same assumption here. Then, without loss of
generality, we can set
$0<\Delta_1\leq\Delta_2\leq1-\Delta_1-\Delta_2<1$ using the
invariance of the eBAEs (\ref{BAE}) under
$\Delta_{1,2}\to\Delta_{1,2}+\mathbb Z$ and
$\Delta_{1,2}\to-\Delta_{1,2}$ (see Appendix \ref{App:C} for
details). Furthermore, we assume
$\Delta_1,\Delta_2,1-\Delta_1-\Delta_2$ take different values and
are not asymptotically close to each other ($\sim \mathcal
O(|\tau|)$ in the Cardy-like limit, for example) to avoid any
potentially complicated behavior near Stokes lines.  Based on this
setup, we find that the non-standard $C=2$ solution falls into two
cases, depending on whether $\Delta_1+\Delta_2\leq\fft12$ or
$\Delta_1+\Delta_2>\fft12$.

%%%%%
\subsubsection*{Case 1. $\Delta_1+\Delta_2\leq\fft12$}\label{case1}
%%%%%

\begin{figure}[t]
    \centering
    \begin{subfigure}[t]{.49\linewidth}
        \centering
        \includegraphics[width=1\linewidth]{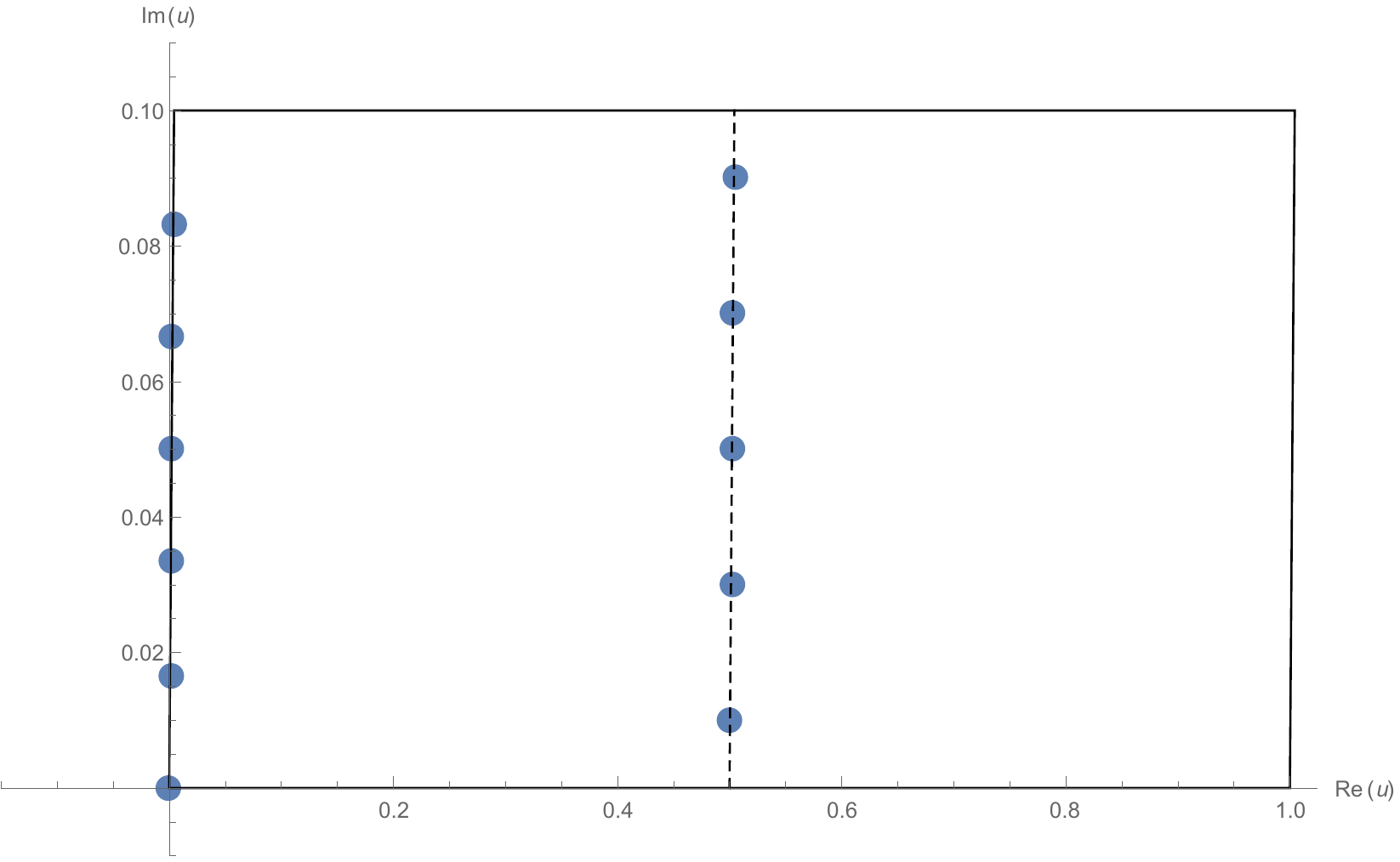}
        \caption{$(\Delta_1,\Delta_2)=(\fft{105}{517},\fft{75}{287})$}
        \label{C=2:highT:case1}
    \end{subfigure}
    \begin{subfigure}[t]{.49\linewidth}
        \centering
        \includegraphics[width=1\linewidth]{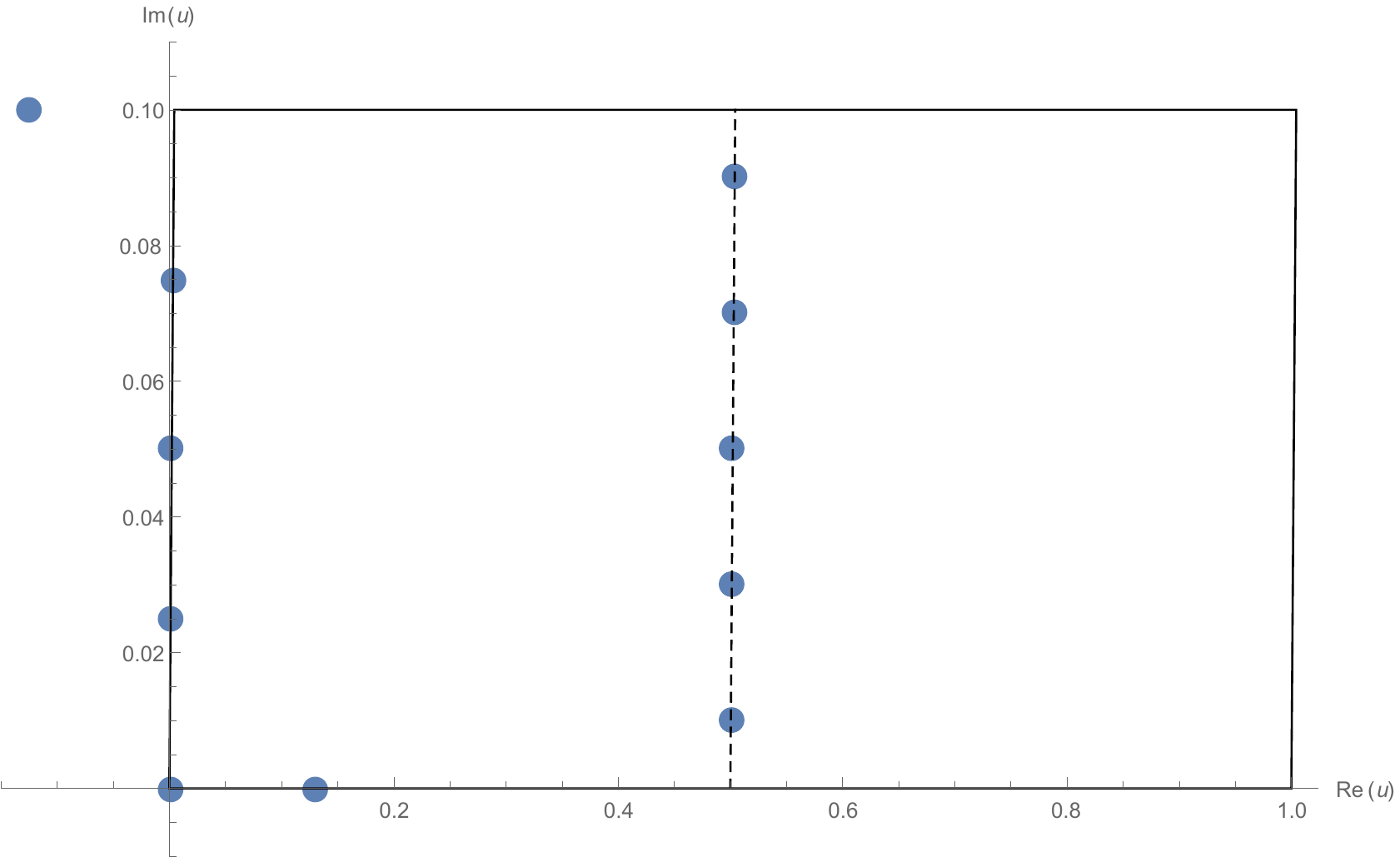}
        \caption{$(\Delta_1,\Delta_2)=(\fft{75}{287},\fft{152}{517})$}
        \label{C=2:highT:case2}
    \end{subfigure}
    \caption{Numerical solutions to the eBAEs (\ref{BAE}) with $N=11$ and $\tau=\fft{1+23i}{230}$. Note that (a) corresponds to Case 1 ($\Delta_1+\Delta_2\leq\fft12$) and (b) corresponds to Case 2 ($\Delta_1+\Delta_2>\fft12$).}
    \label{C=2:highT}
\end{figure}

In appendix~\ref{App:C} we establish a non-standard solution in the Cardy-like limit with
the chemical potentials satisfying $\Delta_1+\Delta_2\leq\fft12$. It is
given explicitly as
\begin{align}
    \hat u=\left\{\fft{i}{(N+1)/2}\tau:i=0,\cdots,\fft{N-1}{2}\right\}\cup\left\{\fft12+\fft{i-1/2}{(N-1)/2}\tau:i=1,\cdots,\fft{N-1}{2}\right\}.\label{non-standard:case1}
\end{align}
This asymptotic non-standard solution satisfies the eBAEs in the
Cardy-like limit (as displayed in (\ref{BAE:highT:2})) up to exponentially suppressed
terms.  Note that this solution corresponds to both packs having
holonomies equally spaced along the periodic $\tau$ direction and
hence can be viewed as a natural odd-$N$ version of the standard
$C=2$ solution.  Although this solution only satisfies the eBAEs
asymptotically, we have demonstrated that similar solutions continue
to exist at finite $\tau$ by numerically solving the exact eBAEs
(\ref{BAE}).  As an example, we present a numerical solution with
$N=11$ in Figure~\ref{C=2:highT:case1}.  This strongly implies that
there is an exact non-standard solution to the eBAEs (\ref{BAE}),
whose asymptotic form is given as (\ref{non-standard:case1}).
Unlike the standard solutions, however, this non-standard $C=2$
solution does not have the holonomies uniformly distributed on the
torus, and moreover generally depends on the potentials $\Delta_a$, except in
the Cardy-like limit.

%%%%
%%%%

It is now straightforward to insert the asymptotic solution,
(\ref{non-standard:case1}), into (\ref{eq:logZcardy}) to obtain the
contribution to the index
\begin{align}
    \log I_{C=2}^{\mathrm{odd\mathchar`-}N}(\Delta_a,\tau)=-\fft{\pi i}{6\tau^2}\sum_{a=1}^3\fft{N^2-1}{8}\kappa(2\Delta_a)+\mathcal O\left(1/\tau\right),\label{eq:Z:non:case1}
\end{align}
which can be compared with the standard $C=2$ solution
\begin{equation}
    \log I_{C=2}^{\mathrm{even\mathchar`-}N}(\Delta_a,\tau)=-\fft{\pi i}{6\tau^2}\sum_{a=1}^3\left(\fft{N^2}8\kappa(2\Delta_a)-\kappa(\Delta_a)\right)+\mathcal O\left(1/\tau\right),
\end{equation}
obtained by taking $C=2$ in (\ref{eq:Zmnr:highT}).  Although these
expressions still demonstrate an even/odd effect at finite $N$, the
leading $\mathcal O({N^2}/{\tau^2})$ behavior is identical in the
large-$N$ limit.

%%%%%
\subsubsection*{Case 2. $\Delta_1+\Delta_2>\fft12$}\label{case2}
%%%%%

We now turn to the second possibility, where the chemical potentials
satisfy $\Delta_1+\Delta_2>\fft12$.  Here the non-standard solution
takes the asymptotic form
\begin{align}
    \hat u&=\left\{\pm\fft{\Delta_1-\epsilon}{2}\right\}\cup\left\{\fft{i}{(N-3)/2}\tau:i=0,\cdots,\fft{N-5}{2}\right\}\cup\left\{\fft12+\fft{i-1/2}{(N-1)/2}\tau:i=1,\cdots,\fft{N-1}{2}\right\}.\label{non-standard:case2}
\end{align}
This corresponds to two isolated holonomies along with packs of
$(N-3)/2$ and $(N-1)/2$ holonomies, respectively.  Numerically, we
can see that, for sufficiently large $\tau$, the first two
holonomies are actually part of a single pack of $(N+1)/2$
holonomies, just as in case~1.  However, as $\Im\tau$ approaches
zero, this pair first moves towards the real axis and then finally
towards $\pm\Delta_1/2$ in the Cardy-like limit.  The exponentially
small deviation $\epsilon$ away from this endpoint is given by
\begin{equation}
    \epsilon=\fft{\tau}{2\pi i}\exp\left[-\fft{\pi i}{\tau}\left(-\fft{N-1}{2}+(N-2)\Delta_1+(N-1)\Delta_2\right)\right],\label{deviation}
\end{equation}
provided the quantity in the parentheses is positive.  This is true
for a generic value of $\Delta_1+\Delta_2>\fft12$ (not
asymptotically close to $\fft12$) and a sufficiently large $N$.
Although (\ref{non-standard:case2}) only holds asymptotically,
numerical solutions of this type can be obtained for finite $\tau$,
and we provide an example in Figure~\ref{C=2:highT:case2}.

Inserting the non-standard solution (\ref{non-standard:case2}) into
the asymptotic expression (\ref{eq:logZcardy}) then gives the
contribution to the index
\begin{align}
    \log I_{C=2}^{\mathrm{odd\mathchar`-}N}(\Delta_a,\tau)&=-\fft{\pi i}{6\tau^2}\left[\sum_{a=1}^3\fft{N^2}{8}\kappa(2\Delta_a)+\fft{3N}{2}(1-2\Delta_2)(1-2\Delta_3)\right.\nn\\
    &\kern5em\left.+\fft34-3\Delta_2+3(\Delta_1+\Delta_2)(-2\Delta_1+\Delta_2+2\Delta_1\Delta_2)\right]+\mathcal O\left(1/\tau\right).\label{eq:Z:non:case2}
\end{align}
Although this expression is more complicated than the corresponding
one for Case 1, it reduces in the large-$N$ limit to the same
leading order behavior as all the other $C=2$ cases, namely
\begin{equation}
    \log I_C(\Delta_a,\tau)=-\fft{\pi iN^2}{\tau^2}\sum_{a=1}^3\fft{\kappa(C\Delta_a)}{6C^3}+\mathcal O(N\log N,1/|\tau|),\label{eq:Z:non-standard}
\end{equation}
with $C=2$.

While we have focused on non-standard solutions for $C=2$, numerical investigations confirm that similar non-standard solutions exist for arbitrary values of $C$ and $N$, at least for $N$ sufficiently large so that it can be divided into nearly equal packs of holonomies.  These non-standard solutions are similar to that with $C=2$ in that they are sensitive to the choice of chemical potentials $\Delta_1$ and $\Delta_2$, with the simpler configuration, corresponding to Case 1 above, occurring only when $\Delta_1+\Delta_2\le1/C$.  In this case, the non-standard solution in the Cardy-like limit is given by
\begin{align}
	\hat u=&\left\{\fft{I}{C}+\fft{i}{(N+C-D)/C}\tau:I=0,\cdots,D-1,~i=0,\cdots,\fft{N-D}{C}\right\}\nn\\&~\cup\left\{\fft{J}{C}+\fft{j-1/2}{(N-D)/2}\tau:J=D,\cdots,C-1,~j=1,\cdots,\fft{N-D}{C}\right\},\label{non-standard:C}
\end{align}
where $N=C\lfloor N/C\rfloor+D~(D=1,\cdots,C-1)$.  This satisfies the eBAEs in the Cardy-like limit (as displayed in (\ref{BAE:highT:2})) up to exponentially suppressed terms.  In principle, this solution can be inserted into (\ref{eq:logZcardy}) to obtain the contribution of a given $C$ non-standard solution to the index.  The resulting expressions simplify in the large-$N$ limit, and reduce to (\ref{eq:Z:non-standard}) as expected.

Although this extension of the Case~1 solution to arbitrary values of $C$ only holds for sufficiently small $\Delta_1+\Delta_2$, we expect that generalizations of the Case~2 solution (where pairs of holonomies may be pulled away from the main packs) exist for other values of the chemical potentials.  We thus conjecture that, at least in the Cardy-like limit, solutions to the eBAEs exist for all values of $C$ and $N$ with $d\le N/C<d+1$. Here, $d$ corresponds to the minimum number of holonomies in a single pack that allows the solution to be categorized as a set of packs instead of individually distributed holonomies.  When $C$ divides $N$, the solution is standard, but otherwise it is non-standard.  For each value of $C$, the contribution to the index then takes the form (\ref{eq:Z:non-standard}), regardless of whether the solution is standard or non-standard.  As a result, the Bethe-ansatz type approach to the index reproduces the improved asymptotic bound (\ref{eq:dSummedLargeN}) found above.

%%%%%
\subsection{Additional non-standard solutions}\label{sec:BA:non-standard}
%%%%%

As we have seen explicitly, the non-standard
solutions (\ref{non-standard:case1}) and (\ref{non-standard:case2}) to
the eBAEs (\ref{BAE}) for $C=2$ and odd $N$ contribute to the index in much the same way as the standard $\{2,N/2,0\}$ solution for even $N$, at least in the large-$N$ limit.  Moreover, such contributions were crucial to reproduce the improved asymptotic bounds (\ref{eq:dSummedLargeN}) with arbitrary integer $C$ that may or may not divide $N$.  This makes us conclude that we cannot ignore the non-standard solutions in the Bethe ansatz approach to the index.

While we have focused on non-standard solutions that are similar to the $\{C,N/C,0\}$ solutions since all standard solutions reduce to this form in the Cardy-like limit, we have also found numerical evidence for the existence of non-standard solutions at finite $\tau$.  This suggests that many if not all standard $\{m,n,r\}$ solutions have generalized counterparts as non-standard solutions.  In addition, there may be additional non-standard solutions that do not fall into any particular classification.  Thus it would be nice to understand the full set of solutions to the eBAEs.

In principle, we would just solve the eBAEs (\ref{BAE}) to obtain a complete set of standard and non-standard solutions.  However, in practice, this is a rather challenging problem, even in the asymptotic limits.  Therefore, to make the problem tractable, we focus on the $N=2$ and $N=3$ cases.  Even so, much of the analysis is rather technical, so we relegate the details to Appendix~\ref{App:C} and only highlight the main results here.

%%%%%
\subsubsection{Non-standard solutions for \texorpdfstring{$N=2$}{N=2}}\label{subsec:SU(2)non-standard}
%%%%%

Since the eBAEs (\ref{BAE}) depend only on the difference $u_{ij}=u_i-u_j$ of the holonomies, and since the free parameter $\lambda$ is unconstrained, the $N=2$ eBAEs reduce to a single equation for a single variable $u_{21}$
\begin{equation}
    1=\prod_{a=1}^3\fft{\theta_1(\tilde\Delta_a+u_{21};\tau)^2}{\theta_1(\tilde\Delta_a-u_{21};\tau)^2}.
    \label{eq:N=2BAE}
\end{equation}
Here for convenience we have used the real chemical potentials $\tilde\Delta_{1,2,3}$, which are essentially the same as $\Delta_{1},\Delta_{2},-\Delta_{1}-\Delta_{2}$, up to simple shifts and reflections such that
\begin{align}
0<\tilde\Delta_a<1,\qquad
\sum_{a=1}^3\tilde\Delta_a=1.\label{eq:deltaTildeCond}
\end{align}
See Appendix \ref{App:C} for more details.

Since each theta function has a first-order zero, this fraction has six zeros.  Furthermore, since it is elliptic, the fraction takes all values, including unity, six times, and thus the eBAE has six solutions.  Four of them are the familiar ones
\begin{equation}
    u_{21}=\left\{0,\ \fft12,\ \fft\tau2,\ \fft{1+\tau}2\right\},
    \label{eq:N=2standard}
\end{equation}
corresponding to the trivial solution and standard $\{2,1,0\}$, $\{1,2,0\}$ and $\{1,2,1\}$ solutions, respectively.  The other two solutions are non-standard, and since the equation is symmetric under $u_{21}\to-u_{21}$, they are negatives of each other
\begin{equation}
    u_{21}=\{u_\Delta,\ -u_\Delta\}\qquad(\mbox{non-standard}).
\end{equation}
It seems challenging to find $u_\Delta$ in closed form, except in the asymptotic limits, so part of the investigation will be numerical.

We denote the `low-temperature' limit to be $|\tau|\gg1$ and the `high-temperature' limit to be $|\tau|\ll1$ where $\arg\tau\,(\neq0,\pi)$ is held fixed for both cases. In either limit, the Jacobi theta function $\theta_1(u;\tau)$ has a straightforward asymptotic expansion, so the equation (\ref{eq:N=2BAE}) can be directly solved.  The asymptotic analysis is presented in Appendix~\ref{App:C}, and we summarize the $N=2$ results here.

\begin{figure}[t]
    \centering
    \includegraphics[width=.4\linewidth]{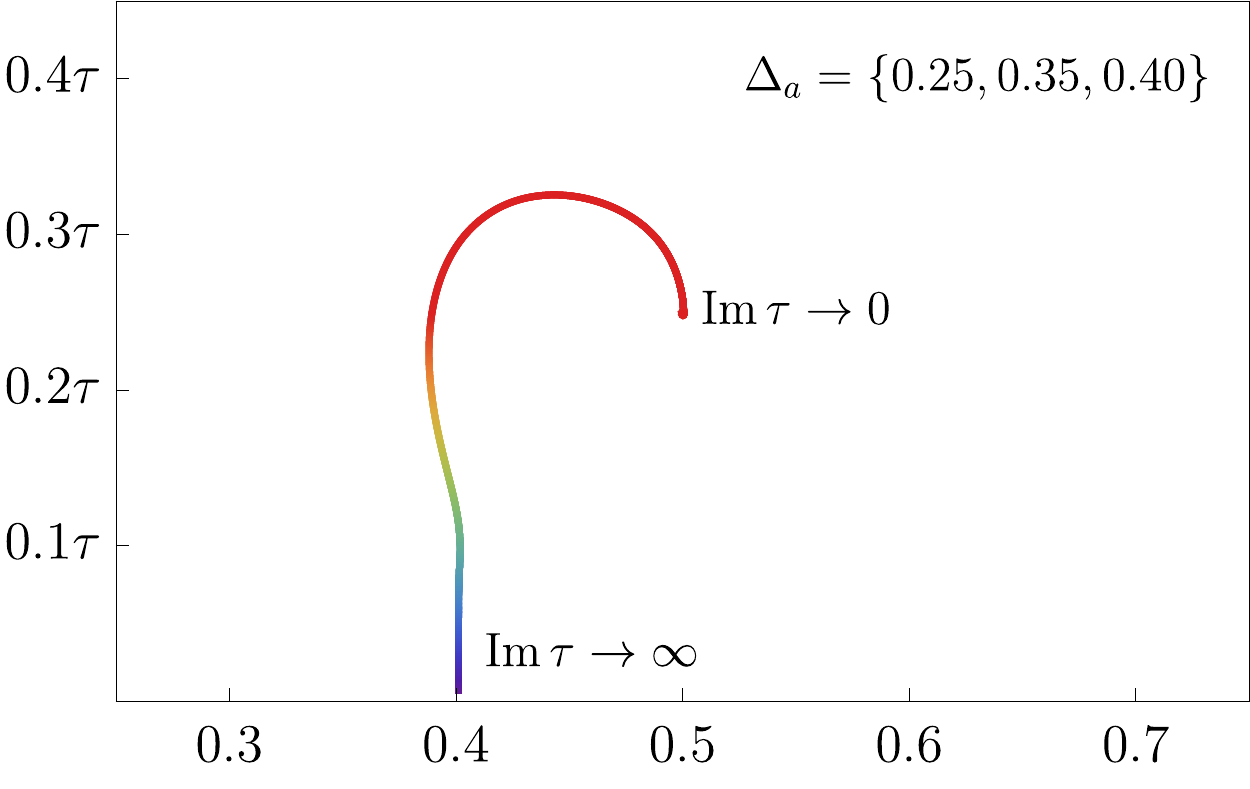}
    \hbox{~}
    \includegraphics[width=.4\linewidth]{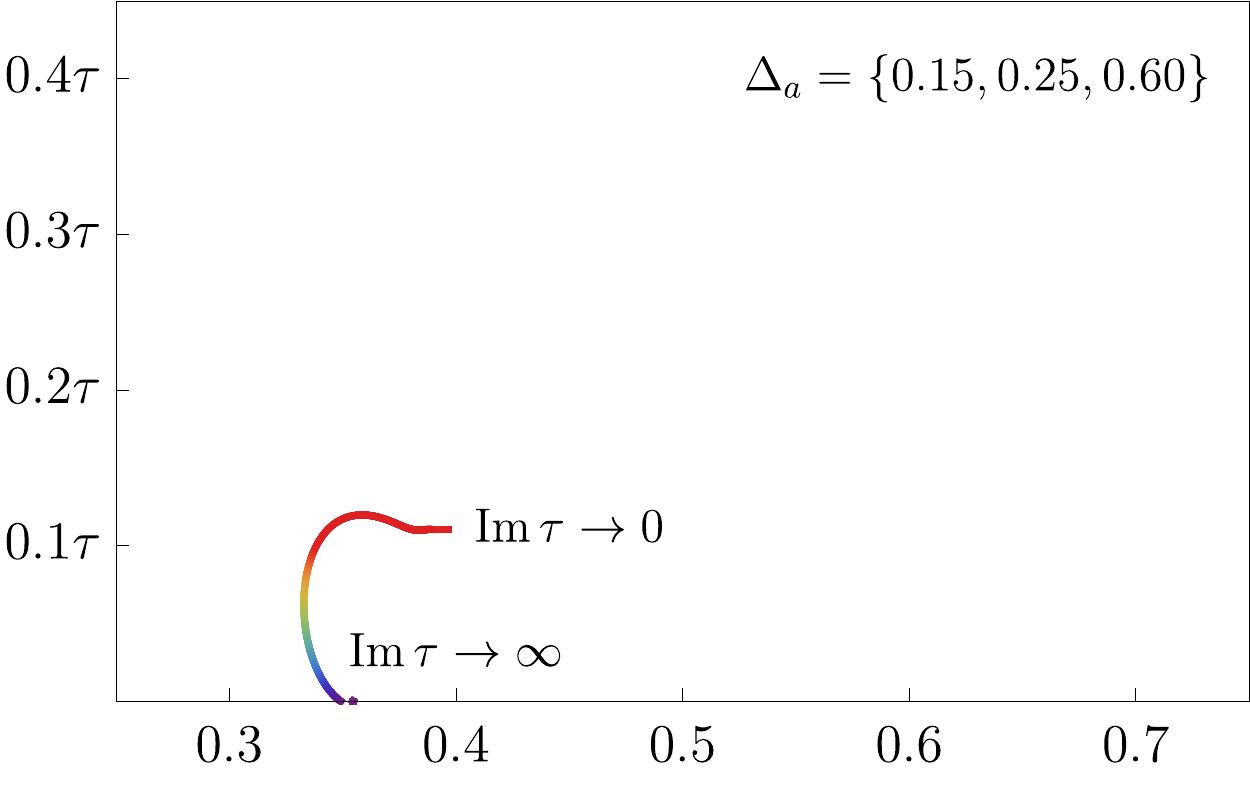}
    \caption{Numerical plots of the non-standard $N=2$ solution $u_\Delta$ with $\arg(\tau)=\pi/4$.  The figure on the left corresponds to $\tilde\Delta_3<1/2$ while that on the right corresponds to $\tilde\Delta_3>1/2$.  Note that the vertical axis is given in units of (complex) $\tau$.}
    \label{N=2:d3<1/2:fig}
\end{figure}

As indicated above, the standard solutions, (\ref{eq:N=2standard}), are exact solutions for any $\tilde\Delta_a$ and for all $\tau$, so we only focus on the non-standard solution specified by $u_\Delta$.  In the low-temperature limit the solution---given in (\ref{BAE:lowT:N=2:sol})---takes the form
\begin{equation}
   u_\Delta(\Im\tau\to\infty)=\fft{1}{2\pi i}\log\left[{-\left(\fft{1-\sum_a\cos 2\pi\tilde\Delta_a}{2}\right)+\sqrt{\left(\fft{1-\sum_a\cos 2\pi\tilde\Delta_a}{2}\right)^2-1}}\right]
    \label{lowT:N=2:flowend}
\end{equation}
The high-temperature asymptotic solution splits into two cases, depending on whether $\tilde\Delta_3$ is less than or greater than $1/2$.  The solution is given by (\ref{BAE:highT:N=2:case1:sol}) and (\ref{BAE:highT:N=2:case2:sol}), which can be summarized as
\begin{equation}
    u_\Delta(\Im\tau\to0)=\begin{cases}\displaystyle
    \fft12+\fft{\tau}4,&\displaystyle\tilde\Delta_3<\fft12;\\[6pt]
    \displaystyle
    1-\tilde\Delta_3+i\fft{\log2}{2\pi}\tau,&\displaystyle\tilde\Delta_3\ge\fft12.
    \end{cases}
    \label{highT:N=2:flowend}
\end{equation}
There does not appear to be a simple analytic solution away from these asymptotic limits.  However, numerical investigations demonstrate that the non-standard solutions (\ref{lowT:N=2:flowend}) and (\ref{highT:N=2:flowend}) are indeed continuously connected.  Figure \ref{N=2:d3<1/2:fig} shows explicit examples of the numerical
solution for $u_\Delta$ with both possibilities of $\tilde\Delta_3$.

To be complete, it should be noted that for $\tilde\Delta_3\ge1/2$, the high-temperature limit also admits a continuous set of solutions
\begin{equation}
    u_\Delta(\Im\tau\to0)=\left\{x+y\,\tau\bigg|x\in(1-\tilde\Delta_3,\tilde\Delta_3),y\in[0,1)\mbox{ and }\tilde\Delta_3\ge1/2\right\},
\end{equation}
which interestingly correspond to the plateaus of $Q_h$ in the integral approach of the previous section. However, these solutions do not appear to be continuously connected to any solutions away from the high-temperature limit, so we believe they are only an artifact of the high-temperature asymptotics, and are not true solutions to the eBAEs once subleading terms are included.

%%%%
%%%%

%%%%%
\subsubsection{Non-standard solutions for finite \texorpdfstring{$N>2$}{N>2}}\label{subsec:SU(3)nonstandard}
%%%%%

The structure of the eBAEs is considerably harder to analyze for $N>2$ as there are more holonomy pairs $u_{ij}$ to consider.  For any $N$, note that the product of the $N$ individual eBAEs (\ref{BAE}) gives the condition $e^{2\pi iN\lambda}=1$ so that $e^{2\pi i\lambda}$ is a root of unity (not necessarily primitive).  Focusing next on a single equation, say $i=1$, then gives the condition
\begin{equation}
    1=\prod_{a=1}^3\prod_{j=2}^N\left(\fft{\theta_1(\tilde\Delta_a+u_{j1};\tau)}{\theta_1(\tilde\Delta_a-u_{j1};\tau)}\right)^N.
\end{equation}
For $N=3$, this means we can solve for $u_{31}$ in terms of $u_{21}$
\begin{equation}
    \prod_{a=1}^3\fft{\theta_1(\tilde\Delta_a+u_{31};\tau)}{\theta_1(\tilde\Delta_a-u_{31};\tau)}=\omega^k\prod_{a=1}^3\fft{\theta_1(\tilde\Delta_a-u_{21};\tau)}{\theta_1(\tilde\Delta_a+u_{21};\tau)},
    \label{eq:cubic}
\end{equation}
where $\omega=e^{2\pi i/3}$ is a primitive cube root of unity and $k=0,1,2$.  We thus look for non-standard solutions by picking a given $u_{21}$ and then (numerically) solving for $u_{31}$.  There are three solutions up to periodicity for each value of $k$, giving nine possible roots $u_{31}$ for a fixed $u_{21}$.  However, not all of these solutions are valid, as they must solve not just the $i=1$ but all of the eBAEs.  Numerically, we find, in addition to the standard solutions which only exist at discrete values of $u_{21}$, that two roots of (\ref{eq:cubic}) with $k=0$ in fact solve the complete set of eBAEs for arbitrary $u_{21}$.  (The third $k=0$ root is the trivial solution $u_{31}=-u_{21}$, but it does not generically solve the remaining eBAEs apart from the standard solutions.)

Although we only obtain solutions numerically for intermediate values of $\tau$, analytic results are possible in the low and high temperature limits.  In the low-temperature limit, we take the two independent holonomies to be $u_{21}$ and $u_{31}$ and write $z_{21}=e^{2\pi iu_{21}}$ and $z_{31}=e^{2\pi iu_{31}}$.  The non-standard solutions then correspond to the two roots of the quadratic expression, (\ref{eq:N=3cont}), which we repeat here for convenience
\begin{equation}
    (1+z_{21}+z_{31})\left(1+\fft1{z_{21}}+\fft1{z_{31}}\right)=3+2\sum_a\cos2\pi\tilde\Delta_a.
    \label{eq:N3low}
\end{equation}
The two roots are related by the map $u_{ij}\to-u_{ij}$, corresponding to taking $z_{ij}\to1/z_{ij}$.  The important feature of this solution is that the eBAEs reduce to a single condition on two complex holonomies.  As a result, we end up with a continuous family of non-standard solutions.

\begin{figure}[t]
    \centering
    \begin{subfigure}[h]{.49\linewidth}
        \centering
        \includegraphics[width=\linewidth]{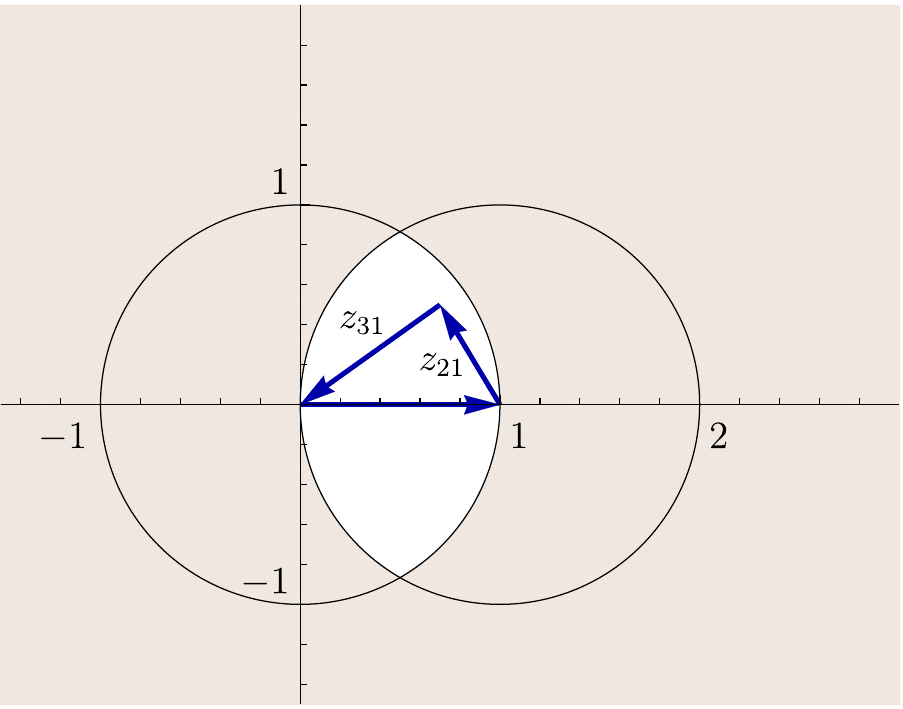}
        \caption{$1+z_{21}+z_{31}=0$}
    \end{subfigure}
    \begin{subfigure}[h]{.49\linewidth}
        \centering
        \includegraphics[width=\linewidth]{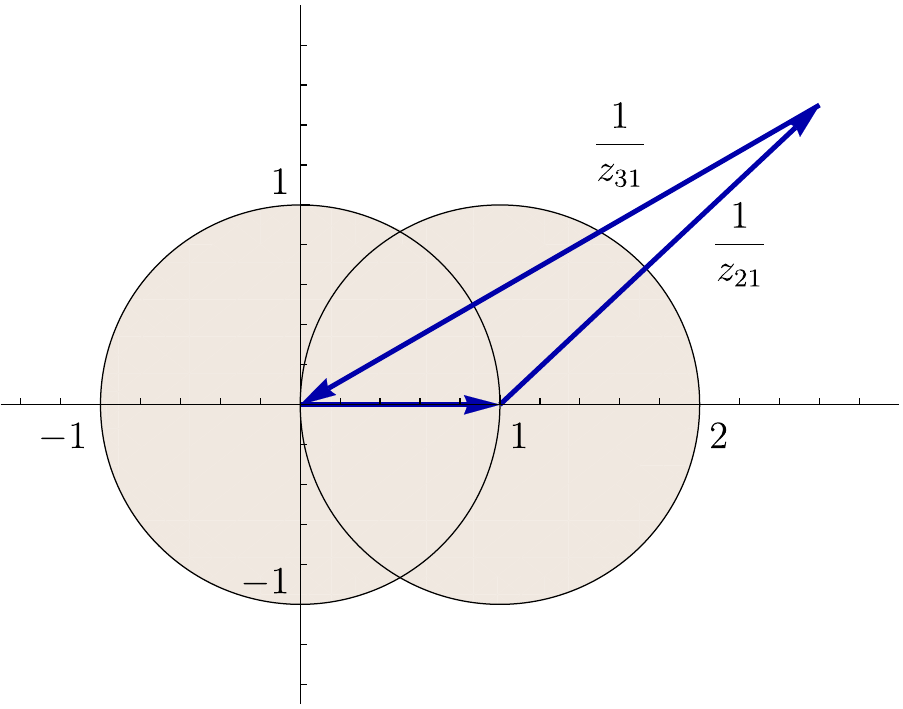}
        \caption{$1+1/z_{21}+1/z_{31}=0$}
    \end{subfigure}    %\includegraphics[width=.47\textwidth]{N3-region-1}\quad
    \caption{The space of non-standard $N=3$ solutions in the low-temperature limit with identical chemical potentials $\tilde\Delta_a=1/3$.  Solutions of the first type, satisfying $1+z_{21}+z_{31}=0$, are shown in (a), while solutions of the second type, satisfying $1+1/z_{21}+1/z_{31}=0$, are shown in (b).}
    \label{fig:N=3space}
\end{figure}

To further explore the nature of the non-standard solutions, we consider, for simplicity, the case where all the chemical potentials are identical to each other, namely $\tilde\Delta_a=1/3$.  In this case, the right-hand side of (\ref{eq:N3low}) vanishes, and we find simply
\begin{equation}
    1+z_{21}+z_{31}=0\qquad\mbox{or}\qquad1+\fft1{z_{21}}+\fft1{z_{31}}=0.
\end{equation}
Since the second case can be obtained from the first by taking $u_i\to-u_i$, we focus on the first case.  The expression $1+z_{21}+z_{31}=0$ has a simple interpretation in the complex plane as a triangle with sides $1$, $z_{21}$ and $z_{31}$.  Since the solution is restricted to $|q|\le|z_{21}|\le1$ and $|q|\le|z_{31}|\le1$, the space of solutions is given by the intersection of two disks of radius one centered at $0$ and $1$, respectively.  The continuous family of non-standard $N=3$ solutions is shown schematically in Figure~\ref{fig:N=3space}.

%%%%
%%%%

\begin{figure}[t]
    \centering
    \includegraphics[height=6cm]{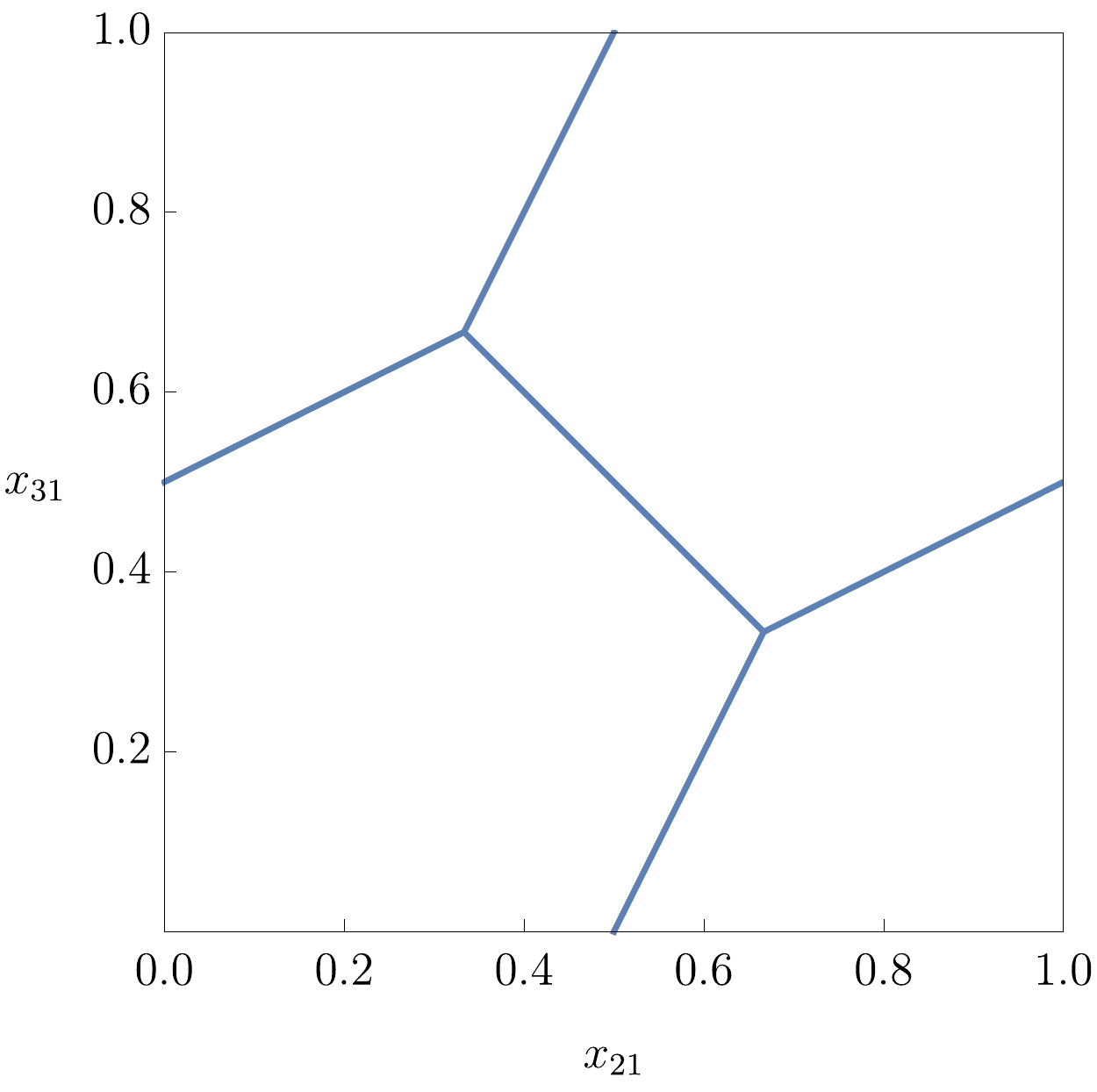}
    \caption{The space of non-standard $N=3$ solutions in the high-temperature limit with identical chemical potentials $\tilde\Delta_a=1/3$.  Here we only show the real components $x_{21}$ and $x_{31}$ under the decomposition $u_{ij}=x_{ij}+y_{ij}\tau$.  The parameter space is actually one-complex dimensional, which would be apparent if we had not suppressed the $y_{21}$-$y_{31}$ plane.}
    \label{fig:N=3highT}
\end{figure}

The analysis of the high-temperature asymptotic solutions is rather involved, so we again focus on the case $\tilde\Delta_a=1/3$.  In addition to the standard solutions, the continuous family of non-standard solutions survives in the high temperature limit, and is given by (\ref{BAE:highT:N=3:sol}) in Appendix~\ref{App:C}, which we rewrite in terms of $u_{21}=x_{21}+y_{21}\tau$ and $u_{31}=x_{31}+y_{31}\tau$
\begin{align}
    u_{21},u_{32}\in&~\left\{u_{21},u_{32}\bigg|0\leq x_{21}\bmod1<\ft13,~u_{21}-2u_{31}=\mathbb Z+\ft13\tau(\mathbb Z+\ft12)\right\}\nn\\
    &~\cup\left\{u_{21},u_{32}\bigg|\ft13<x_{21}\bmod1<\ft23,~u_{21}+u_{31}=\mathbb Z+\ft13\tau(\mathbb Z+\ft12)\right\}\nn\\
     &~\cup\left\{u_{21},u_{32}\bigg|\ft13<x_{21}\bmod1<\ft23,~2u_{21}-u_{31}=\mathbb Z+\ft13\tau(\mathbb Z+\ft12)\right\}\nn\\
     &~\cup\left\{u_{21},u_{32}\bigg|\ft23<x_{21}\bmod1<1,~u_{21}-2u_{31}=\mathbb Z+\ft13\tau(\mathbb Z+\ft12)\right\}.
     \label{highT:N=3:flowend}
\end{align}
This indicates that, up to discrete shifts, there is a one-complex dimensional family of high-temperature solutions.  Although the family appears to be split into four branches, the first and last subset in (\ref{highT:N=3:flowend}) is part of the same branch, as shown in Figure~\ref{fig:N=3highT}.

%%%%
%%%%

We have verified numerically in several examples that the asymptotic low and high temperature solutions are continuously connected with each other.  This suggests that the continuous family of non-standard solutions is in fact generic for any value of $\tau$.  In general, the parameter space and the relation between $u_{21}$ and $u_{31}$ depends on the chemical potentials $\tilde\Delta_a$.  However, there is an interesting point (up to permutations) in this space of non-standard solutions with
\begin{equation}
    u_{21}=\ft12,\qquad u_{31}=\ft12\tau,\label{eq:nonstandardN=3}
\end{equation}
which is $\tilde\Delta_a$-independent. This is an \emph{exact non-standard solution}, whose validity can be checked using (\ref{theta1:property}). We were actually led to it via the correspondence of subsection~\ref{subsec:correspondenceWithN=1*} with $\mathcal{N}=1^\ast$ theory: it corresponds to a Coulomb vacuum of the compactified $\mathcal{N}=1^\ast$ theory with SU$(3)$ gauge group, listed in Dorey's paper \cite{Dorey:1999sj} as the ``fifth'' vacuum.

Despite its simple form somewhat reminiscent of the standard solutions, the solution (\ref{eq:nonstandardN=3}) is part of the continuous family, and not isolated.  To see this, we look for vanishing eigenvalues of the Jacobian matrix $(\partial Q_i/\partial u_j)$ where the $Q_i$ represent the eBAE expressions as given in (\ref{BAE}).  In general, there are $N$ eBAEs along with $N$ holonomies $u_j$.  However, the constraint $\prod_i Q_i=1$ along with the SU$(N)$ constraint $\sum_j u_j\in\mathbb Z+\tau\mathbb Z$ effectively reduces this to an $(N-1)\times(N-1)$ matrix.
\begin{equation}
	\fft{\partial Q_i}{\partial u_j}=-\delta_{ij}\sum_{k=1}^Ng(u_{ki},\tilde\Delta_a;\tau)+g(u_{ji},\tilde\Delta_a;\tau)-g(u_{Ni},\tilde\Delta_a;\tau),\label{component}
\end{equation}
where $i,j=1,\cdots, N-1$ and we have defined
\begin{equation}
	g(u,\tilde\Delta_a;\tau)\equiv\sum_{a=1}^3\left(\fft{\theta_1'(u+\tilde\Delta_a;\tau)}{\theta_1(u+\tilde\Delta_a;\tau)}+\fft{\theta_1'(-u+\tilde\Delta_a;\tau)}{\theta_1(-u+\tilde\Delta_a;\tau)}\right).
\end{equation}
For the present $N=3$ case, we find
\begin{equation}
	\left.\det\left(\fft{\partial Q_i}{\partial u_j}\right)\right|_{(u_{21},u_{31})=(1/2,\tau/2)}=12\left(\prod_{J=2}^4\sum_{a=1}^3\fft{\theta_J'(\tilde\Delta_a;\tau)}{\theta_J(\tilde\Delta_a;\tau)}\right)\sum_{I=2}^4\fft{1}{\sum_{a=1}^3\theta_I'(\tilde\Delta_a;\tau)/\theta_I(\tilde\Delta_a;\tau)},
	\label{det:exnstd}
\end{equation}
where $\theta_{2,3,4}$ are the standard Jacobi theta functions given explicitly in (\ref{theta:different}).  We now make use of the following result.
\begin{lemma}
For any $\tau$ in the upper-half plane, and any complex $\tilde{\Delta}_{1,2,3}$ subject to $\sum_a \tilde{\Delta}_a\in\mathbb Z$, we have
\begin{equation}
	\sum_{I=2}^4\fft{1}{\sum_{a=1}^3\theta_I'(\tilde\Delta_a;\tau)/\theta_I(\tilde\Delta_a;\tau)}=0.
\end{equation}
\end{lemma}

This lemma, combined with (\ref{det:exnstd}), establishes that the Jacobian matrix indeed has a vanishing eigenvalue.  This establishes the existence of a ``zero mode'' taking (\ref{eq:nonstandardN=3}) to nearby eBAE solutions.  Proof%
\footnote{We are indebted to Hjalmar~Rosengren, a mathematician at Chalmers University, for an instrumental correspondence regarding the proof.}
of this lemma can be found in appendix~\ref{App:proof2}.

Although we have yet to perform an analytic investigation of the non-standard solutions for $N>3$, we have investigated some cases numerically (with finite $\tau$).  For $N=4$ and $5$, in addition to the isolated standard solutions, we find solutions on which $\partial Q_i/\partial u_j$ has a single vanishing eigenvalue, suggesting that they are part of one-complex dimensional families of solutions. More interestingly, for $N=6,7,8$, and $9$ we find evidence for both one and \emph{two} complex dimensional families of solutions, the latter signalled by \emph{two} vanishing eigenvalues for $\partial Q_i/\partial u_j$. Similarly for $N=10$ we find evidence for one, two, and \emph{three} complex dimensional families of solutions. These are in agreement with the expectation that the continuous families of solutions correspond to Coulomb vacua of $\mathcal N=1^\ast$ theory. Based on these numerical results and the putative correspondence with  $\mathcal{N}=1^\ast$ theory, we conjecture that at least the dimensionality of the space of the $\mathcal{N}=4$ eBAE solutions is captured correctly by the semi-classical formula for the rank of the Coulomb vacua in SU$(N)$ $\mathcal{N}=1^\ast$ theory.

\begin{conjecture}
For $N\ge(l+1)(l+2)/2$, generic $\tau$ in the upper-half plane, and generic $\Delta_{1,2}\in\mathbb{C}$, the SU$(N)$ $\mathcal{N}=4$ eBAEs in (\ref{BAE}) have $l$-complex-dimensional continua of solutions.
\end{conjecture}

The existence of such flat directions is somewhat unusual, nevertheless, in that it necessarily demands the existence of non-trivial identities among products of Jacobi theta functions.

Of particular significance is that the existence of continuous solutions to the eBAEs poses a serious issue to the full validity of the Bethe ansatz type approach as established in Ref.~\cite{Benini:2018mlo}.  The reason is that the rewriting of the index, (\ref{eq:EHI}), into the Bethe ansatz form relies on the fact that the general solutions to the eBAEs are isolated so that Cauchy's formula may be applied \cite{Benini:2018mlo}.  Hence, if there are continuous families of non-standard solutions, the Bethe ansatz type approach must be reformulated to incorporate such continuous Bethe roots into the sum over solutions to the eBAEs, (\ref{eq:index:BA:2}).

%%%%%
\subsection{The \texorpdfstring{large-$N$}{large-N} limit of the index revisited}\label{sec:BA:large-N}
%%%%%

While we have been motivated by the Cardy-like limit of the index, the appearance of partially deconfined phases as well as non-standard solutions suggests that we re-examine the large-$N$ limit at finite $\tau$ as investigated in Ref.~\cite{Benini:2018ywd}.  In this large-$N$ limit, one or at most a few solutions (if there are any degeneracies) to the BAE would be expected to dominate in the sum over solutions, (\ref{eq:index:BA:2}).  The focus is then on identifying the dominant solution, much as one would search for a dominant saddle point contribution.

The main emphasis in \cite{Benini:2018ywd} was on the \emph{basic}
solution, corresponding to $I_{\{1,N,0\}}(\Delta_a,\tau)$.
In the large-$N$ limit, the contribution was found to be
\begin{align}
    \log I_{\{1,N,0\}}(\Delta_a,\tau)&=\sum_{j,j'=0\,(j\neq j')}^{N-1}\log\fft{\prod_{a=1}^3\tilde\Gamma\big(\fft{j-j'}{N}\tau+\Delta_a;\tau,\tau\big)}{\tilde\Gamma\big(\fft{j-j'}{N}\tau;\tau,\tau\big)}+\mathcal O(N\log N)\nn\\
    &=-\pi iN^2\Theta(\Delta_a,\tau)+\mathcal O(N\log N),\label{eq:index:1N0}
\end{align}
where $\Theta(x_a,\tau)$ is defined as
\begin{align}
    \Theta(x_a,\tau)&\equiv\fft{\sum_{a=1}^3\kappa_\tau(x_a)}{6\tau^2}+\fft{3\sum_{a=1}^3\{x_a\}_\tau(1-\{x_a\}_\tau)-(1-\tau)(1-2\tau)}{3\tau}-{\sum_{a=1}^3}(1-\{x_a\}_\tau),\label{eq:Theta}
\end{align}
where $\{x\}_\tau$ and $\kappa_\tau(x)$ are defined in (\ref{eq:kappa:tau}). Using the constraint $\Delta_3=2\tau-\Delta_1-\Delta_2$, we can rewrite $\Theta(\Delta_a,\tau)$ explicitly as
\begin{align}
    \Theta(\Delta_a,\tau)=\begin{cases}
    \fft{(\{\Delta_1\}_\tau-1)(\{\Delta_2\}_\tau-1)(2\tau+1-\{\Delta_1\}_\tau-\{\Delta_2\}_\tau)}{\tau^2}, & (\{\Delta_1+\Delta_2\}_\tau=\{\Delta_1\}_\tau+\{\Delta_2\}_\tau-1);\\
    \fft{\{\Delta_1\}_\tau\{\Delta_2\}_\tau(2\tau+1-\{\Delta_1\}_\tau-\{\Delta_2\}_\tau)}{\tau^2}-1, & (\{\Delta_1+\Delta_2\}_\tau=\{\Delta_1\}_\tau+\{\Delta_2\}_\tau),
    \end{cases}\label{eq:Theta:simple}
\end{align}
and therefore (\ref{eq:index:1N0}) is in fact a function of
$\Delta_1$, $\Delta_2$ and $\tau$ only.

While it is clear that the basic solution yields a contribution of $\mathcal(N^2)$, there are potentially many other sources of contributions at the same order.  These include other standard $\{m,n,r\}$ solutions as well as isolated and continuous families of non-standard solutions.  The conjecture of Ref.~\cite{Benini:2018ywd} is that only the set of $T$-transformed solutions, $\hat u_{\{1,N,r\}}$, would yield additional contributions at $\mathcal O(N^2)$.  The resulting
large-$N$ index is then argued to take the form
\begin{align}
    \log\mathcal I(q,q,y_{1,2,3})&=\max\left\{\log I_{\{1,N,r\}}(\Delta_a,\tau):r\in\mathbb Z\right\}+\mathcal O(N\log N)\nn\\
    &=\max\left\{-\pi iN^2\Theta(\Delta_a,\tau+r):r\in\mathbb Z\right\}+\mathcal O(N\log N),\label{eq:index:Benini}
\end{align}
unless the $\Delta_a$'s are located along Stokes lines where the
asymptotic expansions of the elliptic functions fail to converge or where
different contributions may compete with each other.

We note that there is some tension between the conjectured large-$N$ index, (\ref{eq:index:Benini}), and our conjectured leading asymptotics in the CKKN limit, namely (\ref{eq:doubleScalingConjecture}).  In particular, the latter, as well as the more rigorous bound, (\ref{eq:dSummedLargeN}), include configurations where the holonomies are split into $C$ nearly equal packs distributed evenly on the circle, and this is not visible in (\ref{eq:index:Benini}).  Of course, it is possible that the Cardy-like limit and the large-$N$ limit do not commute.  Nevertheless, this motivates us to ask whether additional solutions to the eBAEs, either standard or non-standard, may contribute at large $N$.

%%%%%
\subsubsection{A new parametrization of the standard solutions}
\label{not:T-transformed}
%%%%%

In the Cardy-like limit, we have seen that standard solutions obtained from $C$ identical packs of $N/C$ collided holonomies will contribute at $\mathcal O\left(\tau^{-2}(N^2/C^3)\right)$.  Here we are assuming that $C$ divides $N$, so this is a standard solution.  If the limit is smooth, then we expect this behavior to persist even as we move away from the Cardy-like limit.  As in (\ref{eq:CardyC}), the standard $\{m,n,r\}$ solutions that reduce to $C$ packs of holonomies have $\gcd(n,r)=N/C$.  This can be parametrized as the set of standard solutions
\begin{equation}
    \left\{\fft{C}p,p\fft{N}C,q\fft{N}C\right\},\qquad C|N,~p|C,\mbox{ and }\gcd(p,q)=1.
    \label{eq:Cstandard}
\end{equation}
The set of configurations for a given $C$ all have $C$ packs of holonomies with the packs equally spaced $1/C$ apart on the periodic unit circle.  The $N/C$ holonomies within each pack are equally spaced $(C/N)\tau$ apart on the periodic $\tau$ cycle.  To obtain a periodic tiling of $T^2$, each subsequent pack is offset by $(s/N)\tau$ from the previous one where $s=0,1,\ldots,C-1$, so there are precisely $C$ distinct standard solutions in the set of configurations for a given $C$. In fact, this classification of standard solutions by the number of packs $C$ allows for a new (but equivalent) parametrization of the standard solutions (\ref{BAE:sols:standard}), labeled now by
\begin{equation}
    \{C,s\}\quad\mbox{where}\quad C|N\mbox{ and }s=0,1,\ldots,C-1.
\end{equation}
For a given $\{C,s\}$, the holonomies solving the eBAEs are
\begin{equation}
    \hat u_{\{C,s\}}=\left\{u_{\hat a\hat b}=\bar u+\fft{\hat a}C+\fft{\hat bC+\hat as}N\tau\bigg|C|N,0\le s<C,(\hat a,\hat b)\in\mathbb Z_C\times\mathbb Z_{N/C}\right\}.
\end{equation}
Here $\hat a$ labels the pack and $\hat b$ labels a particular holonomy within that pack.

In the large-$N$ limit, provided $C\sim\mathcal O(1)$, the holonomies become dense along the $\tau$ circle, and hence the shift between packs by a fraction $s/N$ does not qualitatively change the appearance of the solution.  This suggests that we can compute $I_{\{C,s\}}$ in the large-$N$ limit in a universal manner, provided $C\sim\mathcal O(1)$ so that $N/C$ remains large.  To do so, we start with the expression for the standard solution given in (\ref{eq:Zmnr}). Ignoring the subleading
contributions from the prefactor $\alpha_N(\tau)$ and the Jacobian
$H(\hat u_{\{C,s\}};\Delta_a,\tau)^{-1}$ as in
\cite{Benini:2018ywd} then gives
\begin{align}
	&\log I_{\{C,s\}}(\Delta_a,\tau)\nn\\&=\sideset{}{'}\sum_{\hat a,\hat a'=0}^{C-1}\sideset{}{'}\sum_{\hat b,\hat b'=0}^{N/C-1}\log\fft{\prod_{a=1}^3\tilde\Gamma\big((\hat a-\hat a')(\fft{1}{C}+\fft{\tau}{N/s})+\fft{\hat b-\hat b'}{N/C}\tau+\Delta_a;\tau,\tau\big)}{\tilde\Gamma\big((\hat a-\hat a')(\fft{1}{C}+\fft{\tau}{N/s})+\fft{\hat b-\hat b'}{N/C}\tau;\tau,\tau\big)}+\mathcal O(N\log N),
\end{align}
where the primes in the summation symbols indicate that the $(\hat
a,\hat b)=(\hat a',\hat b')$ case is excluded. This can be rewritten
as
\begin{align}
	&\log I_{\{C,s\}}(\Delta_a,\tau)\nn\\&=C\sideset{}{'}\sum_{\hat b,\hat b'=0}^{N/C-1}\log\fft{\prod_{a=1}^3\tilde\Gamma\big(\fft{\hat b-\hat b'}{N/C}\tau+\Delta_a;\tau,\tau\big)}{\tilde\Gamma\big(\fft{\hat b-\hat b'}{N/C}\tau;\tau,\tau\big)}\nn\\
	&\quad+\sum_{\hat a=1}^{C-1}\sum_{\hat b,\hat b'=0}^{N/C-1}\left[(C-\hat a)\log\fft{\prod_{a=1}^3\tilde\Gamma\big(\fft{\hat b-\hat b'}{N/C}\tau+\hat a(\fft{1}{C}+\fft{\tau}{N/s})+\Delta_a;\tau,\tau\big)}{\tilde\Gamma\big(\fft{\hat b-\hat b'}{N/C}\tau+\hat a(\fft{1}{C}+\fft{\tau}{N/s});\tau,\tau\big)}\right.\nn\\&\left.\kern7em~+\hat a\log\fft{\prod_{a=1}^3\tilde\Gamma\big(\fft{\hat b-\hat b'}{N/C}\tau+(\hat a-C)(\fft{1}{C}+\fft{\tau}{N/s})+\Delta_a;\tau,\tau\big)}{\tilde\Gamma\big(\fft{\hat b-\hat b'}{N/C}\tau+(\hat a-C)(\fft{1}{C}+\fft{\tau}{N/s});\tau,\tau\big)}\right]+\mathcal O(N\log N),\label{eq:index:pq:1}
\end{align}
where on the first line the $\hat b=\hat b'$ case is
excluded. In fact, the first line is similar to the expression (\ref{eq:index:1N0}) for the basic $\{1,N,0\}$ solution, but with $N$ replaced by $N/C$ and with an extra prefactor of $C$. Note that the replacement $N\to N/C$ in the large-$N$ formula is valid since we have assumed $C\sim\mathcal O(1)$. The second line can be simplified using the formulas
(for $\Delta\neq0$) \cite{Benini:2018ywd},
\begin{subequations}
	\begin{align}
	\sideset{}{'}\sum_{j,j'=0}^{N-1}\log\tilde\Gamma\left(\fft{j-j'}{N}\tau+\Delta;\tau,\tau\right)&=\pi iN^2\fft{(\tau-\{\Delta\}_\tau+1)(\tau-\{\Delta\}_\tau+1/2)(\tau-\{\Delta\}_\tau)}{3\tau^2}+\mathcal O(N),\label{eq:formula:1N0:1}\\
	\sideset{}{'}\sum_{j,j'=0}^{N-1}\log\tilde\Gamma\left(\fft{j-j'}{N}\tau;\tau,\tau\right)&=\pi iN^2\fft{\tau(\tau-1/2)(\tau-1)}{3\tau^2}+\mathcal O(N\log N),
	\end{align}
	\label{eq:formula:1N0}%
\end{subequations}
along with (\ref{eq:index:1N0}). The result is
\begin{align}
	\log I_{\{C,s\}}(\Delta_a,\tau)&=-\fft{\pi iN^2}{C}\left(\sum_{\hat a=0}^{C-1}\Theta(\Delta_a+\fft{\hat a}{C},\tau)-\fft{(C-1)(C(1-3\tau)+1)}{6C\tau}\right)\nn\\
	&\quad\,+\mathcal O(N\log N),\label{eq:index:pq:2}
\end{align}
provided that $\Delta_a$ is not an integer multiple of $1/C$. In particular, this demonstrates $\mathcal O(N^2)$ scaling of the standard $\{C,s\}$ solution for $C\sim\mathcal O(1)$, even away from the the Cardy-like limit.  Moreover, this leading-order behavior is independent of the offset $s$ along the $\tau$ circle between adjacent packs of holonomies, thus confirming the intuition that once the packs are sufficiently dense, the distribution of holonomies within each pack becomes unimportant, at least at leading order.

The analysis of the large-$N$ limit of the standard solutions is still incomplete, as we have not considered the case where $C$ scales with $N$. Nevertheless, the contributions (\ref{eq:index:pq:2}) suggests that the conjecture (\ref{eq:index:Benini}) must be refined by enlarging the holonomy configurations that need to be considered.  In terms of the $\{C,s\}$ labels, the basic $\{1,N,0\}$ solution is included as $\{1,0\}$, while the T-transformed $\{1,N,r\}$ solutions fall under $\{C,s\}$ with $r=qN/C$ where $q$ and $C$ are relatively prime.  Note that this presents a bit of a puzzle as (\ref{eq:index:pq:2}) and (\ref{eq:index:Benini}) appear to be in conflict except for $r=0$.

The resolution of this puzzle is based on two observations.  The first is that, for $r\sim\mathcal O(1)$ in the large-$N$ limit, we must have $C\sim N$, in which case the expression (\ref{eq:index:pq:2}) breaks down.  The second is that if instead $C\sim\mathcal O(1)$ then $r$ must be large.  However, in this case, the map from the first to the second line of (\ref{eq:index:Benini}) breaks down as it is only valid for $r$ scaling as $\mathcal O(N^0)$.  This is because the derivation of $\Theta(\Delta_a,\tau+r)$ from the large-$N$ asymptotics of the standard solution (\ref{eq:Zmnr}) with $\{m,n,r\}=\{1,N,r\}$ was based on making the shift $\tau\to\tau+r$.  However, shifts with $r=\mathcal O(N)$ may lead to a modification to $\log\mathcal I_{\{1,N,r\}}(\Delta_a,\tau)$ that is not captured by simply shifting $\tau$ in (\ref{eq:index:1N0}).

Provided we only take standard solutions into account, the large-$N$ limit of the index will be given by maximizing over all $\{C,s\}$ solutions.  Although we are unable to provide an analytic expression for $C\sim\mathcal O(N)$, we nevertheless expect the large-$N$ index to take a form along the rough lines of
\begin{align}
    \log\mathcal I(q,q,y_{1,2,3})&=\max_{\genfrac{}{}{0pt}{2}{C\sim\mathcal O(1)}{r\sim\mathcal O(1)}}\biggl\{
    -\fft{\pi iN^2}{C}\left(\sum_{j=0}^{C-1}\Theta(\Delta_a+\fft{j}{C},\tau+r)-\fft{(C-1)(C(1-3(\tau+r))+1)}{6C(\tau+r)}\right)\biggr\}\nn\\
    &\qquad+\mathcal O(N\log N),
    \label{eq:largeN}
\end{align}
where $C$ divides $N$ for standard solutions.  The shift of $\tau\to\tau+r$ in the above allows us to include T-transformed versions of the $\{C,s\}$ solutions, and the restriction of this expression to $C=1$ reproduces the conjecture (\ref{eq:index:Benini}) of \cite{Benini:2018ywd}.

Of course, this is still expected to be incomplete, as we have not yet accounted for the non-standard solutions.  We know from the Cardy-like limit that isolated non-standard solutions exist for all integer values of $C$. Numerically, these solutions are continuously connected to low temperature solutions, and hence should exist for arbitrary $\tau$.  Again, in the large-$N$ limit, the isolated non-standard solutions correspond to $C$ packs of either $\lfloor N/C\rfloor$ or $\lfloor N/C\rfloor+1$ holonomies distributed along the $\tau$ circle.  Since we take $C\sim\mathcal O(1)$, the difference between packs shows up as a $\mathcal O(1/N)$ correction, and hence we expect the leading $\mathcal O(N^2)$ behavior to be correctly captured by (\ref{eq:largeN}) where we now drop the restriction that $C$ divides $N$.

Finally, we are able to connect the large-$N$ limit of the index, (\ref{eq:largeN}), to the Cardy-like limit by taking $\tau\to0$.  Looking only at the leading $\mathcal O(1/\tau^2)$ order, it is easy to see that all terms with $r\ne0$ will not contribute.  Making use of
\begin{equation}
    \Theta(\Delta_a,\tau)=\fft{1}{6\tau^2}\sum_{a=1}^3\kappa_\tau(\Delta_a)+\mathcal O(1/|\tau|),
\end{equation}
along with the identity (\ref{eq:simplifyWidentity}) then gives
\begin{equation}
    \log\mathcal I(q,q,y_{1,2,3})=\max_{C\sim\mathcal O(1)}\left\{-\fft{\pi iN^2}{\tau^2}\sum_{a=1}^3\fft{\kappa_\tau(C\Delta_a)}{6C^3}\right\}
    +\mathcal O(N\log N,1/|\tau|),\label{eq:index:CN/C0:3}
\end{equation}
in perfect agreement with the conjecture (\ref{eq:doubleScalingConjecture}) for the leading large-$N$ asymptotics in the Cardy-like limit.  This certainly suggests that the asymptotic behavior of the index is independent of the order of limits between the Cardy-like limit and the large-$N$ limit.  Of course, in order to remove the conjectures and be more rigorous in the Bethe Ansatz approach, we have to consider solutions with $C\sim\mathcal O(N)$, as well as the issue of continuous non-standard solutions.

%%%%%
\section{Discussion}\label{sec:discussion}
%%%%%
\subsection{Summary and relation to previous work}\label{subsec:relationToPrevious}

Several papers investigating asymptotic growth of supersymmetric
indices of the 4d $\mathcal{N}=4$ theory have appeared in the last
year. Below we outline how our findings complement those of the most
closely related recent work.

\begin{itemize}
    \item \textbf{Cardy-like asymptotics of the $\mathcal{N}=4$ index}: Choi-Kim-Kim-Nahmgoong (CKKN) \cite{Choi:2018hmj}, Honda \cite{Honda:2019cio}, and Ardehali \cite{ArabiArdehali:2019tdm}. All three of these papers investigated the Cardy-like limit of the 4d $\mathcal{N}=4$ index $\mathcal{I}(p,q;y_k)$ using its integral representation, and in the limit where $y_k$ approach the unit circle (for $y_k$ not approaching the unit circle the problem is still open---see Problem~2 in \cite{ArabiArdehali:2019tdm} and subsection~\ref{subsec:futureDirections} below). CKKN \cite{Choi:2018hmj} took actually also a large-$N$ limit---\emph{after} the Cardy-like limit---to simplify the analysis, but in \cite{Honda:2019cio,ArabiArdehali:2019tdm} $N$ was left arbitrary. CKKN \cite{Choi:2018hmj} \emph{assumed} that the dominant holonomy configurations in the Cardy-like limit correspond to equal holonomies $x_{ij}=0$; it was later realized in \cite{Honda:2019cio,ArabiArdehali:2019tdm} that this assumption fails in essentially half of the parameter-space (see the Added Note of \cite{ArabiArdehali:2019tdm} for the relation between the findings of \cite{Honda:2019cio} and \cite{ArabiArdehali:2019tdm}). As in \cite{ArabiArdehali:2019tdm} we divide the parameter-space into complementary M wings and W wings. While on the M wings the dominant holonomy configurations indeed correspond to $x_{ij}=0$ and the asymptotics has been understood, on the W wings finding the Cardy-like asymptotics has been an open problem---see Problem~1 of \cite{ArabiArdehali:2019tdm}. In section~\ref{sec:Cardy} we solved this problem for $N\le4$, and also conjectured the formula (\ref{eq:doubleScalingConjecture}) in the large-$N$ limit. In particular, for $N=4$, as well as in the large-$N$ limit, we have discovered regions on the W wings corresponding to partially deconfined phases in the Cardy-like limit of the 4d $\mathcal{N}=4$ index.
    
    \item \textbf{Large-$N$ asymptotics of the $\mathcal{N}=4$ index}: Benini-Milan \cite{Benini:2018ywd}. This reference studied the large-$N$ limit of the 4d $\mathcal{N}=4$ index using its Bethe Ansatz representation. In the present paper we pointed out some difficulties with the Bethe Ansatz representation for $N\ge3$: there seem to be continuous families of Bethe roots, calling for an integration measure which is so far not understood. Setting this difficulty aside, we pointed out that some of the discrete Bethe roots (found in \cite{Hong:2018viz}) are not negligible in the large-$N$ limit of the index as assumed in \cite{Benini:2018ywd}. We moreover discovered new discrete Bethe roots that play an important role in the large-$N$ asymptotics of the index. The new non-negligible contributions that we have found demonstrate partially deconfined phases in the large-$N$ limit of the index as well.
    \item \textbf{Large-$N$ asymptotics of the $\mathcal{N}=4$ index via the density-distribution approach}: CKKN \cite{Choi:2018vbz}. In this reference, following the original work of Kinney-Maldacena-Minwalla-Raju \cite{Kinney:2005ej}, the large-$N$ limit of the index was analyzed using its integral representation, by rewriting the integral in terms of $\rho_n:=\sum_{j=1}^N e^{2\pi i n x_j}$, rather than $x_j$. These new variables are Fourier coefficients of the density distribution $\rho(x)$ for the holonomies in the large-$N$ limit. The drawback of this approach is that the constraint $\rho(x)\ge0$ makes the range of $\rho_n$ difficult to derive. This difficulty hinders an accurate analysis of the saddle-points of the integral over the $\rho_n$ variables. We have thus avoided this approach in the present work.
    
    \item \textbf{Cardy-like asymptotics of the $\mathcal{N}=1$ index on higher Riemann sheets}: Kim-Kim-Song \cite{Kim:2019yrz} and Cabo~Bizet-Cassani-Martelli-Murthy \cite{Cabo-Bizet:2019osg}. Since the fugacities of the index satisfy $y_1 y_2 y_3=pq$, one can ``turn off the flavor fugacities'' by setting $y_k=(pq)^{1/3}$; the resulting function $\mathcal{I}(p,q;(pq)^{1/3})$ is the usual $\mathcal{N}=1$ superconformal index of the $\mathcal{N}=4$ theory, and is known not to have fast asymptotic growth in the \emph{usual} Cardy-like limit $p,q\to1$, due to bose-fermi cancelations. [More precisely, for SU$(N)$ $\mathcal{N}=4$ theory it can be shown from the results in \cite{Ardehali:2015bla} that $\mathcal{I}(p,q;(pq)^{1/3})$ grows like $\beta^{N-1}$, not like $e^{1/\beta^2}$ as expected from the bulk black holes \cite{Hosseini:2017mds}.] These statements are not the end of the story however, as they really apply only to the fundamental Riemann sheet of the function. For $p,q$ inside the punctured unit open disc, the function $\mathcal{I}(p,q;(pq)^{1/3})$, unlike the more well-behaved $\mathcal{I}(p,q;y_k)$, is not meromorphic; therefore besides turning on flavor fugacities there is another way to get fast growth from it---or ``obstruct its bose-fermi cancelations'' if you will---and that is to go to its higher Riemann sheets. Because of the power $1/3$ for $pq$, the non-meromorphic $\mathcal{N}=1$ index $\mathcal{I}(p,q;(pq)^{1/3})$ has in fact three inequivalent sheets, which can be identified along branch cuts at $\arg p=\pi$, and be labeled by $n_0=-1,0,+1$. The papers \cite{Cabo-Bizet:2019osg,Kim:2019yrz} show that on the $n_0=\pm1$ sheets it is possible to get the fast exponential growth associated to the bulk black holes---simply note that $p\to p e^{2\pi i n_0}$ introduces nontrivial phases in $(pq)^{1/3}$, equivalent to ``turning on flavor fugacities'' and setting $y_k=(pq)^{1/3}e^{2\pi i n_0/3}$ on the $n_0=0$ sheet, corresponding to $\Delta_a=n_0/3$ in the CKKN limit, which can obstruct the bose-fermi cancelations as familiar from CKKN's work \cite{Choi:2018hmj}. Now, while for $\mathrm{sign}(\arg\beta)=\pm$ the Cardy-like asymptotics of the index has been obtained from the fully-deconfining holonomy configurations on the $n_0=\pm1$ sheets, on the other ($n_0=\mp1$) sheets we expect that the partially deconfined configurations become significant. Note that here the control-parameters are $n_0,\mathrm{sign}(\arg\beta)$. In particular, for the SU$(4)$ $\mathcal{N}=4$ theory it follows from our results in section~\ref{sec:Cardy} that it is the partially-deconfining $\mathbb{Z}_4\to \mathbb{Z}_2$ holonomy configurations which take over on the $n_0=\mp1$ sheets. Similarly, if our conjecture (\ref{eq:doubleScalingConjecture}) is correct, in the large-$N$ limit the partially-deconfining $\mathbb{Z}_N\to \mathbb{Z}_{N/2}$ holonomy configurations take over on the $n_0=\mp1$ sheets. (As discussed in subsection~\ref{sec:Cardy:large-N}, when $N$ is odd, as $N\to\infty$ we can approximate the problem by replacing $N\to N-1$, and then consider the $\mathbb{Z}_{N-1}\to\mathbb{Z}_{(N-1)/2}$ configurations instead.)
    
    The results of \cite{Kim:2019yrz,Cabo-Bizet:2019osg} also imply universal expressions for the asymptotics of the $\mathcal{N}=1$ indices of large classes of 4d SCFTs on parts of their $n_0=\pm1$ sheets. We expect that on other parts of those sheets, as well as on other sheets when available, partially deconfined phases with different asymptotics might arise. 
    
    \item \textbf{Higher Riemann sheets of the $\mathcal{N}=1$ index with path-integration}: Cabo~Bizet-Cassani-Martelli-Murthy \cite{Cabo-Bizet:2018ehj}. This reference is also related to analytic continuation of the $\mathcal{N}=1$ index to the $n_0\neq0$ sheets, and is in fact the pioneering work on investigation of such higher sheets. However, it adopts a Lagrangian (path-integral) approach, whose analytic continuation is not properly understood. In particular, the path-integral computation in \cite{Cabo-Bizet:2018ehj} suffers from subtleties regarding regularization of the analytically continued supersymmetric Casimir energy: as discussed in footnote 13 of \cite{Cabo-Bizet:2018ehj}, already before analytic continuation the regularization scheme used there does not give the correct result when applied to general theories, and it is somewhat of a coincidence that it works for the $\mathcal{N}=4$ theory. There seems to be no reason to believe that the scheme keeps being coincidentally correct in the analytically continued case, which is relevant to black hole counting. [Such subtleties can be overcome when working with the path-integral version of the $\mathcal{N}=4$ index $\mathcal{I}(p,q;y_k)$ in the CKKN limit though, which would be relevant to the analysis here; see section~4 of \cite{ArabiArdehali:2019tdm} where this issue was addressed.]
    
    \item \textbf{Large-$N$ asymptotics of the $\mathcal{N}=1$ index on higher Riemann sheets}: Cabo~Bizet-Murthy \cite{Cabo-Bizet:2019eaf}. This paper studies the large-$N$ asymptotics of the $\mathcal{N}=1$ index (with $p=q$) of the $\mathcal{N}=4$ theory on its $n_0\neq0$ sheets in the density-distribution approach. It adopts a variational perspective and looks for saddle-point configurations of $\rho(x)$, instead of working with the subtle Fourier coefficients $\rho_n$. While the focus of \cite{Cabo-Bizet:2019eaf} is on the saddle-point configurations corresponding to the standard $\{m,n,r\}$ eBAE solutions with $\gcd(m,n,r)=1$, our results imply that detecting the partially deconfined phases in their approach requires investigating more general saddle-point configurations corresponding to the standard eBAE solutions with $\gcd(m,n,r)>1$, as well as configurations corresponding to non-standard eBAE solutions.
    
    \item \textbf{The topologically-twisted $\mathcal{N}=4$ index}: Hosseini-Nedelin-Zaffaroni \cite{Hosseini:2016cyf} and Hong-Liu \cite{Hong:2018viz}. Besides the superconformal index, there is the rich and interesting topologically-twisted index that one can compute exactly for the $\mathcal{N}=4$ theory. This index has been analyzed using a Bethe Ansatz expression with precisely the same eBAEs that feature in the 4d $\mathcal{N}=4$ superconformal index. Our results in this paper thus imply that the expressions analyzed in \cite{Hosseini:2016cyf} and \cite{Hong:2018viz} too, for rank $\ge2$, suffer from the difficulty of continuously connected Bethe roots. Furthermore, in some regions of the parameter-space this index exhibits a well-understood, fast asymptotic growth of type $e^{1/\beta}$ in the Cardy-like limit, which is associated to black strings in AdS$_5$ \cite{Hosseini:2016cyf}. In the rest of the parameter-space finding the asymptotics is more challenging however \cite{Hong:2018viz}. In those regions, we expect that the new discrete solutions that we introduced in this paper are of significance in the Cardy-like asymptotics of the topologically twisted index.
\end{itemize}

\subsection{Future directions}\label{subsec:futureDirections}

We conclude by outlining some of the interesting open problems and
exciting prospects related to this work.

\begin{itemize}
    \item \textbf{The Bethe Ansatz approach}. There are several important open questions regarding the Bethe Ansatz approach. Already at rank 1 (i.e. for SU$(2)$) where the Bethe Ansatz formula seems valid, not all the solutions of the eBAE are known; as discussed in section~\ref{sec:BA}, besides the four standard solutions \cite{Hong:2018viz} $0,1/2,\tau/2,(1+\tau)/2$, there are two more solutions of the form $\pm u_\Delta$, and it would be nice to find $u_\Delta$ in closed form.

    For rank $\ge2$ we have given numerical evidence that there is a continuously connected set of solutions, undermining the Bethe Ansatz formula in its current form as a finite sum over eBAE vacua; it would be nice to have a proof for existence of uncountably many Bethe roots for rank $\ge2$. It would also be important to find ways of reformulating the BA approach to incorporate the continuous sets of Bethe roots into account. It would moreover be necessary for black hole-counting applications (as in \cite{Benini:2018ywd}) to characterize all the Bethe roots contributing to the leading asymptotics of the index in the large-$N$ limit. (We expect similar challenges facing the analyses in \cite{Lezcano:2019pae,Lanir:2019abx} as well.)

    \item \textbf{General Cardy-like asymptotics and black holes with unequal charges}. As remarked above, the Cardy-like asymptotics of the 4d $\mathcal{N}=4$ index for general complex $\Delta_{1,2}$ is still an open problem. This  problem is relevant to the microstate counting of the bulk black holes with unequal charges, because it is only for the equal-charge black holes that $\Delta_{1,2}\in\mathbb{R}$ as we have assumed \cite{ArabiArdehali:2019tdm}.  While it might be quite challenging to address this problem for arbitrary $N$, in the SU$(2)$ case this does not seem out of reach: finding the asymptotic form of $u_\Delta$ would lead to the desired asymptotics using the Bethe Ansatz formula.

    \item \textbf{The gravity dual of the partially deconfined phases}. The bulk duals of the partially-deconfined phases have not been constructed as far as we are aware. They should be new black objects and their accurate interpretation is not clear to us at this point. It is tempting to speculate that they correspond to multi-center black holes, which in cases with lens-space horizon topology are sometimes referred to as black lenses \cite{Kunduri:2014kja,Tomizawa:2019yzb}. (We expect partially-deconfined phases in the indices analyzed in \cite{Lezcano:2019pae,Lanir:2019abx,Amariti:2019mgp} as well, with new possibly multi-center black objects associated to them in the bulk.)

    \item \textbf{The black hole operators}. One of the most exciting aspects of the recent advances in AdS$_5$/CFT$_4$ microstate counting is the prospect they open for explicit construction of the operators dual to the bulk BPS black holes. The operators dual to the bulk KK multi-particle states have long been known of course: they are the multi-trace operators, generated by the famous single-trace operators nicely reviewed in Table~7 of \cite{DHoker:2002nbb} for instance. The black hole operators on the other hand, have been elusive even in AdS$_3$/CFT$_2$. We hope that the emerging refined understanding of the asymptotic growth of the 4d $\mathcal{N}=4$ index can guide future pursuits of these long-sought operators in the $\mathcal{N}=4$ theory.\\

\end{itemize}

\noindent\textbf{Note added.} After the first version of the present paper appeared on arXiv, we became aware that the latest version of \cite{Cabo-Bizet:2019osg} contains the contribution of a divisor holonomy configuration to the Cardy-like asymptotics of the $\mathcal{N}=1$ index on its higher Riemann sheets. In particular, Eq.~(3.50) in that work is essentially the analog of our (\ref{eq:dSummedLowerBound}) in that context.  Our conjecture (\ref{eq:doubleScalingConjecture}) implies that for SU($N$) $\mathcal{N}=4$ theory in the large-$N$ limit, irrespective of whether $N$ is even or odd, Eq.~(3.50) of \cite{Cabo-Bizet:2019osg} with $K=2$ would give the leading Cardy-like asymptotics in the parameter-regimes which were previously unexplored (i.e.  $\mathrm{Re}(i/\tau\sigma)>0$, $n_0=-1$, or $\mathrm{Re}(i/\tau\sigma)<0$, $n_0=+1$).

%%%%%%%%%%%%%%%%%%%%
\begin{acknowledgments}
We thank F.~Benini, D.~Cassani, T.~Hansen, Y.~L\"{u}, R.~de~Mello~Koch, and J.~Minahan for related discussions, D.~Speyer for correspondence on Lemma~1, and H.~Rosengren for correspondence on Lemma~2. AAA would like to thank P.~Milan for suggesting numerical investigation of the Cardy-like limit. JTL wishes to thank B.~Willett for raising the issue of non-standard solutions to the eBAEs. This work was supported in part by
the Knut and Alice Wallenberg Foundation under grant Dnr KAW
2015.0083 and by the U.S.~Department of Energy under grant DE-SC0007859. JH is supported in part by a Leinweber Graduate Fellowship from the University of Michigan.
\end{acknowledgments}

\appendix

%%%%%
\section{Proof of Lemma~1}
\label{App:proof}
%%%%%
Cover the torus $\mathbb{R}^2/\mathbb{Z}^2$ with balls of radius
$\varepsilon/2$. By the pigeonhole principle, there are two integers
$A < B$ such that $(\{Ax\},\{Ay\})$ and $(\{Bx\},\{By\})$ are in the
same ball. Then $(\{(B-A) x\}, \{(B-A) y\})$ is  in the ball of
radius $\varepsilon$ around 0 (mod $\mathbb{Z}^2$).

Now, if $\{(B-A) x\}+ \{(B-A) y\}>1$, we are done by taking $C=B-A$.

If on the other hand $\{(B-A) x\}+ \{(B-A) y\}<1$, then the relation
\begin{equation}
\{\alpha-\beta\}=\begin{cases}
\{\alpha\}-\{\beta\},\quad & \{\alpha\}\ge\{\beta\};\\
\{\alpha\}-\{\beta\}+1,\quad & \{\alpha\}<\{\beta\},
\end{cases}\label{eq:fracPartSubtract}
\end{equation}
guarantees that for $\varepsilon$ small enough we have $(\{(B-A-1)x\}+ \{(B-A-1)y\})>1$, so we are done by taking $C=B-A-1$. {\it Q.E.D.}\\

Let us see how things work in an example. Take $x=y=1/3$. The two
values $B=6$ and $A=3$ are acceptable. Now, since $\{(B-A) x\}+
\{(B-A) y\}=\{1\}+\{1\}=0<1$, we can take $C=B-A-1=2$. Indeed
$\{2\cdot\frac{1}{3}\}+\{2\cdot\frac{1}{3}\}=\frac{4}{3}>1$ as
desired. As explained above (\ref{eq:oneTermBound}), this implies
that for $\arg\beta<0$, at the point
$(\Delta_1,\Delta_2)=(1/3,1/3)$, the bound in
(\ref{eq:oneTermBound}) with $C=2$ guarantees a partially deconfined
behavior for $\mathcal{I}_{N\to\infty}$.

%%%%%
\section{Proof of Lemma~2}\label{App:proof2}
%%%%%

Using the identity
\begin{equation}
\theta'_1(n;\tau)=\theta_2(n;\tau)\theta_3(n;\tau)\theta_4(n;\tau)
\end{equation}
for an arbitrary integer $n\in\mathbb Z$, we can generalize Theorem 2.1 of \cite{Men:2016} as
\begin{equation}
	\sum_{a=1}^3\fft{\theta_I'(\tilde\Delta_a;\tau)}{\theta_I(\tilde\Delta_a;\tau)}=\begin{cases}
	+\fft{\theta_1'(\tilde n;\tau)}{\theta_I(\tilde n;\tau)}\prod_{a=1}^3\fft{\theta_1(\tilde\Delta_a;\tau)}{\theta_I(\tilde\Delta_a;\tau)} & (I=3); \\
	-\fft{\theta_1'(\tilde n;\tau)}{\theta_I(\tilde n;\tau)}\prod_{a=1}^3\fft{\theta_1(\tilde\Delta_a;\tau)}{\theta_I(\tilde\Delta_a;\tau)} & (I=2,4),
	\end{cases}\label{eq:Theorem2point1}
\end{equation}
where $\sum_a\tilde\Delta_a=\tilde n\in\mathbb Z$. Here $\theta_{2,3,4}$ are related to $\theta_1(u;\tau)$ defined in (\ref{theta1:product}) via
\begin{subequations}
\begin{align}
	\theta_2(u;\tau)&=\theta_1(u+1/2;\tau),\\
	\theta_3(u;\tau)&=e^{\fft{\pi i\tau}{4}}e^{\pi iu}\theta_1(u+(1+\tau)/2;\tau),\\
	\theta_4(u;\tau)&=-ie^{\fft{\pi i\tau}{4}}e^{\pi iu}\theta_1(u+\tau/2;\tau).
\end{align}\label{theta:different}%
\end{subequations}
These theta functions are basically obtained from $\theta_1(u;\tau)$ by shifting the first argument $u$ by three different half-periods $\fft12,\fft{1+\tau}{2},\fft{\tau}{2}$ respectively. They satisfy the so-called Jacobi's formula together with $\theta_1(u;\tau)$, namely \cite{WW:1927}
\begin{align}
	2\theta_1(u_0,u_1,u_2,u_3;\tau)&=\theta_1(u_0',u_1',u_2',u_3';\tau)+\theta_2(u_0',u_1',u_2',u_3';\tau)\nn\\&\quad-\theta_3(u_0',u_1',u_2',u_3';\tau)+\theta_4(u_0',u_1',u_2',u_3';\tau),\label{Jacobi}
\end{align}
where $2u'_\alpha=\sum_{\beta=0}^3u_\beta-2u_\alpha~(\alpha=0,1,2,3)$, and we have used the abbreviations
\begin{equation}
	\theta_I(u_0,u_1,u_2,u_3;\tau)\equiv\prod_{\alpha=0}^3\theta_I(u_\alpha;\tau).
\end{equation}
Since $\theta_1(n;\tau)=0$ for an arbitrary integer $n\in\mathbb Z$, the following special case of Jacobi's formula (\ref{Jacobi}) is valid for $\sum_a\tilde\Delta_a=\tilde n\in\mathbb Z$:
\begin{equation}
	0=\theta_2(\tilde n,\tilde\Delta_1,\tilde\Delta_2,\tilde\Delta_3;\tau)-\theta_3(\tilde n,\tilde\Delta_1,\tilde\Delta_2,\tilde\Delta_3;\tau)+\theta_4(\tilde n,\tilde\Delta_1,\tilde\Delta_2,\tilde\Delta_3;\tau).\label{eq:specialJacobiFormula}
\end{equation}
Combining (\ref{eq:specialJacobiFormula}) and (\ref{eq:Theorem2point1}) establishes the lemma. {\it Q.E.D.}\\

As discussed in the main text, the lemma proves that the SU$(3)$ eBAEs have a ``zero mode'' at the exact non-standard solution (\ref{eq:nonstandardN=3}).

%%%%%
\section{The elliptic Bethe Ansatz equations in the asymptotic regions}
\label{App:C}
%%%%%

In this appendix, we investigate the eBAEs (\ref{BAE}) in the
asymptotic regions, including both the low-temperature
($|\tau|\gg1$) and the high-temperature ($|\tau|\ll1$) limits. Then
we look for asymptotic, non-standard solutions for $N=2$ and $N=3$.

Before getting into details, first we rewrite (\ref{BAE}) as
\begin{equation}
1=Q_i=e^{2\pi
i\lambda}\prod_{j=1}^N\prod_{a=1}^3\fft{\theta_1(u_{ji}+\tilde\Delta_a;\tau)}{\theta_1(u_{ij}+\tilde\Delta_a;\tau)},\label{BAE:tilde}
\end{equation}
where, for notational convenience, we have introduced $\tilde\Delta_a$
as
\begin{equation}
\tilde\Delta_a=\begin{cases}
\Delta_a & (a=1,2); \\
-\Delta_1-\Delta_2 & (a=3).
\end{cases}\label{chemical:tilde}
\end{equation}
This is to avoid using a complex $\Delta_3=2\tau-\Delta_1-\Delta_2$
and to keep all the chemical potentials $\tilde\Delta_a$ real. We now use quasi-periodicity and the oddness of $\theta_1(u;\tau)$, namely (\ref{theta1:property}),
to rearrange the chemical potentials $\tilde\Delta_a$ and the holonomies $u_k$ for convenience when writing down the asymptotic expansions.  Note that we always assume $\Delta_{1,2}\in\mathbb R$, or equivalently $\tilde\Delta_a\in\mathbb R$. 

We start with shifting $\tilde\Delta_a$ by integers using
(\ref{theta1:property}) so that $0\leq\Re\tilde\Delta_a<1$ is
satisfied. This yields $\sum_a\tilde\Delta_a\in\{0,1,2\}$. Then
since the eBAEs (\ref{BAE:tilde}) are invariant under
$\tilde\Delta_a\to1-\tilde\Delta_a$ and $\lambda\to-\lambda$ due to
(\ref{theta1:property}), redefining $\tilde\Delta_a$ as
$\tilde\Delta_a\to1-\tilde\Delta_a$ does not change the eBAE
solutions $\{u_i\}$. Therefore, we can redefine $\tilde\Delta_a$ as
$\tilde\Delta_a\to1-\tilde\Delta_a$ whenever
$\sum_a\tilde\Delta_a=2$ and consequently we have
$\sum_a\tilde\Delta_a\in\{0,1\}$. Finally, we assume
$\tilde\Delta_a\in\mathbb R$, then
$0\leq\mathrm{Re}\tilde\Delta_a<1$ together with
$\sum_a\tilde\Delta_a=0$ leads to $\tilde\Delta_a=0$, which is
forbidden for the index to converge. So we have
\begin{align}
0<\tilde\Delta_a<1,\qquad
\sum_{a=1}^3\tilde\Delta_a=1.\label{chemical}
\end{align}
The final expression of $\tilde\Delta_a$ shifted from (\ref{chemical:tilde}) is given in terms of $\Delta_a$ as
\begin{equation}
\begin{split}
    \tilde\Delta_{1,2}&=\begin{cases}
    \{\Delta_{1,2}\} & (\{\Delta_1\}+\{\Delta_2\}+\{-\Delta_1-\Delta_2\}=1); \\
    1-\{\Delta_{1,2}\} & (\{\Delta_1\}+\{\Delta_2\}+\{-\Delta_1-\Delta_2\}=2),
    \end{cases}\\
    \tilde\Delta_3&=\begin{cases}
    \{-\Delta_1-\Delta_2\} & (\{\Delta_1\}+\{\Delta_2\}+\{-\Delta_1-\Delta_2\}=1);\\
    1-\{-\Delta_1-\Delta_2\} & (\{\Delta_1\}+\{\Delta_2\}+\{-\Delta_1-\Delta_2\}=2),
    \end{cases}
\end{split}\label{eq:tilde:Delta}
\end{equation}
where as usual $\{\cdot\}=\cdot-\lfloor \cdot\rfloor$, and we have assumed $\Delta_{1,2}\in\mathbb R\setminus\mathbb{Z}$.

We can also specify the range of holonomies $u_k$ using the same
properties of $\theta_1(u;\tau)$ given in (\ref{theta1:property}). The key
is that the eBAEs (\ref{BAE:tilde}) are invariant under $u_k\to
u_k+p_k+q_k\tau$ for arbitrary integers $p_k,q_k\in\mathbb Z$. Using
this invariance, we can set $u_k$ as
\begin{equation}
u_k=x_k+y_k\tau\quad(0\leq x_k,y_k<1),\label{BAE:sols}
\end{equation}
with $y_i\leq y_j$ for $i\leq j$ without loss of generality.

Following the setup (\ref{chemical}) and (\ref{BAE:sols}), now we
investigate the eBAEs (\ref{BAE:tilde}) in asymptotic regions and
look for non-standard solutions there.

%%%%%
\subsection{Low-temperature asymptotic solutions}
%%%%%
We start with the low-temperature ($|\tau|\gg1$) asymptotic region.
First we rewrite the infinite product form of $\theta_1(u;\tau)$ (\ref{theta1:product}) as
\begin{align}
	\theta_1(u;\tau)=ie^{\fft{\pi i\tau}{4}}e^{\pi i(p+p(p+1)\tau-u(2p+1))}\prod_{k=1}^\infty(1-e^{2\pi ik\tau})(1-e^{2\pi i((k-1)\tau+(u-\tau p))})(1-e^{2\pi i(k\tau-(u-\tau p))}),\label{lowT:expansion}
\end{align}
where we have defined the integer $p$ as
\begin{equation}
	p=\left\lfloor\fft{\mathrm{Im}\,u}{|\tau|\sin\theta}\right\rfloor\qquad(\tau=|\tau|e^{i\theta}).\label{angle}
\end{equation}
Substituting the form of eBAE solutions $u_k$ (\ref{BAE:sols}) and
the product form (\ref{lowT:expansion}) into the eBAEs
(\ref{BAE:tilde}) then gives
\begin{equation}
e^{\pi
i(N-2\lambda-1)}=\prod_{a=1}^3\prod_{j=1}^N\prod_{k=1}^\infty\fft{(1-e^{2\pi
i(-x_{ij}+(k-\{y_{ij}\})\tau+\tilde\Delta_a)})(1-e^{2\pi
i(x_{ij}+(k-1+\{y_{ij}\})\tau-\tilde\Delta_a)})}{(1-e^{2\pi
i(x_{ij}+(k-1+\{y_{ij}\})\tau+\tilde\Delta_a)})(1-e^{2\pi
i(-x_{ij}+(k-\{y_{ij}\})\tau-\tilde\Delta_a)})},\label{BAE:lowT:1}
\end{equation}
where the curly bracket denotes mod $\mathbb Z$ as in
(\ref{eq:kappa}). Under the low-temperature limit, the contributions
from $k\geq2$ are exponentially suppressed and therefore
(\ref{BAE:lowT:1}) reduces to
\begin{equation}
e^{\pi i(N-2\lambda-1)}=\prod_{a=1}^3\prod_{j=1}^N\fft{(1-e^{2\pi
i(x_{ij}+\{y_{ij}\}\tau-\tilde\Delta_a)})(1-e^{2\pi
i(-x_{ij}+(1-\{y_{ij}\})\tau+\tilde\Delta_a)})}{(1-e^{2\pi
i(x_{ij}+\{y_{ij}\}\tau+\tilde\Delta_a)})(1-e^{2\pi
i(-x_{ij}+(1-\{y_{ij}\})\tau-\tilde\Delta_a)})}\times
\left(1+\mathcal
O(e^{-2\pi|\tau|\sin\theta})\right).\label{BAE:lowT:2}
\end{equation}
These are the eBAEs reduced under the low-temperature limit, which
yield $N-1$ independent algebraic equations. They are still involved
to solve in general, however, so we consider simple cases: $N=2$ and
$N=3$.

%%%%%
\subsubsection{\texorpdfstring{$N=2$}{N=2}}
%%%%%
For $N=2$, the reduced eBAEs (\ref{BAE:lowT:2}) yield a single
algebraic equation,
\begin{align}
\pm1=\prod_{a=1}^3\fft{1-e^{2\pi
i(u_{21}-\tilde\Delta_a)}}{1-e^{2\pi
i(u_{21}+\tilde\Delta_a)}}\fft{1-e^{2\pi
i(\tau-u_{21}+\tilde\Delta_a)}}{1-e^{2\pi
i(\tau-u_{21}-\tilde\Delta_a)}}\times \left(1+\mathcal
O(e^{-2\pi|\tau|\sin\theta})\right).\label{BAE:lowT:N=2}
\end{align}
The complete set of solutions to (\ref{BAE:lowT:N=2}) is given as
\begin{equation}
u_{21}\in\left\{0,\fft12,\fft{\tau}{2},\fft{1+\tau}{2},\pm\fft{1}{2\pi
i}\log\left[{\scriptstyle-\fft{1-\sum_a\cos2\pi\tilde\Delta_a}{2}+\sqrt{\left(\fft{1-\sum_a\cos2\pi\tilde\Delta_a}{2}\right)^2-1}}\right]\right\}.\label{BAE:lowT:N=2:sol}
\end{equation}
Here the first
solution is trivial and the next three solutions correspond to the
standard $\{2,1,0\}$, $\{1,2,0\}$, and $\{1,2,1\}$ solutions
respectively. The other two low-temperature asymptotic solutions are
non-standard and completely distinguished from the trivial solution
and the standard $\{m,n,r\}$-type solutions.

%%%%%
\subsubsection{\texorpdfstring{$N=3$}{N=3}}
%%%%%
For $N=3$, we define
\begin{equation}
    z_{21}=e^{2\pi iu_{21}},\qquad z_{31}=e^{2\pi iu_{31}},\qquad y_a=e^{2\pi i\tilde\Delta_a}\qquad\mbox{and}\qquad q=e^{2\pi i\tau}.
\end{equation}
Note that the low-temperature limit corresponds to $|q|\ll1$.  In addition, the setup (\ref{BAE:sols}) restricts
\begin{equation}
    |q|\le|z_{31}|\le|z_{21}|\le1.
\end{equation}
Under this restriction, the reduced eBAEs (\ref{BAE:lowT:2}) yield two algebraic
equations,
\begin{align}
&\prod_{a=1}^3\fft{(1-y_az_{21})(1-y_a^{-1}\fft{q}{z_{21}})(1-y_az_{31})(1-y_a^{-1}\fft{q}{z_{31}})}{(1-y_a^{-1}z_{21})(1-y_a\fft{q}{z_{21}})(1-y_a^{-1}z_{31})(1-y_a\fft{q}{z_{31}})}\nn\\
&\qquad=\prod_{a=1}^3\fft{(1-y_a^{-1}z_{21})(1-y_a\fft{q}{z_{21}})(1-y_a\fft{z_{31}}{z_{21}})(1-y_a^{-1}q\fft{z_{21}}{z_{31}})}{(1-y_az_{21})(1-y_a^{-1}\fft{q}{z_{21}})(1-y_a^{-1}\fft{z_{31}}{z_{21}})(1-y_aq\fft{z_{21}}{z_{31}})}\nn\\
&\qquad=\prod_{a=1}^3\fft{(1-y_a^{-1}z_{31})(1-y_a\fft{q}{z_{31}})(1-y_a^{-1}\fft{z_{31}}{z_{21}})(1-y_aq\fft{z_{21}}{z_{31}})}{(1-y_az_{31})(1-y_a^{-1}\fft{q}{z_{31}})(1-y_a\fft{z_{31}}{z_{21}})(1-y_a^{-1}q\fft{z_{21}}{z_{31}})},\label{BAE:lowT:N=3}
\end{align}
up to exponentially small terms of at most $O(|q|)$.

If we assume $|z_{31}|$ is strictly greater than $|q|$, then we can simply set $q=0$ in (\ref{BAE:lowT:N=3}) in the log-temperature limit.  The resulting equations then simplify and admit a discrete set of solutions
\begin{equation}
    (z_{21},z_{31})=(1,1)\cup(\omega,\omega^{-1})\cup(\omega^{-1},\omega)\cup(-1,0),
    \label{eq:N=3disc}
\end{equation}
where $\omega=e^{2\pi i/3}$, as well as a continuous family of solutions satisfying a single condition
\begin{equation}
    (1+z_{21}+z_{31})\left(1+\fft1{z_{21}}+\fft1{z_{31}}\right)=3+2\sum_a\cos2\pi\tilde\Delta_a,
    \label{eq:N=3cont}
\end{equation}
which reduces to the finitely many non-standard solutions for $N=2$ given in (\ref{BAE:lowT:N=2:sol}) under $z_{31}=z_{21}$. Starting with the discrete solutions, the first three are based on cube roots of unity and correspond to the high temperature limit of the standard solutions.  In particular, the first entry in (\ref{eq:N=3disc}) corresponds to the $C=1$ case, and the next two correspond to the $C=3$ cases. The final solution in (\ref{eq:N=3disc}) is novel, as it is a non-standard solution that is nevertheless independent of the chemical potentials $\tilde\Delta_a$.  Note that, strictly speaking, this solution violates the assumption made above that $|z_{31}|>|q|$.  However, it is in fact easy to see that this extends to an exact non-standard solution
\begin{equation}
    (u_{21},u_{31})=(1/2,\tau/2),
    \label{eq:exnstd}
\end{equation}
for arbitrary $\tau$.

Perhaps more interestingly, we have found a continuous family of solutions given by (\ref{eq:N=3cont}).  This one complex dimensional family of solutions is obviously non-standard since it depends on the chemical potentials $\tilde\Delta_a$.  In addition, the limiting form of (\ref{eq:N=3cont}) for $z_{21}\to-1$ and $z_{31}\to0$ encompasses the non-standard solution (\ref{eq:exnstd}).  This suggests that (\ref{eq:exnstd}) is not a discrete solution but rather a part of the continuous family. We confirm this explicitly in Appendix \ref{App:proof2} by demonstrating that the transformation matrix $\partial Q_i/\partial u_j~(i,j=1,2)$ where $Q_i$ represents the $i$-th BAE contains a zero mode corresponding to a flat direction.

%%%%%
\subsection{High-temperature asymptotic solutions}
%%%%%
Next we consider the high-temperature ($|\tau|\ll1$) asymptotic
region. Using the infinite product form of $\theta_1(u;\tau)$ in
(\ref{theta1:product}) together with the $S$-transformation 
\begin{align}
	S:&\quad \theta_1(u/\tau;-1/\tau)=-i\sqrt{-i\tau}e^{\fft{\pi iu^2}{\tau}}\theta_1(u;\tau),\label{S-transf}
\end{align}
we can derive the high-temperature expansion of $\theta_1(u;\tau)$,
\begin{align}
	\theta_1(u;\tau)&=(-i\tau)^{-\fft12}e^{-\fft{\pi i}{4\tau}}e^{\fft{\pi i}{\tau}u(1-u)}(1-e^{-\fft{2\pi i}{\tau}u})\prod_{k=1}^\infty(1-e^{-\fft{2\pi i}{\tau}k})(1-e^{-\fft{2\pi i}{\tau}(k-u)})(1-e^{-\fft{2\pi i}{\tau}(k+u)})\nn\\
&=(-i\tau)^{-\fft12}e^{-\fft{\pi i}{4\tau}}(-1)^{p}e^{\fft{\pi
i}{\tau}\{u\}_\tau(1-\{u\}_\tau)}\prod_{k=1}^\infty(1-e^{-\fft{2\pi
i}{\tau}k})(1-e^{-\fft{2\pi i}{\tau}(k-\{u\}_\tau)})(1-e^{-\fft{2\pi
i}{\tau}(k-1+\{u\}_\tau)})\label{highT:expansion}
\end{align}
where we have used the notation $\{u\}_\tau$ introduced in
(\ref{eq:kappa:tau}) and defined the integer $p$ as
\begin{equation}
	p=\lfloor\mathrm{Re}u-\cot\theta\,\mathrm{Im}u\rfloor\qquad(\tau=|\tau|e^{i\theta}).
\end{equation}
Substituting the form of eBAE solutions $u_k$
(\ref{BAE:sols}) and the product form (\ref{highT:expansion}) into
the eBAEs (\ref{BAE:tilde}) then gives
\begin{align}
e^{-2\pi i\lambda}&=\prod_{a=1}^3\prod_{j=1}^N\Bigg[e^{\fft{\pi
i}{\tau}\left(\{-x_{ij}+\tilde\Delta_a\}(1-\{-x_{ij}+\tilde\Delta_a\})-\{x_{ij}+\tilde\Delta_a\}(1-\{x_{ij}+\tilde\Delta_a\})\right)}\nn\\&\kern5em+e^{\pi
i(\lfloor-x_{ij}+\tilde\Delta_a\rfloor-\lfloor
x_{ij}+\tilde\Delta_a\rfloor-2y_{ij}(1-\{-x_{ij}+\tilde\Delta_a\}-\{x_{ij}+\tilde\Delta_a\}))}\nn\\&\kern5em\times\prod_{k=1}^\infty\fft{(1-e^{-\fft{2\pi
i}{\tau}(k-\{-x_{ij}+\tilde\Delta_a\}+y_{ij}\tau)})(1-e^{-\fft{2\pi
i}{\tau}(k-1+\{-x_{ij}+\tilde\Delta_a\}-y_{ij}\tau)})}{(1-e^{-\fft{2\pi
i}{\tau}(k-\{x_{ij}+\tilde\Delta_a\}-y_{ij}\tau)})(1-e^{-\fft{2\pi
i}{\tau}(k-1+\{x_{ij}+\tilde\Delta_a\}+y_{ij}\tau)})}\Bigg].\label{BAE:highT:1}
\end{align}
where the curly bracket denotes mod $\mathbb Z$ as in
(\ref{eq:kappa}). Under the high-temperature limit, the
contributions from $k\geq2$ are exponentially suppressed and
therefore (\ref{BAE:highT:1}) reduces to
\begin{align}
e^{-2\pi i\lambda}&=\prod_{a=1}^3\prod_{j=1}^N\Bigg[e^{\fft{\pi i}{\tau}\left(\{-x_{ij}+\tilde\Delta_a\}(1-\{-x_{ij}+\tilde\Delta_a\})-\{x_{ij}+\tilde\Delta_a\}(1-\{x_{ij}+\tilde\Delta_a\})\right)}\nn\\
&\kern5em+e^{\pi i(\lfloor-x_{ij}+\tilde\Delta_a\rfloor-\lfloor x_{ij}+\tilde\Delta_a\rfloor-2y_{ij}(1-\{-x_{ij}+\tilde\Delta_a\}-\{x_{ij}+\tilde\Delta_a\}))}\nn\\
&\kern5em\times\fft{(1-e^{-\fft{2\pi i}{\tau}(1-\{-x_{ij}+\tilde\Delta_a\}+y_{ij}\tau)})(1-e^{-\fft{2\pi i}{\tau}(\{-x_{ij}+\tilde\Delta_a\}-y_{ij}\tau)})}{(1-e^{-\fft{2\pi i}{\tau}(1-\{x_{ij}+\tilde\Delta_a\}-y_{ij}\tau)})(1-e^{-\fft{2\pi i}{\tau}(\{x_{ij}+\tilde\Delta_a\}+y_{ij}\tau)})}\Bigg]\nn\\
&\qquad\times\left(1+\mathcal
O(e^{-\fft{2\pi\sin\theta}{|\tau|}})\right).\label{BAE:highT:2}
\end{align}
These are the eBAEs reduced under the high-temperature limit, which
yield $N-1$ independent algebraic equations. They are still involved
to solve in general, however, so we consider simple cases: $N=2$ and
$N=3$.

%%%%%
\subsubsection{\texorpdfstring{$N=2$}{N=2}}
%%%%%
For $N=2$, the reduced eBAEs (\ref{BAE:highT:2}) yield a single
algebraic equation (take logarithmic function on both sides),
\begin{align}
\mathbb Z\pi i=&~\fft{\pi
i}{\tau}\sum_{a=1}^3\left(\{x_{21}+\Delta_a\}(1-\{x_{21}+\Delta_a\})-\{-x_{21}+\Delta_a\}(1-\{-x_{21}+\Delta_a\})\right)\nn\\&+\pi
i\sum_{a=1}^3\left(\lfloor x_{21}+\Delta_a\rfloor-\lfloor
-x_{21}+\Delta_a\rfloor+2y_{21}(1-\{x_{21}+\Delta_a\}-\{-x_{21}+\Delta_a\})\right)\nn\\&+\sum_{a=1}^3\log\left[\fft{(1-e^{-\fft{2\pi
i}{\tau}(1-\{x_{21}+\Delta_a\}-y_{21}\tau)})(1-e^{-\fft{2\pi
i}{\tau}(\{x_{21}+\Delta_a\}+y_{21}\tau)})}{(1-e^{-\fft{2\pi
i}{\tau}(1-\{-x_{21}+\Delta_a\}+y_{21}\tau)})(1-e^{-\fft{2\pi
i}{\tau}(\{-x_{21}+\Delta_a\}-y_{21}\tau)})}\right]+\mathcal
O(e^{-\fft{2\pi\sin\theta}{|\tau|}}).\label{BAE:highT:N=2}
\end{align}
Solving this equation is quite involved in general mainly because
the last logarithmic term may yield a large contribution for some
special values of $x_{21}$. We therefore solve this equation under
appropriate assumptions on $x_{21}$.

\subsubsection*{CASE 1. $\{\pm x_{21}+\tilde\Delta_a\}$ are NOT asymptotically close to 0 or 1.}
First, we consider the solutions where $\{\pm
x_{21}+\tilde\Delta_a\}$ are not asymptotically close to the end
points 0 and 1. In this case, the last logarithmic term in
(\ref{BAE:highT:N=2}) is exponentially suppressed as $\sim\mathcal
O(e^{-\fft{2\pi\sin\theta}{|\tau|}x})$ where we have defined a real
number $x\in(0,1)$ as
\begin{equation}
x\equiv\min(\{x_{21}+\tilde\Delta_a\},1-\{x_{21}+\tilde\Delta_a\},\{-x_{21}+\tilde\Delta_a\},1-\{-x_{21}+\tilde\Delta_a\}:\,a=1,2,3).
\end{equation}
Then (\ref{BAE:highT:N=2}) is simplified into the following system
of algebraic equations,
\begin{subequations}
    \begin{align}
    0&=\sum_{a=1}^3\left(\{x_{21}+\tilde\Delta_a\}(1-\{x_{21}+\tilde\Delta_a\})-\{-x_{21}+\tilde\Delta_a\}(1-\{-x_{21}+\tilde\Delta_a\})\right),\\
    \mathbb Z&=\sum_{a=1}^3\left(\lfloor x_{21}+\tilde\Delta_a\rfloor-\lfloor -x_{21}+\tilde\Delta_a\rfloor+2y_{21}(1-\{x_{21}+\tilde\Delta_a\}-\{-x_{21}+\tilde\Delta_a\})\right),
    \end{align}\label{BAE:highT:N=2:case1}%
\end{subequations}
up to $\mathcal O(e^{-\fft{2\pi\sin\theta}{|\tau|}x})$. If we set
$0<\tilde\Delta_1\leq\tilde\Delta_2\leq\tilde\Delta_3<1$ without
loss of generality, the complete set of solutions to
(\ref{BAE:highT:N=2:case1}) can be written as
\begin{align}
u_{21}=x_{21}+y_{21}\tau\in&~\left\{0,\fft12,\fft{\tau}{2},\fft{1+\tau}{2}\right\}\cup\left\{\fft12+\fft{\tau}{4},\fft12+\fft{3\tau}{4}\bigg|\tilde\Delta_3<\fft12\right\}\nn\\&~\cup\left\{x_{21}+y_{21}\tau\bigg|x_{21}\in[1-\tilde\Delta_3,\tilde\Delta_3],~y_{21}\in[0,1),~\tilde\Delta_3\geq\fft12\right\},\label{BAE:highT:N=2:case1:sol}
\end{align}
up to $\mathcal O(e^{-\fft{2\pi\sin\theta}{|\tau|}x})$. The complete
set of solutions to (\ref{BAE:highT:N=2}) corresponding to the CASE
1 is therefore given as (\ref{BAE:highT:N=2:case1:sol}) whose subset
violating the assumption ``$\{x_{21}\pm\tilde\Delta_a\}$ are not
asymptotically close to 0 or 1'' is excluded. For example,
$u_{21}=\tilde\Delta_3+y_{21}\tau~(\tilde\Delta_3\geq1/2)$ must be
excluded from the set of solutions to (\ref{BAE:highT:N=2})
corresponding to the CASE 1 even though it's included in
(\ref{BAE:highT:N=2:case1:sol}), since
$\{-x_{21}+\tilde\Delta_3\}=0$ violates the assumption.

\subsubsection*{CASE 2. Some $\{\pm x_{21}+\tilde\Delta_a\}$ are asymptotically close to 0 or 1.}
Next, we consider the solutions where some of the $\{\pm
x_{21}+\tilde\Delta_a\}$ are asymptotically close to 0 or 1. In this
case, the last logarithmic term in (\ref{BAE:highT:N=2}) may yield a
contribution which is not exponentially suppressed. Therefore it is
hard to reduce (\ref{BAE:highT:N=2}) further as we did in CASE 1,
and consequently it becomes difficult to find the complete set of
solutions to (\ref{BAE:highT:N=2}) in CASE 2.

We can still find, however, a particular solution in CASE 2 by using
an appropriate ansatz. Setting
$0<\tilde\Delta_1\leq\tilde\Delta_2\leq\tilde\Delta_3<1$ without
loss of generality, we use the following ansatz,
\begin{equation}
u_{21}=\tilde\Delta_3+\epsilon|\tau|+y_{21}\tau\quad\left(\tilde\Delta_3\geq\fft12\right),\label{case2:ansatz}
\end{equation}
where $\epsilon$ is a positive real number. Note that this ansatz
has $\{-x_{21}+\tilde\Delta_3\}$ asymptotically close to 0 so breaks
the assumption for CASE 1; hence, using this ansatz may yield a new
asymptotic solution that is not covered by
(\ref{BAE:highT:N=2:case1:sol}).

Substituting (\ref{case2:ansatz}) into (\ref{BAE:highT:N=2}), we get
\begin{equation}
\pm1=\fft{e^{-2\pi
i\left(\epsilon\fft{|\tau|}{\tau}+y_{21}\right)}}{1-e^{-2\pi
i\left(\epsilon\fft{|\tau|}{\tau}+y_{21}\right)}}\times\left(1+\mathcal
O(e^{-\fft{2\pi\sin\theta}{|\tau|}\min(\tilde\Delta_1-\epsilon|\tau|,2\tilde\Delta_3-1+2\epsilon|\tau|)})\right),
\end{equation}
after taking exponential function on both sides. Ignoring the
exponentially suppressed term, it is clear that the LHS must take
$+1$ to yield a regular solution and consequently we have
\begin{equation}
\epsilon=\fft{\log
2}{2\pi\sin\theta}~~\mbox{and}~~y_{21}=\left\{-\fft{\log2\cot\theta}{2\pi}\right\}\quad\Leftrightarrow\quad
u_{21}=\tilde\Delta_3-i\fft{\log2}{2\pi}\tau~(\mbox{mod}~\mathbb
Z+\mathbb Z\tau).
\end{equation}
Recall that $u_{21}\to-u_{21}$ does not affect that a given solution
satisfies the eBAEs (\ref{BAE:tilde}) so we found two examples of
high-temperature asymptotic solutions in CASE 2, namely
\begin{equation}
u_{21}\in\left\{\pm\left(-\tilde\Delta_3+i\fft{\log2}{2\pi}\tau\right)\bigg|\tilde\Delta_3\geq\fft12\right\}.\label{BAE:highT:N=2:case2:sol}
\end{equation}

%%%%%
\subsubsection{\texorpdfstring{$N=3$}{N=3}}
%%%%%
For $N=3$, the reduced eBAEs (\ref{BAE:highT:2}) yield two algebraic
equations. Since they are more involved than the ones for $N=2$, we
only consider the case where $\{x_{ij}+\tilde\Delta_a\}~(1\leq i\neq j\leq3)$ are not
asymptotically close to 0 or 1. Then the two algebraic equations
from the eBAEs (\ref{BAE:highT:2}) are simplified as
\begin{equation}
F_1(u_{21},u_{32})=F_2(u_{21},u_{32})=F_3(u_{21},u_{32})\label{BAE:highT:N=3}
\end{equation}
up to exponentially suppressed terms of order $\mathcal
O(e^{-\fft{2\pi\sin\theta}{|\tau|}X})$, where we have defined
\begin{align}
F_i(u_{21},u_{32})\equiv&~\fft{\pi i}{\tau}\sum_{a=1}^3\sum_{j=1}^3\left(\{-x_{ij}+\tilde\Delta_a\}(1-\{-x_{ij}+\tilde\Delta_a\})-\{x_{ij}+\tilde\Delta_a\}(1-\{x_{ij}+\tilde\Delta_a\})\right)\nn\\&+\pi i\sum_{a=1}^3\sum_{j=1}^3\left(\lfloor -x_{ij}+\tilde\Delta_a\rfloor-\lfloor x_{ij}+\tilde\Delta_a\rfloor-2y_{ij}(1-\{-x_{ij}+\tilde\Delta_a\}-\{x_{ij}+\tilde\Delta_a\})\right),\\
X\equiv&~\min(\{x_{ij}+\tilde\Delta_a\},1-\{x_{ij}+\tilde\Delta_a\}:\,a=1,2,3,~1\leq
i\neq j\leq3).
\end{align}

It is still difficult to solve (\ref{BAE:highT:N=3}) for general
chemical potentials and therefore we assume that all the chemical
potentials $\tilde\Delta_a$ are identical as $\tilde\Delta_a=1/3$.
Then the complete set of solutions to (\ref{BAE:highT:N=3}) is given
as
\begin{align}
&\left\{x_{21}=x_{32}=\fft{\mathbb Z}{3},~y_{21},y_{32}=\fft{\mathbb
Z}{3}\right\}\cup\left\{0\leq
x_{21}<\fft13,~x_{32}=\fft{1-x_{21}}{2},~y_{21}+2y_{32}=\fft{2\mathbb
Z+1}{6}\right\}\nn\\&\cup\left\{\fft13<x_{21}\leq\fft12,~x_{32}=1-2x_{21},~2y_{21}+y_{32}=\fft{2\mathbb
Z+1}{6}\right\}\cup\left\{\fft13<x_{21}=x_{32}<\fft23,~y_{21}-y_{32}=\fft{2\mathbb
Z+1}{6}\right\}\nn\\&\cup\left\{\fft12<x_{21}<\fft23,~x_{32}=2(1-x_{21}),~2y_{21}+y_{32}=\fft{2\mathbb
Z+1}{6}\right\}\nn\\&\cup\left\{\fft23<x_{21}<1,~x_{32}=1-\fft{x_{21}}{2},~y_{21}+2y_{32}=\fft{2\mathbb
Z+1}{6}\right\}.\label{BAE:highT:N=3:sol}
\end{align}
up to $\mathcal O(e^{-\fft{2\pi\sin\theta}{|\tau|}X})$. Note that
the first subset of (\ref{BAE:highT:N=3:sol}) actually breaks the assumption that $\{x_{ij}+\tilde\Delta_a\}$ are not asymptotically close to 0 or 1, but we know it corresponds to the
standard $\{m,n,r\}$ solutions. The others represent continuous
families of non-standard solutions in the high-temperature limit.

%%%%
\bibliographystyle{JHEP}
\bibliography{indexrefs}

\end{document}